\documentclass[11pt,a4paper]{article}
\usepackage{jheppub}
\usepackage[utf8]{inputenc}
\usepackage{multirow}
\usepackage{amsmath}
\usepackage{tabularx}
\usepackage{amssymb}
\usepackage[dvipsnames]{xcolor}
    \definecolor{darkgreen}{rgb}{0,0.5,0}
    \definecolor{darkblue}{rgb}{0,0,0.6}
    \definecolor{purple}{rgb}{0.4,.2,0.7}
    \definecolor{medblue}{RGB}{0,23,127}
    \definecolor{teal}{RGB}{58,144,100}
\usepackage{graphicx}
\usepackage{subcaption}
\usepackage{caption}
\usepackage{slashed,cancel}
\usepackage[compat=1.1.0]{tikz-feynman}
\usepackage{graphicx,subcaption}
\usepackage[section]{placeins}

\graphicspath{ {./figures/} }
    
\newcommand{\blue}[1]{\textcolor{medblue}{#1}}
\newcommand{\teal}[1]{\textcolor{teal}{#1}}
\newcommand{\violet}[1]{\textcolor{purple}{#1}}

\newcommand{\be}{\begin{equation}}
\newcommand{\ee}{\end{equation}}
\newcommand{\bea}{\begin{eqnarray}}
\newcommand{\eea}{\end{eqnarray}}
\def\ba{\begin{eqnarray}}
\def\ea{\end{eqnarray}}

	\newcommand{\bes}{\begin{equation} \begin{split} }	
	\newcommand{\ees}{\end{split} \end{equation} }

	\usepackage{physics}

\def\Tr{\,{\rm Tr}\,}

\def\b{\beta}

\def\m{\mu}

\newcommand\aNLO{{\sc\small MadGraph5\_\-aMC@NLO}}

\newcommand{\tev}{\,\textrm{TeV}}
\newcommand{\gev}{\,\textrm{GeV}}

\newcommand\UFO{{\sc\small UFO}}

\def\ket{\rangle}

 \usepackage{hhline}
\usepackage{mathtools}
\usepackage{makecell}

\numberwithin{equation}{section}
\renewcommand{\arraystretch}{1.25}
\begin{document}
\title{Quantum properties of $H\to VV^*$: precise predictions in the SM and sensitivity to new physics}
\author[a]{Morgan Del Gratta,}
\author[a,b]{Federica Fabbri,}
\author[a,b]{Priyanka Lamba,}
\author[a,b,c]{Fabio Maltoni,}
\author[\,b]{Davide Pagani}
\affiliation[a]{Dipartimento di Fisica e Astronomia, Universit\`{a} di Bologna, Via Irnerio 46, 40126 Bologna, Italy}
\affiliation[b]{INFN, Sezione di Bologna, Via Irnerio 46, 40126 Bologna, Italy}
\affiliation[c]{Centre for Cosmology, Particle Physics and Phenomenology (CP3), Universit\'{e} Catholique de Louvain, B-1348 Louvain-la-Neuve, Belgium}

\emailAdd{morgan.delgratta@studio.unibo.it}
\emailAdd{federica.fabbri@cern.ch}
\emailAdd{priyanka.lamba2@unibo.it}
\emailAdd{fabio.maltoni@unibo.it}
\emailAdd{davide.pagani@bo.infn.it}

\abstract{We study the quantum properties of the Higgs-boson decays into four fermions via two vector bosons $(H\to VV^*\to 4f)$. In particular, we focus on the case of two different-flavour lepton pairs $(H\to ZZ^*\to \mu^+\mu^- e^+ e^-)$. We compute the quantum-information observables for the corresponding two-qutrit system $(ZZ)$  at next-to-leading order electroweak (NLO EW) accuracy in the SM. We find that NLO EW corrections lead to giant (order 1) effects in some specific cases, significantly altering the extraction of observables quantifying the quantum correlations.  We identify observables that are robust and can be used to extract reliable information. Finally, we discuss possible new physics (NP) effects, parametrised via an effective-field-theory approach. We show how quantum observables can increase the sensitivity to NP also for the process considered in this study.}  
\preprint{\begin{flushright}
COMETA-2025-14
\end{flushright}
}


 \maketitle

\clearpage

\section{Introduction}
\label{sec:Intro}
\label{sec:intro}

The study of quantum information (QI)-inspired  observables in the analysis of particle physics phenomena, particularly in high-energy collisions, has sparked the interest of a rapidly growing community of theorists and experimentalists. Following an initial proposal~\cite{2003.02280}, numerous phenomenological studies exploring quantum correlations across a variety of final states accessible at present and future high-energy colliders have been conducted~\cite{Maltoni:2024tul,Barr:2024djo,Aoude:2022imd,Severi:2022qjy,Fabbrichesi:2024xtq,Fabbrichesi:2022ovb,Altakach:2022ywa,Aoude:2023hxv,Bernal:2023ruk,Fabbrichesi:2023jep,Cheng:2025cuv}. Remarkably, in the past two years the first experimental results on detection of entanglement between the spins of top-antitop pairs were published by the ATLAS and CMS collaborations~\cite{ATLAS:2023fsd, CMS:2024pts, CMS:2024zkc}. These experimental results demonstrate the feasibility of such measurements at colliders and pave the way for similar analyses in other final states. 

Among the many intriguing possibilities, a particularly promising process is Higgs ($H$) decay into a pair of weak bosons, either $ZZ^*$ or $WW^*$ (collectively denoted $VV^*$), which subsequently decay into fermions, resulting in a four-fermion final state. The spins of the massive weak bosons from Higgs decay can be viewed as an (entangled) pair of qutrits. In the SM, therefore, qutrits arise naturally, and their quantum properties can be studied in different final states. It is interesting to note that qutrits hold a unique place in quantum computing due to their higher dimensionality compared to qubits, which gives them special properties such as enhanced error-correction capabilities, greater noise resilience, and longer coherence times~\cite{PhysRevA.64.052109,PhysRevLett.88.040404,PhysRevLett.85.4418, PhysRevA.64.024101,PhysRevA.65.032118}. The first proposal to study QI observables  in this final state, focusing specifically on an operator representing a Bell-type inequality, was presented for $ H \rightarrow WW^{*}  \rightarrow \ell^{+} \nu \ell^{-} \bar{\nu} $ in Ref.~\cite{Barr:2021zcp}. Due to the scalar nature of the Higgs boson and its relatively low mass, the vector bosons are expected to exhibit strong entanglement across the entire phase space—a feature explored in detail in Refs.~\cite{Fabbrichesi:2023cev,Aguilar-Saavedra:2022wam}. Moreover, recent studies indicate that in this scenario, the presence of entanglement is both a necessary and sufficient condition for the violation of Bell inequalities~\cite{Bernal:2024xhm}.
In addition to theoretical efforts, phenomenological studies have examined the feasibility of measuring QI observables at the LHC in $ H \rightarrow WW^{*} $,   including semileptonic final states~\cite{Fabbri:2023ncz}, and in  $ H \rightarrow ZZ^{*} $~\cite{Bernal:2023ruk}. This final state has also been proposed for the investigation of other characteristic property of quantum matter, as the identical particle behaviour~\cite{Aguilar-Saavedra:2024jkj}. These studies also assess the sensitivity of these processes to (heavy) physics Beyond the Standard Model (BSM). Although the studies assume somewhat optimistic detector effects and simplified evaluations of systematic uncertainties, they all clearly suggest that evidence for the entanglement of the diboson state originating from $H$ decay could be established in the current LHC data set, with potential for full observation at the high-luminosity LHC. 

Given that directly measuring the spin of particles produced at the LHC is not feasible, QI-observable measurements rely on quantum tomography (QT) techniques. In these techniques, spin information is statistically inferred by measuring the directions of decay products in specific reference frames. Through multiple measurements, the spin density matrix of the original bipartite state can be reconstructed to a given accuracy. The process for QT in generic diboson final states has been studied in various works ({\it e.g.}, Ref~\cite{Ashby-Pickering:2022umy,Grossi:2024jae,Sullivan:2024wzl,Aguilar-Saavedra:2024whi,Aguilar-Saavedra:2022wam,Aguilar-Saavedra:2024jkj,Fabbri:2023ncz,Bernal:2023ruk,Bernal:2024xhm,Fabbrichesi:2023cev,Fabbrichesi:2023jep,Barr:2021zcp,Barr:2022wyq,Subba:2024mnl,Subba:2024aut}). Notably, the Higgs decay into vector boson pairs allows for a significant simplification of this procedure due to the constraints on the helicity states of the vector bosons, leading to a straightforward spin density matrix structure~\cite{Aguilar-Saavedra:2022wam}. Consequently, simple relations can be established to determine whether the final-state spins in
$H \rightarrow VV^*$ are separable or entangled, greatly simplifying the general analysis of qutrit pairs.

From an experimental perspective, among all possible final state signatures, the $H \rightarrow ZZ^* \rightarrow e^+ e^- \mu^+ \mu^- $ channel stands out due to several advantages. First, this system can be fully and, in general, very accurately reconstructed using the four charged leptons in the final state. Second, it has relatively low background, primarily from non-resonant $ZZ$ production. 

An LHC analysis aimed at measuring QI observables in this final state would start by reconstructing the four charged leptons, without direct access to the intermediate states. The reconstructed events would include several factors beyond the leading-order $H \rightarrow ZZ^* \rightarrow e^+ e^- \mu^+ \mu^- $ process, such as higher-order diagrams with real photon emissions as well as loop contributions, which can also not feature any $Z$ intermediate state.

All previous work on the $H\rightarrow ZZ^{*}$ process have considered only the tree-level contributions, with the exception of Ref.~\cite{Grossi:2024jae},  where higher-order effects coming from electroweak (EW) corrections have been determined. 
As a first step, we quantify the NLO effects on the coefficients needed to determine the spin density matrix and the observables related to the entanglement and Bell inequality operator. Observing uneven and large effects on some terms of the matrix, the goal of the paper is also to investigate the cause of this behaviour, the solidity of the QT approach, and to suggest observables robust against NLO EW corrections.

A second goal of this work is to establish the sensitivity of QI observables to new physics (NP), as is done in Refs.~\cite{Fabbrichesi:2023jep,Sullivan:2024wzl}.  To this aim, we study the structure of the spin density matrix obtained from the most general effective interaction between the Higgs boson and two vector bosons. This allows us to estimate the sensitivity of elements of the spin density matrix and QI-inspired observables to various modifications of the $HVV$ couplings. We also compare NP effects with the NLO EW corrections on the same quantities, to assess the need to include these corrections in potential NP interpretations of measurements performed in this sector.

This paper is organised as follows: in Sec.~\ref{sec:QIintro}, we introduce the parametrisation used to study the diboson spin state using QI principles and techniques, including QT and the definition of a suitable set of QI observables. Section~\ref{sec:Lodetails} describes the numerical simulation setup, analysis procedures, and LO results. In Sec.~\ref{sec:NLOeffects}, we compute the NLO EW corrections and study their impact on the spin density matrix and QI observables. Section~\ref{sec:NP} examines the structure of the spin density matrix in the presence of BSM effects, focusing on generic modifications of the $HVV$ couplings that are induced by new heavy NP states. We summarise our findings and present our conclusions in Sec.~\ref{sec:conclusions}.

\section{QI analysis of $H\rightarrow VV^*$}
\label{sec:QIintro}
\label{QIintro}

As anticipated in Sec.~\ref{sec:intro}, in this work we study the $ H \rightarrow ZZ^* \rightarrow e^+ e^- \mu^+ \mu^- $ decay channel and its properties in the context of QI. In this section we describe the theoretical framework already used in, {\it e.g.}, Refs.~\cite{Aguilar-Saavedra:2022wam,Grossi:2024jae}  for modelling this process and more in general the  $ H \rightarrow VV^* \rightarrow f_1 \bar f_2 f_3 \bar f_4$ class. We restrict the discussion to the LO picture, which is based  on only tree-level diagrams. The complications arising at NLO are then discussed in Sec.~\ref{sec:NLOeffects}.

In Sec.~\ref{sec:formal} we summarise theoretical aspects related to a composite spin-0 system stemming from two qutrits, {\it i.e.} the $H\to VV^*$ intermediate step of the process of interest.  In  Sec.~\ref{QMT} we discuss how to derive information on the polarisation of the $V$ boson and we connect the formalism of Sec.~\ref{sec:formal} with quantities extracted via the QT, which relies on the measurement of angular distributions of the final state $f_1 \bar f_2 f_3 \bar f_4$. In Sec.~\ref{sec:entan} we discuss the necessary/sufficient conditions for entanglement, and we provide the definitions for the upper and lower bounds on the concurrence. In Sec.~\ref{sec:IntroBell} we describe Bell-type inequalities for our system and we define a suitable operator to test them.

\subsection{The $VV^*$ system at LO}
\label{sec:formal}

In the $H\to VV^*$ decay the spin of each $V$ boson is a physical realisation of a qutrit, which is a quantum object characterised by three possible states. In this case they correspond to the spin-one projections on a given axis. First of all, we notice that in such decay, at least one of the two $V$ bosons has to be off-shell since the mass of the Higgs boson ($m_H$) is smaller than the sum of the two $V$-boson masses ($m_V$), for both the cases $V=W$ and $V=Z$. In principle, an off-shell vector boson carries a spin-0 component, see {\it e.g.}~Ref.~\cite{Korner:2014bca}. However, when it is coupled to massless fermions such component vanishes. Since in the paper we will focus on the $ H \rightarrow ZZ^* \rightarrow e^+ e^- \mu^+ \mu^- $ channel and in general we will treat always the fermions as massless, we can consider both  on-shell and off-shell bosons as spin-1. 

\medskip

The bipartite quantum system formed by a pair of massive vector bosons can be described by a density operator $\rho$ acting on the Hilbert space defined by the tensor product of spin states of each vector boson.  
The density operator $\rho$ can be parametrised in several different ways for a composite state formed by pair of particles with generic spin $S$. In the polarisation operator basis, which is employed in this work, the basic objects are the polarisation operators $T_{L,M}(S)$. They are irreducible tensors of rank $L$, with indices $L=0, ..., 2S$ and $M=-L, ..., L$. For a given $S$ there are therefore $\sum_{L=0}^{2S}(2L+1)=(2S+1)^2$ of such tensors acting on spin functions, and their general form can be derived from the equation  \cite{Varshalovich:1988ifq,Bertlmann:2008jgt,kryszewski2006alternativerepresentationntimes}
\be
T_{L,M}(S)=\sqrt{\frac{2L+1}{2S+1}}\sum_{m,m'=S,\dots,-S}C^{Sm}_{Sm',LM}|Sm\rangle \langle Sm'| \,,
\label{tope}
\ee 
where  $C^{Sm}_{Sm',LM}$ are Clebsch-Gordan coefficients.

In the case $S = 1$, {\it i.e.} the one corresponding to qutrits, the $T_{L,M}(1)$ tensors, which from now on we will simply denote as   $T_{L,M}$, can be written as $3\times3$ matrices via the spin-1 operators  $J_x$, $J_y$ and $J_z$. In particular, 
\begin{equation} 
T_{1,\,\pm 1} = \mp \sqrt{\frac{3}{2}} (J_x \pm i J_y)\,,\qquad T_{1,\,0} = \sqrt{\frac{3}{2}} J_z\,,
\end{equation}
where a factor $\sqrt{3}$ was added for normalisation purposes, and
\begin{gather}
T_{2, \,\pm 2} = \frac{2}{\sqrt{3}} (T_{1,\,\pm 1})^2\,, \\
T_{2,\,\pm 1} = \sqrt{\frac{2}{3}} \left[ T_{1,\,\pm 1} T_{1,\,0} + T_{1,\,0} T_{1,\,\pm 1} \right]\,, \\
T_{2,\,0} = \frac{\sqrt 2}{3} \left[ T_{1,\,1} T_{1,\,-1} + T_{1,\,-1} T_{1,\,1} + 2 (T_{1,\,0})^2 \right] \, .
\end{gather}
Explicitly, the $T_{L,M}$ matrices can be written as
\begin{equation}
T_{1,\,1} = \sqrt{\frac{3}{2}}
\begin{pmatrix}
0 & -1 & 0 \\
0 & 0 & -1 \\
0 & 0 & 0
\end{pmatrix}, \quad
T_{1,\,0} = \sqrt{\frac{3}{2}}
\begin{pmatrix}
1 & 0 & 0 \\
0 & 0 & 0 \\
0 & 0 & -1
\end{pmatrix}, \quad
T_{1,\,-1} = \sqrt{\frac{3}{2}}
\begin{pmatrix}
0 & 0 & 0 \\
1 & 0 & 0 \\
0 & 1 & 0
\end{pmatrix},
\end{equation}
\begin{equation}
T_{2,\,2} = \sqrt{3}
\begin{pmatrix}
0 & 0 & 1 \\
0 & 0 & 0 \\
0 & 0 & 0
\end{pmatrix}, \quad
T_{2,\,-2} = \sqrt{3}
\begin{pmatrix}
0 & 0 & 0 \\
0 & 0 & 0 \\
1 & 0 & 0
\end{pmatrix}, \quad
T_{2,\,1} = \sqrt{\frac{3}{2}}
\begin{pmatrix}
0 & -1 & 0 \\
0 & 0 & 1 \\
0 & 0 & 0
\end{pmatrix}, \nonumber
\end{equation}
\begin{equation}
T_{2,\,-1} = \sqrt{\frac{3}{2}} \begin{pmatrix}
0 & 0 & 0 \\
1 & 0 & 0 \\
0 & -1 & 0
\end{pmatrix}, \quad
T_{2,\,0} = \frac{1}{\sqrt{2}} \begin{pmatrix}
1 & 0 & 0 \\
0 & -2 & 0 \\
0 & 0 & 1
\end{pmatrix} \, .
\end{equation}
\smallskip
As can be noted, the $T_{L,M}$ matrices are normalised such that 
\be
\text{Tr}\,[ T_{L,M} (T_{L',M'})^\dagger ] = 3\,, \label{eq:trace3}
\ee 
and they satisfy the relation 
\be 
(T_{L,M})^\dagger = (-1)^M T_{L,-M}\,.
\ee
The term $T_{0,0}$ corresponds to the $3\times 3$ identity matrix $\mathbf{1}_3$. 

\medskip

Using these operators, the spin density matrix of the two qutrits can be parameterised as 
\begin{equation}
\rho = \frac{1}{9} \left[\mathbf{1}_3 \otimes \mathbf{1}_3 + A^a_{L,M} (T_{L,M} \otimes \mathbf{1}_3 )+ A^b_{L,M} (\mathbf{1}_3 \otimes T_{L,M}) + C_{L_a,M_a,L_b,M_b} (T_{L_a, M_a} \otimes T_{L_b, M_b}) \right],
\label{eq:rhoexp}
\end{equation}
where in Eq.~\eqref{eq:rhoexp} a sum in $L = 1, 2$ and $-L \leq M \leq L$, and similarly for $L_{a}(L_{b})$ and $M_{a}(M_{b})$, is understood. 
Since $\mathbf{1}_3 = T_{0,0}$, it is easy to see that the $A$ coefficients can also be denoted as $A^a_{L,M}=C_{L,M,0,0}$,  $A^b_{L,M}=C_{0,0,L,M}$, and that $A^a_{0,0}=A^b_{0,0}=C_{0,0,0,0}=1$, which is fixed by the requirement $\rm Tr[\rho]=1$. 

The general form of $\rho$, expressed in terms of $A_{L,M}$ and $C_{L,M,L,M}$, can be found in Appendix \ref{sec:rhogeneral}. Before calculating the explicit expression of $\rho$ for the $H\to VV^*$ decay at LO it is interesting to see the relations imposed by symmetries or other properties that must be satisfied by $\rho$.
First of all, the spin density matrix has to be hermitian, and this condition leads to the following constraints on the coefficients of the expansion: 
\begin{eqnarray}
(A_{L,M}^{j})^*&=& (-1)^M A^j_{L,-M} \quad{\rm with}\quad j=a,b\,, \nonumber \\
C_{L_a,M_a,L_b,M_b}&=&(-1)^{M_a+M_b}(C_{L_a,-M_a,L_b,-M_b})^{*} \, .
\label{hermitianprop}
\end{eqnarray}
With such constraints there are altogether 80 independent real parameters (81 parameters minus $C_{0,0,0,0}$) for the $9 \times 9$ matrix.  

Then, we notice that the $VV^*$ pair, stemming from a scalar, cannot have an orbital angular momentum component along the axis formed by the two $V$-boson tri-momenta, see also Ref.~\cite{Aguilar-Saavedra:2024whi}. This is precisely the axis defining the helicity basis (see later Sec.~\ref{sec:LO}). In addition, the projections of the spins of the two $V$ bosons have to sum to zero, {\it i.e.}, $M_a+M_b=0$. From this condition we find 
\be
C_{L_a, M_a, L_b, M_b} \neq 0 \quad \Longleftrightarrow \quad M_a=-M_b \, ,
\ee
and
\be
A_{L, M} \neq 0 \quad \Longleftrightarrow \quad M=0 \, .
\ee
This reduces the number of non-vanishing independent real parameters to 18, corresponding to 9 $C$ coefficients, of which 5 complex and 4 real, and 4 $A$ coefficients, all real. 
The condition is equivalent to assuming a cylindrical symmetry along the $V$-pair axis, which is also observed in the angular distributions of the decay products of the $V$ bosons.

Other constraints can be imposed if the system, and therefore the $\rho$ matrix, is symmetric under the exchange of one $Z$-boson with the other. We will denote in the following such symmetry as ``up-down'' symmetry, which leads to 
\be
C_{L_a, M_a, L_b, M_b} =C_{L_b, M_b, L_a, M_a}\,,
\ee
and
\be
A^a_{L, M} =A^b_{L, M}\,. 
\ee
These two relations additionally reduce the number of non-vanishing independent real parameters to 13, associated to 7 $C$ coefficients, of which 4 complex and 3 real, and 2 $A$ coefficients, both real.

Furthermore, if parity is conserved, the sum of $L_1+L_2$ must always be even or odd, depending on the intrinsic parity eigenstate of the Higgs boson, respectively either 1 or $-1$.
In the former case, corresponding to the SM Higgs, this implies that the number of non-vanishing independent real parameters further reduces to 9, and we obtain 5 $C$ coefficients ($C_{1,1,1,-1}$, $C_{1,0,1,0}$, $C_{2,0,2,0}$, $C_{2,-1,2,1}$, $C_{2,2,2,-2}$), of which 3 complex and 2  real, and 1 $A$ coefficient ($A^2_{2,0}$) that is real. In the latter case, the number of independent parameters becomes 4, 2 $C$ coefficients  ($C_{1,1,2,-1},C_{2,0,1,0}$), of which one complex and one real, and $A^2_{1,0}$, that is real.

The requirement that CP is conserved then leads to only real coefficients and therefore, in the case of $H$ being $P$-even, to  6 non-vanishing independent coefficients: 5 $C$ coefficients  and 1 $A$ coefficient, all real.

\medskip

When considering the $H\rightarrow V V^*$ decay in the SM, the number of non-vanishing independent coefficients is very small and the texture of $\rho$ very simple. As shown in Ref.~\cite{Aguilar-Saavedra:2022wam}, the state of the $V V^*$ system, for fixed values of the invariant mass of $V_a$ and $V_b$, can be written for the LO $H\rightarrow V V^*$ decay in the SM as
\be  
\label{state}
|\psi\rangle= a_L |0 \,0\rangle + a_T \frac{|+ -\rangle + |- +\rangle}{\sqrt{2}}\,,
\ee 
where ($+, 0, -$) are three spin-polarisation of the two $V$ bosons and 
\ba
\label{betadef}
a_L= \frac{-\beta}{\sqrt{2+\beta^2}} \,,\quad\quad a_T=\frac{\sqrt{2}}{\sqrt{2+\beta^2}}\,,\quad \quad \beta=1+\frac{m_H^2-(m_a+m_b)^2}{2m_a m_b}\,.
\ea
The quantity $\beta$ is related to the velocities of the two $V$ bosons with invariant masses $m_a$ and $m_b$. \footnote{$\beta$ is the scalar product of the two four-velocities for the vectors $V_a$ and $V_b$, generalised to the off-shell case:  $p_a\cdot p_b/(m_a m_b)= \frac{\vec{p}^2}{m_a m_b}+\frac{E_a E_b}{m_a m_b}$, where $\vec{p}$ is the three-momentum of $V_a$ or $V_b$ in the Higgs rest frame and $E_a$ and $E_b$ are their energies, respectively. It can be rewritten in terms of the standard velocities  $v_i$ as $\beta=\frac{1+v_a v_b}{\sqrt{(1-v_a^2)(1-v_b^2)}}$.} We notice that for $m_a+m_b\rightarrow m_H$ the system is at rest, $\beta \rightarrow 1$, and the system is maximally entangled, $|\psi\rangle= 1/\sqrt{3} ( -|0 \,0\rangle + |+ -\rangle + |- +\rangle)$.

The $\rho$ matrix can then be simply constructed as the operator $|\psi\rangle \langle \psi|$, which leads to:

\be
\label{beta_rho}
\rho_{\rm LO}(\beta)=\left(
\setlength\arraycolsep{6.5pt}\begin{array}{ccccccccc}
 \cdot & \cdot & \cdot & \cdot & \cdot & \cdot & \cdot & \cdot & \cdot \\
 \cdot & \cdot & \cdot & \cdot & \cdot & \cdot & \cdot & \cdot & \cdot \\
 \cdot & \cdot & \frac{a_T^2}{2} & \cdot &\frac{a_L a_T}{\sqrt 2} & \cdot & \frac{a_T^2}{2} & \cdot & \cdot \\
 \cdot & \cdot & \cdot & \cdot & \cdot & \cdot & \cdot & \cdot & \cdot \\
 \cdot & \cdot & \frac{a_L a_T}{\sqrt 2}
   & \cdot &a_L^2 & \cdot &\frac{a_L a_T}{\sqrt 2} & \cdot & \cdot \\
 \cdot & \cdot & \cdot & \cdot & \cdot & \cdot & \cdot & \cdot & \cdot \\
 \cdot & \cdot & \frac{a_T^2}{2} & \cdot & \frac{a_L a_T}{\sqrt 2} & \cdot &\frac{a_T^2}{2} & \cdot & \cdot \\
 \cdot & \cdot & \cdot & \cdot & \cdot & \cdot & \cdot & \cdot & \cdot \\
 \cdot & \cdot & \cdot & \cdot & \cdot & \cdot & \cdot & \cdot & \cdot \\
\end{array}
\right)\,,
\ee
\renewcommand{\arraystretch}{1}
where the dots correspond to entries that are equal to zero.
 
In other words, if $m_a$ and $m_b$ are fixed there is only one independent non-vanishing coefficient and it depends on $\beta$. All the other can be derived from it.  Moreover, the matrix $\rho_{\rm LO}(\beta)$ corresponds to a pure state, and has indeed only one non-vanishing eigenvalue equal to one.

In fact, the situation is a bit more involved than this. First, in the $ H \rightarrow VV^* \rightarrow e^+ e^- \mu^+ \mu^- $ decay different pair of values $(m_a,m_b)$ contribute and with different weights ($w(\beta)$), such that
\be
\rho_{\rm LO}=\int \rho_{\rm LO}(\beta) w(\beta) d\beta\,.
\ee
For this reason, the system is not pure anymore, and the coefficients $C$ and $A$ also receive contributions from different values of $\beta$ with different weights $w(\beta)$.
Second, and most importantly, the spin of the $V$ bosons cannot be directly measured at the LHC and in the envisaged future colliders. This information has to be extracted via the analysis of the decay product of the $V$ bosons. This method, called QT, is discussed in the next section for the process studied in this work.

Nevertheless, we can already infer the form of $\rho_{\rm LO}$ noticing that $a_T^2$ and $a_L^2$ in $\rho_{\rm LO}(\beta)$ are related via the linear equation $a_T^2+a_L^2=1$, which is required to ensure that $\rm Tr[\rho]=1$. This means that such relation is preserved in $\rho_{\rm LO}$, after integrating over $\beta$, for all matrix entries that depend solely on $a_T^2$ or $a_L^2$. In contrast, the quantity $\frac{a_L a_T}{\sqrt{2}}$ is not linearly related to either $a_L^2$ or $a_T^2$, and therefore the identities $a_L a_T = a_L \sqrt{1 - a_L^2} = a_T \sqrt{1 - a_T^2}$ are not expected to hold after integration for the corresponding matrix elements. As a result, the density matrix $\rho_{\rm LO}$ takes the form

\renewcommand{\arraystretch}{0.8} 
\be
\label{rhovv}
\rho_{\rm LO}=\left(
\setlength\arraycolsep{6pt}
\begin{array}{ccccccccc}
\cdot & \cdot & \cdot & \cdot & \cdot & \cdot & \cdot & \cdot & \cdot \\
 \cdot & \cdot & \cdot & \cdot & \cdot & \cdot & \cdot & \cdot & \cdot \\
 \cdot & \cdot & {x} & \cdot & {y}  & \cdot & {x}  & \cdot & \cdot \\
 \cdot & \cdot & \cdot & \cdot & \cdot & \cdot & \cdot & \cdot & \cdot \\
 \cdot & \cdot & {y} 
   & \cdot & {1-2x}  & \cdot &{y}  & \cdot & \cdot \\
 \cdot & \cdot & \cdot & \cdot & \cdot & \cdot & \cdot & \cdot &
 \cdot \\
 \cdot &
 \cdot & {x}  & \cdot & {y}  & \cdot &{x}  & \cdot & \cdot \\
 \cdot & \cdot & \cdot & \cdot & \cdot & \cdot & \cdot & \cdot & \cdot \\
 \cdot & \cdot & \cdot & \cdot & \cdot & \cdot & \cdot & \cdot & \cdot \\

\end{array}
\right)\,,
\ee
and depends only on two  parameters.
 The system is not pure anymore, and the number of non-vanishing eigenvalues increases from one to two. The purity condition is equivalent to $y^2=x(1-2x)$, as in $\rho_{\rm LO}(\beta)$. Moreover, in order to have $\rho_{\rm LO}$ semi-positive defined, {\it i.e.} with eigenvalues equal or larger than zero, the condition $y^2\leq x(1-2x) \Rightarrow 0\leq x\leq 1/2$  has to be satisfied.

\subsection{Quantum tomography} \label{QMT}
As already anticipated, the spin of particles produced in high-energy collisions cannot be measured by the multi-purpose detectors located around the interaction points. The current detectors are designed to provide a precise measurement of the final-state particles' momenta, especially for isolated charged particles.
However, due to the nature of the electroweak interaction, the information on the spin of the vector boson is either entirely or partially transmitted to the direction of the daughter particles. As a consequence, the spin of the parent particle can be reconstructed on a statistical basis by measuring angular distributions of the final state particles and averaging on multiple events.
This procedure is called, as already said, Quantum-State-Tomography or simply Quantum-Tomography (QT), and allows to reconstruct the full polarization/spin density matrix of the pair of vector bosons, when it is only possible to measure directly the momenta and charge of the final state fermions. In the case of $H\to VV^*$,  the cleanest possible final state is $H\rightarrow ZZ^{*}\rightarrow\ell^{+}\ell^{-}\ell^{+}\ell^{-}$, with $\ell=\mu^\pm $ or $e^\pm$. In particular, as already mentioned, in this paper we consider the case of different-flavour lepton pairs: $ H \rightarrow ZZ^* \rightarrow e^+ e^- \mu^+ \mu^- $. 
This final state simplifies the task of pairing the leptons in order to reconstruct the associated $Z$ boson, as it will be discussed in Sec.~\ref{sec:LO}. Considering the same-flavour case, the pairing of the leptons is ambiguous and additional quantum effects appear, as discussed in Ref.~\cite{Aguilar-Saavedra:2024jkj}.

QT techniques have been discussed extensively in the literature~\cite{White_1999,PhysRevA.64.052312,Thew:2002fom}.  Employing specific parameterisations, they allow to extract the density matrix of a quantum state from experimental data. In particular, in the polarisation operator basis the coefficients $A$ and $C$ of Eq.~\eqref{eq:rhoexp} can be extracted via QT from the angular distributions of the decay products of the two $Z$ bosons.

This approach can be used in general for any $H\rightarrow V V^* \to 4f$ decay, therefore in the following we refer to generic vector bosons and use the symbols $V_{a}$ and $V_{b}$. The normalised joint angular distribution of the final decay products can be written as \cite{Rahaman:2021fcz,Boudjema:2009fz}:
\ba
\label{differntialdistribution}
\frac{1}{\sigma}\frac{d\sigma}{d\Omega_a d\Omega_b}&=&\frac{2S_a+1}{4\pi}\frac{2S_b+1}{4\pi}\sum_{\lambda_a,\lambda^{'}_{a},\lambda_b,\lambda^{'}_{b}}\rho(\lambda_a,\lambda^{'}_{a},\lambda_b,\lambda^{'}_{b})\Gamma_{a}(\lambda_a,\lambda^{'}_{a})\Gamma_{b}(\lambda_b,\lambda^{'}_{b})\nonumber\\
&=&\left(\frac{3}{4\pi}\right)^2 \Tr[\rho(\Gamma_a\otimes\Gamma_b)^T]\,.
\ea 
The quantities entering Eq.~\eqref{differntialdistribution} are defined as follows.
The term $S_{a}(S_{b})$ is the spin of particle $`a(b)$', $d\Omega_a=\sin\theta_a d\theta_a d\phi_a$ and the angles $(\theta_a, \phi_a)$ are the polar coordinates of the three-momentum of $f_1$ from the $V_a$ decay ($V_a\rightarrow f_1 \bar f_2$), in the $V_a$ rest frame. Similarly, $(\Omega_b,\theta_b, \phi_b)$ correspond to the coordinates of three-momentum of $f_3$ from the $V_b$ decay ($V_b\rightarrow f_3 \bar f_4$) in the $V_b$ rest frame.
The sum runs over all possible helicities ($-S_{a(b)}\leq\lambda_{a(b)}$,$\lambda_{a(b)}'\leq S_{a(b)}$) of the particles $V_a,V_b$,  and $\sigma$ is the total cross section of $V_{a}$ and $V_{b}$ production, via the Higgs decay, followed by their decays themselves. It can be written as
\ba
\label{eq:fromsigmatoGamma}
\sigma=\sigma_{pp \rightarrow H}\times {\rm Br}(H\rightarrow V_a V_b \rightarrow f_1 \bar f_2 f_3 \bar f_4)\,,
\ea
where $\sigma_{pp \rightarrow H}$ is the Higgs production mode considered and ${\rm Br}(H\rightarrow V_a V_b \rightarrow f_1 \bar f_2 f_3 \bar f_4)$ the branching ratio of the Higgs into the fermions considered.
Finally, the term denoted as $\rho(\lambda_a,\lambda^{'}_{a},\lambda_b,\lambda^{'}_{b})$ is the normalised production spin density matrix and $\Gamma_{a}(\lambda_a,\lambda^{'}_{a})$ is the decay density matrix of $V_{a}$, normalised to unit trace. 

To align with the common notation already adopted in the literature and avoid a redundancy of symbols, in this section we refer to the cross section $\sigma$ in Eq.~\eqref{differntialdistribution}. However, the focus of the present work is solely on the Higgs decay. In fact, as can be easily seen in Eq.~\eqref{eq:fromsigmatoGamma},\footnote{Equation \eqref{eq:fromsigmatoGamma} is derived in the Narrow-Width-Approximation, which is a very efficient approximation for the case of the Higgs boson, since the  $\Gamma_H/m_H$ ratio,  where $\Gamma_H$ is the total Higgs decay width, is very small.  } because the Higgs boson is a scalar, the quantity $\sigma_{pp \rightarrow H}$ simplifies in the ${\rm d}\sigma/\sigma$ ratio, which can therefore be rewritten as ${\rm d Br}/ {\rm Br} = \text{d}\Gamma_H/\Gamma_H$. Here, $\Gamma_H$ refers \textit{not} to the decay density matrix of either $V$ boson, but to the partial decay width of the full process $H \rightarrow V_a V_b \rightarrow 1,2,3,4$. This explains the use of the symbol $\sigma$ in this section. Starting from Sec.~\ref{sec:LO}, we will instead adopt the notation $\Gamma_H$ directly.

\medskip

At LO the general form of $\Gamma_{a}(\lambda_a,\lambda^{'}_{a})$, {\it i.e.}~the density matrix for a spin-1 particle decaying into two spin-1/2 particles, is known, and reads~\cite{Rahaman:2021fcz,Boudjema:2009fz}:
\begin{eqnarray}
&&(\Gamma_a(\lambda_a,\lambda_a^{'}))_{V_a\rightarrow f_1 \bar f_2}= \label{pmatrix}\\
&&\begin{pmatrix}
       \frac{1+\eta+(1-3\eta)\cos^2\theta_a+2\alpha\cos\theta_a}{4}  & \frac{\sin\theta_a(\alpha+(1-3\eta)\cos\theta_a)}{2\sqrt{2}}e^{i\phi_a} & \frac{(1-3\eta)(1-\cos^2\theta_a)}{4} e^{2i\phi_a}\\
      \frac{\sin\theta_a(\alpha+(1-3\eta)\cos\theta_a)}{2\sqrt{
2}}e^{-i\phi_a} &  \eta+(1-3\eta)\frac{\sin^2\theta_a}{2} & \frac{\sin\theta_a(\alpha-(1-3\eta)\cos\theta_a)}{2\sqrt{2}}e^{i\phi_a}\\
     \frac{(1-3\eta)(1-\cos^2\theta_a)}{4} e^{-2i\phi_a} & \frac{\sin\theta_a(\alpha-(1-3\eta)\cos\theta_a)}{2\sqrt{2}}e^{-i\phi_a}   & \frac{1+\eta+(1-3\eta)\cos^2\theta_a-2\alpha\cos\theta_a}{4}
\end{pmatrix}\,. \nonumber
\end{eqnarray}
In Eq.~\eqref{pmatrix} the angle $\theta_a$ and $\phi_a$ are the polar and azimuthal angle of the decay particle $f_1$ in the rest frame of the parent particle $V_a$, while $\eta$ and $\alpha$ are two additional parameters,\footnote{The quantity $\alpha$ corresponds to $-\eta_\ell$ in Ref.~\cite{Aguilar-Saavedra:2022wam}. } whose definition can be found in Refs.~\cite{Rahaman:2021fcz,Boudjema:2009fz}. In particular, assuming dimension-4 interactions, parameterised as $\Bar{f}_1\gamma_\m(c_L P_L+c_R P_R)f_2 \,V_a^\m$ with $P_{R/L}\equiv (1\pm\gamma_5)/2$, and massless fermions, the two parameters read~\cite{Rahaman:2021fcz}:
\ba
\label{alphavalue}
\eta=0\,,\qquad \alpha=\frac{c_R^2-c_L^2}{c_R^2+c_L^2}\,.
\ea
The quantity $\alpha$ is also called the spin-analysing power of $f_1$. If one considers $V=W^{\pm}$, then $\alpha=-1$\footnote{The spin-analysing power for the anti-particle $\bar f_2$ is $\alpha=1$.  }. 
In the case $V=Z$, the value of $\alpha$ depends on the fermion considered. Different cases, for massless fermions, are reported for the SM in Tab.~\ref{tab:alphavalue}, together with the values of  $c_R^2$ and $c_L^2$.\footnote{Numerical values have been obtained by setting $\sin^2\theta_W=$ 0.222247, consistent with the input parameters (see Eq.~\eqref{eq:inputLOandNLO}) used for LO and NLO results presented in this paper.} 
  
We want to stress here a point that will be crucial in the discussion of the NLO results in Sec.~\ref{sec:NLOeffects}. In the case of electrons, muons and taus the value of $\alpha$ is particularly small due to accidental cancellations. Indeed, $\sin^2\theta_W\simeq 0.25$ and in particular for $\sin^2\theta_W = 0.25$ we obtain $c_L^2=c_R^2$ and in turn $\alpha=0$.  For the same reason, which is accidental and not protected by any symmetry, $\alpha$ is very sensitive to variation of the input parameters or to possible corrections. For instance, it is strongly dependent on the value of  $\sin^2\theta_W$: 1\% difference around the value used in this work leads to $\approx 11$\%  effects on $\alpha$. As already anticipated, this fact will have consequences for the case of NLO EW corrections.

\begin{table}[!t]
\renewcommand{\arraystretch}{1.5}
    \centering
    \begin{tabular}{c|c|c|c}
       $f$  &  $c_L$ & $c_R$ &$\alpha$ \\
       \hline
       $\nu$  & $\frac{1}{2}$ & 0 & $-1$\\
       \hline
      $e$   & $\frac{-1+2\sin^2\theta_W}{2}$ & $\sin^2\theta_W$ & $-0.219$\\
      \hline
      $u$   &$\frac{1}{2}-\frac{2}{3}\sin^2\theta_W$  &$-\frac{2}{3}\sin^2\theta_W$  & $-0.699$ \\
      \hline
      $d$   & $-\frac{1}{2}+\frac{1}{3}\sin^2\theta_W$  & $\frac{1}{3}\sin^2\theta_W$ & $-0.941$\\
    \end{tabular}
    \caption{Spin-analysing power $\alpha$ of the $Z$ decay for different fermions in the SM.}
    \label{tab:alphavalue}
\end{table}

In the case of massless fermions the density matrix $\Gamma_a$ can be written in terms of the standard spherical harmonics  $Y_{L_a,M_a}(\theta_a,\phi_a)$ as
\be
\Gamma_a= \frac{1}{3} \left[\mathbf{1}_3 + B^a_1 (T_{1,M_a}  Y_{1,M_a}) + B^a_2 (T_{2,M_a}  Y_{2,M_a}) \right] \, ,\label{eq:GammaYs}
\ee
where we have dropped the explicit dependence on $(\theta_a,\phi_a)$ and the quantities $B^a_1$ and $B^a_2$ are defined as
\be
 B^a_1=  \sqrt{2\pi} \alpha_{a} \,,\qquad B^a_2 = \sqrt{\frac{2\pi}{5}}\, \label{eq:Bcoeffs}.
\ee
Only $B_1^a$ depends on the value of $\alpha$ for the specific type of fermions $f_1\bar f_2$ associated to $V_a$, and has been denoted as $\alpha_{a}$ in Eq.~\eqref{eq:Bcoeffs}. Using the same convention, the value of $\alpha$ for $f_3\bar f_4$ associated to $V_b $ can be denoted as $\alpha_{b}$. Via Eq.~\eqref{eq:trace3} we obtain 
\be
\label{tarces}
\Tr[{\mathbf 1}_3 \Gamma^T]=2\sqrt{\pi}Y_{0,0}\,,\quad \Tr[T_{1,M_a} \Gamma^T]=B^a_1 Y_{1,M_a}\,,\quad \Tr[T_{2,M_a} \Gamma^T]=B^a_2 Y_{2,M_a}\,.
\ee
The identity matrix in Eq.~\eqref{eq:GammaYs} can also be written as 
\be
\mathbf{1}_3 = B^a_0 (T_{0,0}Y_{0,0})\, ,
\ee
where
\be
 B^a_0=2\sqrt{\pi}=1/Y_{0,0} \,.
 \ee

It is therefore possible to rewrite Eq.~\eqref{differntialdistribution} in terms of  spherical harmonics  and $A$, $C$ and $B$ coefficients using Eqs.~\eqref{eq:rhoexp} and \eqref{tarces} :
\begin{multline}
\frac{1}{\sigma}\frac{d\sigma}{d\Omega_a d\Omega_b}=\frac{1}{(4\pi)^2}[1+A^a_{L,M} B_L^a  Y_{L,M}(\theta_a, \phi_a)+A^b_{L,M}  B_L^b Y_{L,M}(\theta_b,\phi_b) \\
+C_{L_a, M_a, L_b, M_b} B_{L_a}^a B_{L_b}^b Y_{L_a, M_a}(\theta_a,\phi_a)Y_{L_b, M_b}(\theta_b,\phi_b)]\,. 
\label{eqn:sigma}
\end{multline}
The above equation shows the relation between the $A$ and $C$ coefficients and the joint angular distribution of the decay particles ($f_1, f_3$). This relation allows to compute the full spin density matrix integrating over the spherical harmonics, that are functions of the measurable angles $\theta$ and $\phi$.
In particular, the fundamental property of the harmonics is that they are an orthonormal basis 
\be
\int d\Omega \, Y_{L,M}Y^*_{L',M'}=\delta_{L,L'} \delta_{M,M'} \,.
\ee 
Exploiting this property we obtain
\begin{gather}
\int \frac{1}{\sigma} \frac{d\sigma}{d\Omega_a d\Omega_b} Y^{*}_{L,M} (\Omega_j) \, d\Omega_a d\Omega_b = \frac{B_L^j}{4\pi} A^j_{L,M} \quad{\rm with}\quad j=a,b \label{eqn:coeff_defA}\,,\\
\int \frac{1}{\sigma} \frac{d\sigma}{d\Omega_a d\Omega_b} Y^{*}_{L_a,M_a}(\Omega_a) Y^{*}_{L_b,M_b}(\Omega_b) \, d\Omega_a d\Omega_b = \frac{B_{L_a}^a B_{L_b}^b}{(4\pi)^2} C_{L_a, M_a, L_b, M_b}\,, 
\label{eqn:coeff_defC}
\end{gather}
where in Eq.~\eqref{eqn:coeff_defA} we have used the fact that $\int d\Omega=4\pi=\int d\Omega |Y_{0,0}|^2 (B_0)^2$. \footnote{In a realistic experimental scenario the integral in $d\Omega$ cannot be performed over $4\pi$ due to the unavoidable cuts that are present in the analysis. This aspect is beyond the scope of the paper and therefore left for future studies. }

At this point we clearly see how the QT works. First, by measuring the angular distributions of $H\rightarrow V_a V_b \rightarrow f_1 \bar f_2 f_3 \bar f_4$ we can calculate the l.h.s.~of Eq.~\eqref{eqn:coeff_defA} and \eqref{eqn:coeff_defC}. Then, {\it assuming} that the decay is the one parameterised by the density matrix $\Gamma$, we can derive the coefficients $A$ and $C$ and finally build the $\rho_{\rm LO}$ matrix via Eq.~\eqref{eq:rhoexp}. However, we know already the structure that $\rho_{\rm LO}$ must have, see Eq.~\eqref{rhovv}, and this implies that there are only two independent coefficients and that
\ba
\label{vecrho}
x&\equiv&\frac{C_{2,2,2,-2}}{3}\, , \label{eq:xexplicit}\\
y&\equiv&\frac{C_{2,1,2,-1}}{3} \,. \label{eq:yexplicit}
\ea

One finds that there are only ten $A$ and $C$ coefficients that are non-vanishing, and all are real. They are related as:
\begin{gather}
A^a_{2,0} = A^b_{2,0} \neq 0\,, \label{eq:veccoeff}\\
C_{1,-1,1,1}=C_{1,1,1,-1}=-C_{2,-1,2,1}=-C_{2,1,2,-1}\,\neq 0\,,  \\
 C_{2,-2,2,2}=C_{2,2,2,-2} = - C_{1,0,1,0}=2- C_{2,0,2,0}\,\neq 0\,, 
 \end{gather}
consistently with the properties discussed in the previous section (up-down symmetry, cylindrical symmetry, {\it etc.}). Moreover,
\begin{gather}
 \frac{A^a_{2,0}}{\sqrt{2}} + 1 = C_{2,2,2,-2}\,.
\label{eq:coeff_LOrelation1ok}
\end{gather}

At this point we want to stress  an aspect that will be of particular relevance when this framework will be extended to the NLO level in Sec.~\ref{sec:NLOeffects}. The quantity in the l.h.s. of Eqs.~\eqref{eqn:coeff_defA} and \eqref{eqn:coeff_defC} depends on the entire process $H \rightarrow V_a V_b \rightarrow 1,2,3,4$, while by construction the $A$ and $C$ coefficients depend only on the  $H\rightarrow VV^*$ decay, and therefore cannot depend on the interactions among the $V$ bosons and the fermions. Since $B^a_1$ and $B^b_1$ are functions of, respectively, $\alpha_a$ and $\alpha_b$, this means that for the case $C_{1,-1,1,1}$ (or $C_{1,0,1,0}$)  the l.h.s. of Eq.~\eqref{eqn:coeff_defC} does depend on $\alpha_a$ and $\alpha_b$, and that $B^a_1$ and $B^b_1$ cancel this dependence when solving the equation for $C_{1,-1,1,1}$ (or $C_{1,0,1,0}$). 
This can be made explicit looking at their analytical expression at a fixed $\beta$, which takes the general form 
\begin{equation}
    C_{L_a,M_a,L_b,M_b} (\beta) = \frac{(4\pi)^2}{B_{L_a}^{a}B_{L_b}^{b}}{\cal G}_{L_a,M_a,L_b,M_b}\,.
\end{equation}
Here ${\cal G}_{L_a,M_a,L_b,M_b}$ is the l.h.s.~of Eq.~\eqref{eqn:coeff_defC} at fixed $\beta$, and the only two independent and non-vanishing ${\cal G}_{1,M_a,1,M_b}$ read
\begin{align}
    {\cal G}_{1,1,1,-1} &= \alpha_{a} \alpha_{b}  \left(\frac{-3\beta}{8\pi(2+\beta^2)}\right)\,,  \label{eq:G111-1}\\
    {\cal G}_{1,0,1,0} &= \alpha_{a} \alpha_{b} \left(\frac{-3}{8\pi(2+\beta^2)}\right)\, ,   \label{eq:G1010}
\end{align}
while the other non-vanishing ${\cal G}_{L_a,M_a,L_b,M_b}$ coefficients with $L_a \neq 1$ and $L_b \neq 1$ do not depend on the spin-analysing powers $\alpha_a$ and $\alpha_b$. It is therefore manifest that the correct extraction of the $C_{1,-1,1,1}$ and $C_{1,0,1,0}$ coefficients, and in turn on the whole $\rho$ density matrix, critically depends on the value of $\alpha_a$ and $\alpha_b$, and thus on the correct modelling of the $V$ decays.

\medskip

In conclusion, at LO, the $\rho$ density matrix for the process $H \rightarrow VV^*$ in the SM depends on only one independent parameter at fixed $\beta$. However, when integrated over different values of $\beta$—which itself depends on the two invariant masses $m_a$ and $m_b$—the matrix instead depends on two independent parameters. The $\rho$ matrix cannot be accessed directly via experimental measurements, but can be extracted via QT. In doing so, we exploit the fact that we know the density matrix $\Gamma$ of the $V\to f\bar f$ decay considered, which, at LO and for massless fermions, depends only on the parameter $\alpha$, {\it i.e.}, the spin-analysing power of the fermion considered. It is important to note that at LO  $\Gamma$ does  {\it not} depend on the value of $m_a$ or $m_b$.
As we will discuss in Sec.~\ref{sec:NLOeffects}, several of the aforementioned properties cannot be trivially extended at the NLO EW level.

\subsection{Entanglement}
\label{sec:entan}
Knowledge of the spin density matrix elements allows the determination of  entanglement markers, {\it e.g.}, through the logarithmic negativity~\cite{PhysRevLett.95.090503,Vidal:2002zz} or the Peres-Horodecki criterion~\cite{Peres:1996dw,Horodecki:1996nc}. Knowing the full $\rho$ matrix, specific entanglement measures, including concurrence~\cite{Bennett_1996, Wootters:1997id, Rungta_2001, Mintert_2005, Mintert_2007} and quantum entropy~\cite{Horodecki:2009zz}, can be calculated. These measures quantify the degree of entanglement and can be used to define conditions for non-separability.
In the case of pure states such quantities are often analytically calculable, whereas for mixed states this is generally not feasible \cite{Mintert_2007}.

For fixed $\beta$, the spin density matrix for the process $H\rightarrow VV^{*}$ (Eq.~\eqref{beta_rho}) represents a pure state, as also clear from Eq.~\eqref{state}. In this case, any non-zero off-diagonal term indicates a superposition of states and is a direct indicator of entanglement. However, integrating over $\beta$ (Eq.~\eqref{rhovv}) the state is not pure anymore. Nonetheless, we can still use the Peres-Horodecki criterion, due to the specific structure of the $\rho$ matrix in Eq.~\eqref{rhovv}, which leads to a very simple condition for entanglement (see also Ref.~\cite{Aguilar-Saavedra:2022wam}),
\be
\label{entanglement_vv}
C_{2,2,2,-2}\neq 0\quad\quad \mathrm{or}\quad\quad C_{2,1,2,-1}\neq 0 \, .
\ee
In this specific case, the sufficient condition for entanglement is also a necessary one. As a result, the spin density matrix in Eq.~\eqref{rhovv} corresponds to a separable state if and only if both coefficients in Eq.~\eqref{entanglement_vv} vanish simultaneously; otherwise, the state is entangled. This simple condition remains valid as long as the spin density matrix retains the structure given in Eq.~\eqref{rhovv}. However, if the matrix structure becomes more complex and involves more than two independent parameters, identifying a similarly straightforward entanglement condition using the Peres-Horodecki criterion becomes significantly more challenging.

As we will discuss in Sec.~\ref{sec:NLOeffects}, higher-order corrections to the $H\rightarrow \ell^{+}\ell^{-}\ell^{+}\ell^{-}$ process result in a spin density matrix with a structure that differs significantly from the LO one. Consequently, to find evidence of entanglement in such cases, we will employ the most commonly used entanglement witnesses for a mixed bipartite qutrit state: the lower and upper bounds on the squared concurrence. These quantities can be derived analytically~\cite{PhysRevA.78.042308}:
\be
\label{concurrencedef}
({\cal C(\rho)})^2\geq 2\max(\Tr\rho^2-\Tr\rho_a^2,\Tr\rho^2-\Tr\rho_b^2)\,,\quad\quad ({\cal C(\rho)})^2\leq 2\min(1-\Tr\rho_a^2,1-\Tr\rho_b^2)\,,
\ee
where $\rho$ is defined in Eq.~\eqref{rhovv} and $\rho_{a}$ and $\rho_{b}$ are  the reduced density matrices respectively evaluated by ``tracing out" the $b$ and $a$  subsystem:
\be
\label{rhoab}
\rho_{j}=\frac{1}{3}({\mathbf 1}_3 + A^{j}_{L_j,M_j}T_{L_j,M_j})\quad{\rm with}\quad j=a,b \,.
\ee

We can calculate the previous expression in terms of the $A$ and $C$ coefficients, using the above definition and Eq.~(\ref{hermitianprop}):
\bea 
\label{rhosquaretraces}
\Tr \rho^2&=& \frac{1}{9}\left(1+\sum_{j=a,b}\sum_{L,M} |A^{j}_{L,M}|^2+\sum_{L_a,L_b,M_a,M_b}|C_{L_a,M_a,L_b,M_b}|^2\right)\,,\\
\Tr\rho_{j}^2&=& \frac{1}{3}\left(1+\sum_{L,M}|A^{j}_{L,M}|^2)\right)\quad{\rm with}\quad j=a,b\,.
\eea

 We can thus rewrite the lower $(\mathcal{C} _{LB})$ and upper $(\mathcal{C} _{UB})$ bounds on the squared concurrence as following: 
{\small
\bea 
\textstyle ({\cal C(\rho)})^2&\geq&
\frac{2}{9}\max \Bigg[\Big(-2-2\sum_{L,M} |A^{a}_{L,M}|^2+\sum_{L,M}|A^{b}_{L,M}|^2+\sum_{L_a,L_b,M_a,M_b}| C_{L_a,M_a,L_b,M_b}|^2\Big), \nonumber \\
 &&\Big(-2+\sum_{L,M} |A^{a}_{L,M}|^2-2\sum_{L,M}|A^{b}_{L,M}|^2+\sum_{L_a,L_b,M_a,M_b} |C_{L_a,M_a,L_b,M_b}|^2\Big)\Bigg]\,, \label{eqn:clb} \\
 ({\cal C(\rho)})^2&\leq& \frac{2}{3}\min\Bigg[\Big(2-\sum_{L,M}|A^{a}_{L,M}|^2\Big),\Big(2-\sum_{L,M}|A^{b}_{L,M}|^2\Big)\Bigg]\,.
 \label{eqn:cub} 
 \eea
 }

 While these quantities by themselves do not provide a precise measure of entanglement, they can be used to establish the presence of entanglement: $\mathcal{C} _{LB} > 0$ implies a non vanishing concurrence and consequently an entangled state,  on the other hand a null  $\mathcal{C} _{UB}$ implies a separable state.

\subsection{Bell inequality violation}
\label{sec:IntroBell}
Bell-type inequalities, introduced by J.~Bell in 1964~\cite{PhysicsPhysiqueFizika.1.195}, arose as a way to formulate a measurable observable in reply to the thought experiment originally proposed by Einstein, Podolsky, and Rosen (EPR) to argue that quantum mechanics (QM) might be incomplete \cite{Einstein:1935rr}. Bell inequalities establish upper bounds on correlations predicted by local hidden variable theories; any observed violation of these bounds is taken as evidence of non-local behaviour, directly contradicting the EPR argument. Experimental measurements of Bell inequality violations have been conducted across various systems — photons, ions, solid-state, and superconducting systems~\cite{PhysRevLett.28.938,PhysRevLett.49.1804,Rowe2001,Ansmann2009} — with recent studies aiming to address the communication, detection, and freedom-of-choice loopholes in experimental setups~\cite{Hensen:2015ccp,PhysRevLett.115.250401,PhysRevLett.115.250402}.

At very high energies, or equivalently at very small scales, the situation becomes more nuanced. Quantum field theory has proven extremely powerful in accurately describing fundamental interactions, with quantum mechanics playing an essential role in countless instances. However, only recently have high-energy collider experiments begun to focus on observables specifically designed to be sensitive to quantum entanglement or violations of Bell-type inequalities. The interest in such observables is multifaceted, ranging from testing the very foundations of quantum mechanics to employing entanglement as a resource for probing fundamental interactions more deeply.

In high-dimensional quantum systems, such as pair of qutrits, various Bell-type inequalities have been formulated. Among these, the Collins-Gisin-Linden-Massar-Popescu (CGLMP) inequality~\cite{PhysRevLett.88.040404} is especially notable. This inequality has been shown to yield the largest deviation from the locality bound in diboson systems generated from scalar decays at rest or with small boosts~\cite{Barr:2022wyq}.

The CGLMP inequality can be illustrated as follows: consider a scenario in which a Higgs boson decays into two vector bosons, $V_a$ and $V_b$, measured independently by two observers, Alice and Bob. Alice performs two distinct measurements, $a_1$ and $a_2$, on particle $V_a$, while Bob performs two measurements, $b_1$ and $b_2$, on particle $V_b$. Alice and Bob’s measurements are conducted without communication, preserving the locality condition. For a qutrit system, each measurement outcome is one of three possible values: 0, 1, or 2. The CGLMP inequality is then evaluated by analyzing the joint probabilities of Alice and Bob’s measurement outcomes~\cite{PhysRevLett.88.040404,PhysRevLett.115.250401,PhysRevLett.115.250402}:
 \bea
\label{bellinequprob}
{\cal I}_3&=&[P(a_1=b_1)+P(b_1=a_2+1)+P(a_2=b_2)+P(b_2=a_1)]\nonumber\\
&-&[P(a_1=b_1-1)+P(b_1+a_2)+P(a_2=b_2-1)+P(b_2=a_1-1)]\,,
 \eea 
where $P(a_i=b_j+k)$ is the probability that the outcome $a_i$ differs from the outcome $b_j$ by an integer $k$, modulo 3. The maximum value of ${\cal I}_3$ for local variable theories is 2, and for non-local theories, still respecting causality, is 4~\cite{PhysRevLett.88.040404}.   A relevant aspect of all Bell inequalities is that the violation depends on the type of measurement performed. If we take the measurement as the projection of the spin on a certain axis, then the choice of the axis is essential to observe a violation of the Bell inequality.  The maximum value that can be reached by ${\cal I}_3$, according to quantum mechanics, for maximally entangled states is ${\cal I}_3\approx 2.8729$~\cite{PhysRevLett.88.040404}. Noticeably, for mixed qutrit systems this does not correspond to the maximal violation of the inequality \cite{Acin:2002zz}.

\begin{figure}[h!]
    \centering
    \begin{subfigure}[b]{0.46\textwidth}
        \centering
        \includegraphics[width=\textwidth]{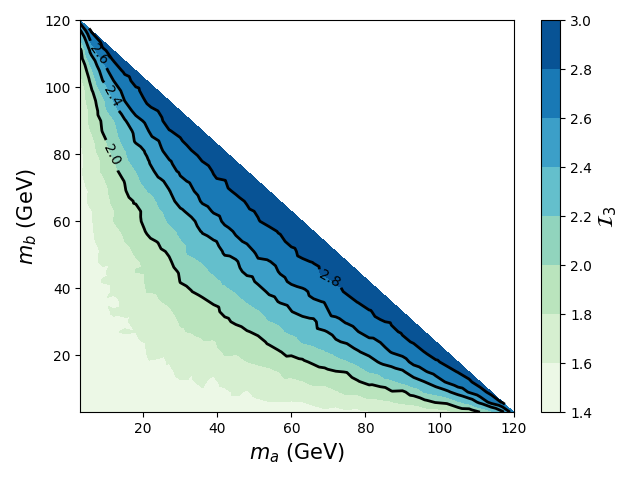}
        \caption{}
        \label{figI3a}
    \end{subfigure}
    \centering
    \begin{subfigure}[b]{0.46\textwidth}
        \centering
        \includegraphics[width=\textwidth]{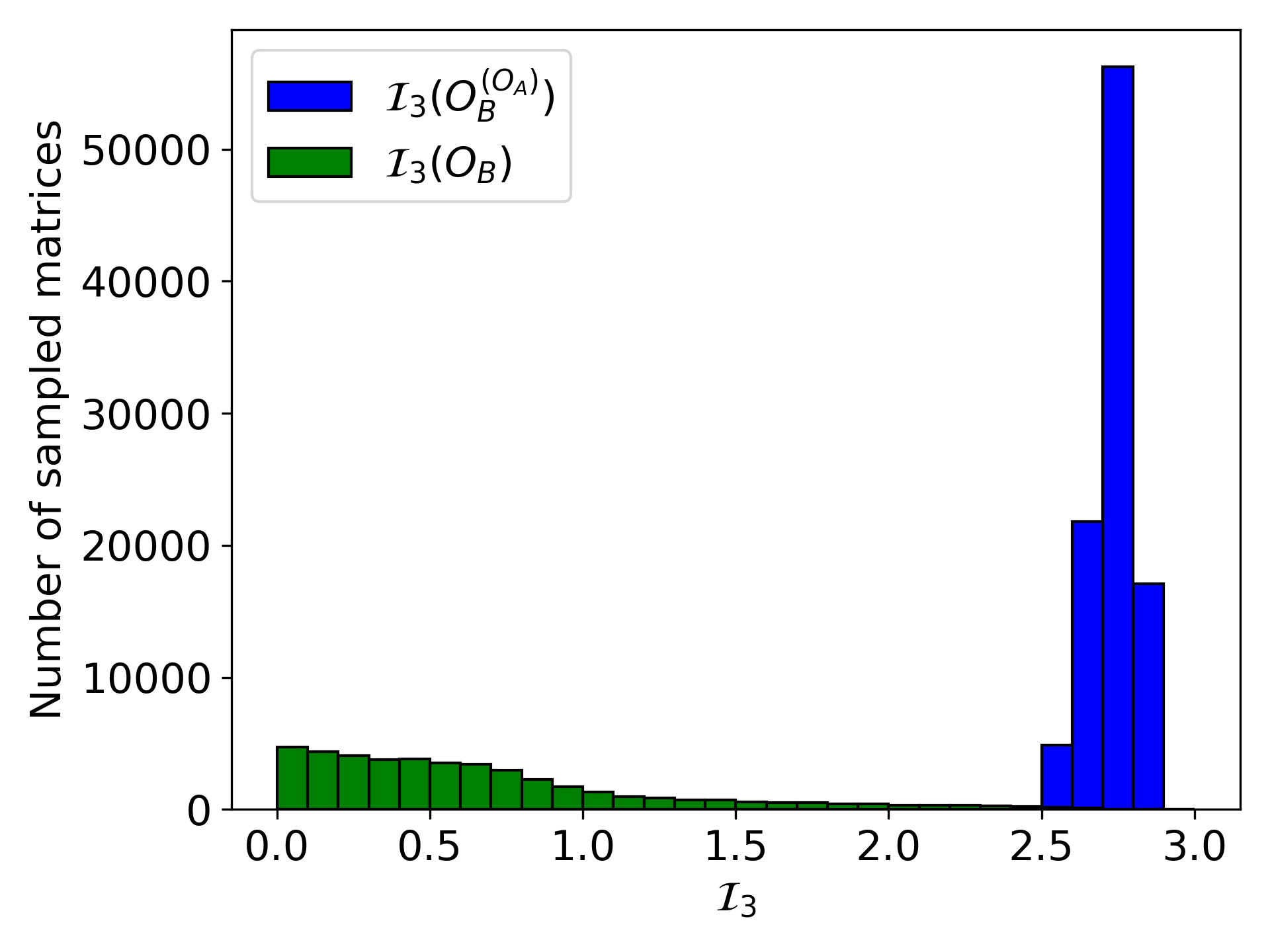}
        \caption{}
        \label{figI3b}
    \end{subfigure}
   \caption{(a) \(\mathcal{I}_3\) calculated for the decay \(H \to VV^{*}\) as a function of the masses of the two vector bosons \(V\). The quantity \(\mathcal{I}_3\) is maximised by sampling random three-dimensional unitary matrices \(U\) and \(V\), which are used as defined in \(O_B\). (b) \(\mathcal{I}_3\) values obtained by sampling only the unitary matrix \(U\) according to the \(O_B^{(O_A)}\) definition, compared to those obtained using the $O_B$ definition. The density matrix \(\rho\) is evaluated with \(\beta\) fixed at 1.5.}
    \label{I3fig1}
\end{figure}

The CGLMP inequality can be expressed in terms of an operator acting on the spin density matrix~\cite{Acin:2002zz} by defining
\be
\label{i3def}
{\cal I}_3=\Tr[\rho\, O_{\rm Bell} ]\, ,
\ee
where $O_{\rm Bell}$ can be expressed in different ways, corresponding to different $a$ and $b$ measurements.
In this work we use the Bell operator presented in Ref.~\cite{Acin:2002zz}, that can be derived using irreducible tensor operators as following:
\bea
\label{belloperatordef}
    O_{\rm Bell}&=&\frac{4}{3\sqrt{3}}(T_{1,1} \otimes T_{1,1} + T_{1,{-1}} \otimes T_{1,{-1}})+\frac{2}{3}(T_{2,2} \otimes T_{2,2} + T_{2,{-2}} \otimes T_{2,{-2}})\,,
\eea 
leading to:
\be
\label{O_Bell}
O_{\rm Bell}=\left(
\setlength\arraycolsep{4.5pt}\begin{array}{ccccccccc}
 0 & 0 & 0 & 0 & \frac{2}{\sqrt{3}} & 0 & 0 & 0 & 2 \\
 0 & 0 & 0 & 0 & 0 & \frac{2}{\sqrt{3}} & 0 & 0 & 0 \\
 0 & 0 & 0& 0 & 0 & 0 & 0 & 0 & 0 \\
 0 & 0 & 0 & 0 & 0 & 0 & 0 & \frac{2}{\sqrt{3}} & 0 \\
 \frac{2}{\sqrt{3}} & 0 & 0 & 0 & 0 & 0 & 0 & 0 & \frac{2}{\sqrt{3}} \\
 0 & \frac{2}{\sqrt{3}} & 0 & 0 & 0 & 0 & 0 & 0 & 0 \\
 0 & 0 & 0& 0 & 0 & 0 & 0 & 0 & 0 \\
 0 & 0 & 0 &  \frac{2}{\sqrt{3}} & 0 & 0 & 0 & 0 & 0 \\
 2 & 0 & 0 & 0 &  \frac{2}{\sqrt{3}} & 0 & 0 & 0 & 0 \\
\end{array}
\right)\,. 
\ee 
The above operator yields the maximal violation of the Bell inequality according to QM  (${\cal I}_3 \approx 2.9149$)~\cite{Acin:2002zz} for any generic two-qutrit system, and such violation is realised for the entangled qutrit state $|\psi\ket = \frac{1}{\sqrt{2+\delta^2}} ( |11\ket + \delta|22\ket + |33\ket)$, where $\delta \approx 0.7923$. This state slightly deviates  from the maximally entangled qutrit state $|\psi\ket = \frac{1}{\sqrt{3}} ( |11\ket + |22\ket + |33\ket)$, which in the case of spins for a qutrit pair, and using our notation,  corresponds to  $|\psi\ket = \frac{1}{\sqrt{3}} ( |0 \,0\ket + |+ -\ket + |- +\ket)$. Notably, there is a sign difference compared to Eq.~\eqref{state}.
To reach the maximal value of ${\cal I}_3$, according to the state at hand, the Bell operator can be modified by local unitary transformations:
\begin{eqnarray}
\label{eqn:i3def}
O_{B}&=&(V^\dag \otimes U^\dag) O_{\rm Bell}(V\otimes U) \,, \\
{\cal I}_3&=&\Tr[\rho\, O_{B} ]\,,
\end{eqnarray}
where $U$ and $V$ are three-dimensional unitary matrices.
Few studies have proposed optimal transformations  to maximise ${\cal I}_3$ ~\cite{Fabbrichesi:2023cev,Acin:2002zz} for the $H \rightarrow VV^*$ state across the whole phase space. Notably, no single unitary transformation that can universally maximise ${\cal I}_3$ for all values of $m_a$, $m_b$ has been found; instead, the optimal matrices vary depending on the (invariant) masses of the two $V$ bosons. 
The value of ${\cal I}_3$ as a function of $m_a$ and $m_b$, obtained by maximising ${\cal I}_3$ using random $3\times 3$ unitary  matrices  $U$ and $V$, is shown in Figure~\ref{figI3a}.
The region where ${\cal I}_3$ reaches its maximum is along the primary diagonal of the $m_a$ and $m_b$ plane, corresponding to the region where the sum of the two masses is precisely the Higgs boson mass. ${\cal I}_3$ is above two on a large region of the phase space, dropping below the local realism bound only when both $V$ bosons are produced significantly off-shell ($\approx \beta > 6$), which is a very suppressed configuration.

In the following we would like to focus on a method to evaluate the Bell inequality violations that could be applied both to numerical results, and to real data collected in collider experiments. An optimisation of $O_B$ dependent on $\beta$ is currently un-feasible in the case of real measurements, as it would require the extraction of the spin density matrix across a finely-grained $\beta$ spectrum, which is very limited by statistics. As a consequence we will focus on an inclusive optimisation, based on the $\rho$ extracted in the region(s) of interest.

A convenient way of maximising ${\cal I}_3$, in the region close to $\beta$ = 1, was introduced in Ref.~\cite{Aguilar-Saavedra:2022wam}. Here, instead of a generic unitary transformation depending on two matrix $U,V$, a basis change was introduced to write the singlet state representing the system $H \rightarrow VV^*$ at $\beta$ = 1 ($|\psi\ket = \frac{1}{\sqrt{3}} ( |-+\ket - |00\ket + |+-\ket)$) directly in the form $|\psi\ket = \frac{1}{\sqrt{3}} ( |-+\ket + |00\ket + |+-\ket)$.
This transformation can be written as: $|\psi\ket \rightarrow U O_A \otimes U^* |\psi\ket$; it depends on a single three-dimensional unitary matrix $U$ that can be selected to maximise ${\cal I}_3$, while $O_A$ is fixed:
\be
\label{O_{A}}
O_{A}=\left(
\setlength\arraycolsep{4.5pt}\begin{array}{ccccccccc}
 0 & 0 & 1 \\
 0 & -1 & 0 \\
 1 & 0 & 0 \\
\end{array}
\right)\,. 
\ee
The transformation can be translated on the Bell operator as
\be
O_B^{(O_A)} = (U O_A \otimes U)^{\dag} O_{\rm Bell} (UO_A \otimes U)\, .
\ee

We have seen that, in the range $\beta < 5$, the maximal value of ${\cal I}_3$ achieved using the transformation involving $O_A$ (i.e., $O_B^{(O_A)}$) is very similar to that obtained using two completely random unitary matrices ($O_B$), where no assumptions are made about the state. The main difference lies in the advantage of having $O_B^{(O_A)}$ in the numerical optimisation, as shown in Figure~\ref{figI3b}. Here, the optimisation of the two methods is compared, using the $\rho$ matrix for $H\rightarrow VV^{*}$ evaluated for $\beta$=1.5 and 100000 random extractions of one or two unitary matrices. While the maximum of ${\cal I}_3$ is very similar in the two cases, the region with ${\cal I}_3 > 2$ represents only a tail of the ${\cal I}_3$ values extracted with $O_B$. On the contrary, all the ${\cal I}_3$ derived using $O_B^{(O_A)}$ lie in this region.
As a consequence, in the rest of the paper we will exploit the method $O_B^{(O_A)}$ to evaluate ${\cal I}_3$.

In Ref.~\cite{Aguilar-Saavedra:2022wam} a specific value of the unitary matrix $U$ that maximises ${\cal I}_3$ for a large range of $\beta$ is identified (see Eq.~(40) in Ref.~\cite{Aguilar-Saavedra:2022wam}), still considering $\beta <5$. This matrix, which we denote as $U_\mathrm{fix}$, is employed in conjunction to $O_A$ in order to transform the Bell operator, which is then applied to the analytical form of the matrix in Eq.~\eqref{rhovv}. As a result, a simple analytical expression for ${\cal I}_3$ is derived:
\bea
    {\cal I}_3 \rightarrow \Tr[\rho\, O_B^{(O_A, U_\mathrm{fix})} ]\equiv  \Tr[\rho\,  (U_\mathrm{fix} O_A \otimes U_\mathrm{fix})^{\dag} O_{\rm Bell} (U_\mathrm{fix} O_A \otimes U_\mathrm{fix}) ]=\nonumber \\ 
    \frac{1}{36} (18 + 16\sqrt{3} - \sqrt{2}(9 - 8\sqrt{3})A^1_{2,0} - 8(3+2\sqrt{3})C_{2,1,2,-1} + 6 \, C_{2,2,2,-2}) \ .
    \label{eqn:I3JA}
\eea

This formula is however valid only if the spin-density matrix has the form in Eq.~\eqref{rhovv}. In the following, the results indicated by $O_B^{(O_A, U_\mathrm{fix})}$ are simply obtained by applying Eq.~\eqref{eqn:I3JA}, even in instances where the spin density matrix deviates from the form described in Eq.~\eqref{rhovv} and therefore Eq.~\eqref{eqn:I3JA} is in principle not valid. 

\section{$ H \rightarrow ZZ^* \rightarrow e^+ e^- \mu^+ \mu^- $ at LO}
\label{sec:Lodetails}
\label{sec:LO}

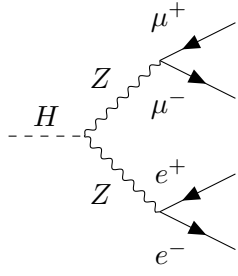
\begin{figure}[!t]
    \centering
\begin{tikzpicture}
  \begin{feynman}
    \centering
    \vertex (a1);
    \vertex[right=1.0cm of a1] (a2);
    \vertex[above right=1.0cm and 1.0cm of a2] (b1);
    \vertex[below right=1.0cm and 1.0cm of a2] (b2);
    \vertex[above right=0.5cm and 1.0cm of b1] (c1);
    \vertex[below right=0.5cm and 1.0cm of b1] (c2);
    \vertex[above right=0.5cm and 1.0cm of b2] (d1);
    \vertex[below right=0.5cm and 1.0cm of b2] (d2);
    
    \diagram* {
      (a1) -- [dashed, edge label=\(H\)] (a2),
      (a2) -- [boson, edge label=\(Z\)] (b1),
      (a2) -- [boson, edge label'=\(Z\)] (b2),
      (b1) -- [anti fermion,edge label=\(\mu^+\)] (c1),
      (b1) -- [fermion,edge label'=\(\mu^-\)] (c2),
      (b2) -- [anti fermion, edge label=\(e^+\)] (d1),
      (b2) -- [fermion, edge label'=\(e^-\)] (d2),
    };
  \end{feynman}
\end{tikzpicture}
\caption{The only  Feynman diagram that contributes to the  $H \rightarrow e^+ e^- \mu^+ \mu^- $  tree-level amplitude in the SM.}
\label{fig:born}
\end{figure}

In this section we present numerical results based on the theoretical framework described in Sec.~\ref{QIintro}. We focus on the $ H \rightarrow ZZ^* \rightarrow e^+ e^- \mu^+ \mu^- $ process at LO (see Figure~\ref{fig:born} for the relevant Feynman diagram), while NLO EW corrections are discussed in Sec.~\ref{sec:NLOeffects}.

Although limited by the small statistics available, the $e^+ e^- \mu^+ \mu^- $ final state is the one among those emerging from the $H \rightarrow VV^{*}$ decays that can be more easily separated from the background and reconstructed.
The numerical evaluation of the spin density matrix in this work is performed by simulating the $H \rightarrow  e^+ e^- \mu^+ \mu^- $ process using {\aNLO} \cite{Alwall:2014hca, Frederix:2018nkq} and performing QT. As already mentioned in Sec.~\ref{QMT}, we consider only the (on-shell) Higgs boson decay; 
our simulations do not take into account the $H$ production mechanism. We remind the reader that the main subject of this work is indeed the impact of NLO EW corrections on the $\rho$ density matrix and in turn on its interpretation in the context of QI. 

As can be explicitly seen in Eqs.~\eqref{eqn:coeff_defA}--\eqref{eqn:coeff_defC}, the key observables that need to be reconstructed for the QT approach are the polar ($\theta$) and azimuthal ($\phi$) angles of the leptons in the parent $Z$ rest frame. As a consequence, the first step of the analysis consists in reconstructing the two $Z$ bosons. This is done simply by combining the four-momentum of the four leptons in pairs, requiring the same flavour as condition for the combination. The two $Z$ bosons reconstructed with this approach are then used to define the reference frames. This reconstruction technique is applied independently of the order of the simulation. 

\begin{figure}[t!]
     \centering
     \begin{subfigure}[b]{0.45\textwidth}
         \centering
         \includegraphics[width=\textwidth]{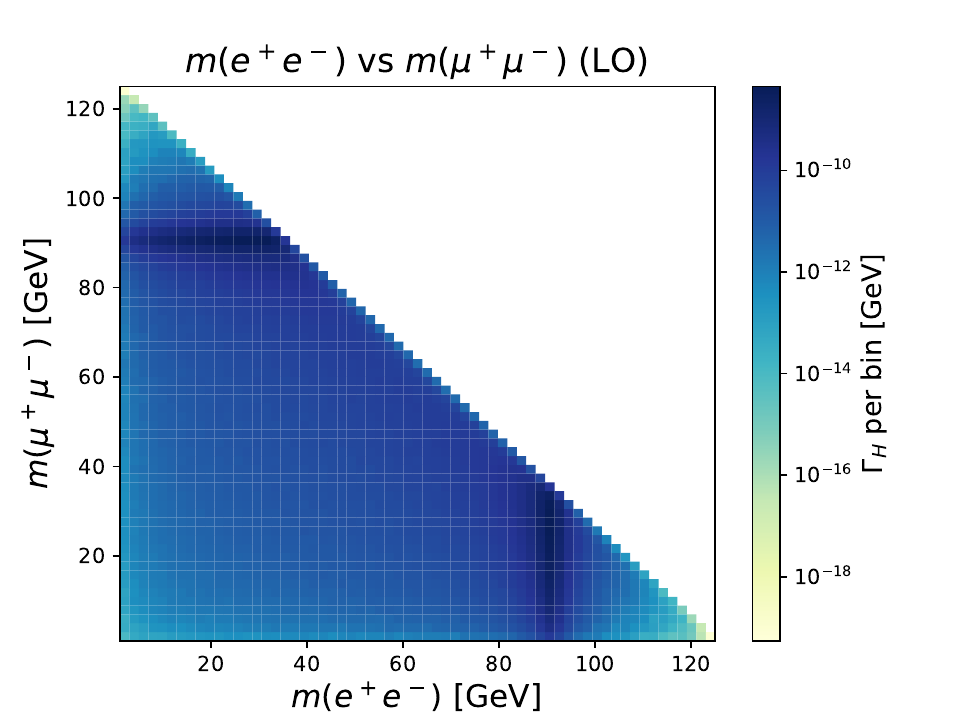}
         \caption{}
         \label{figGa}
     \end{subfigure}
     \hfill
     \begin{subfigure}[b]{0.54\textwidth}
         \centering
         \includegraphics[width=\textwidth]{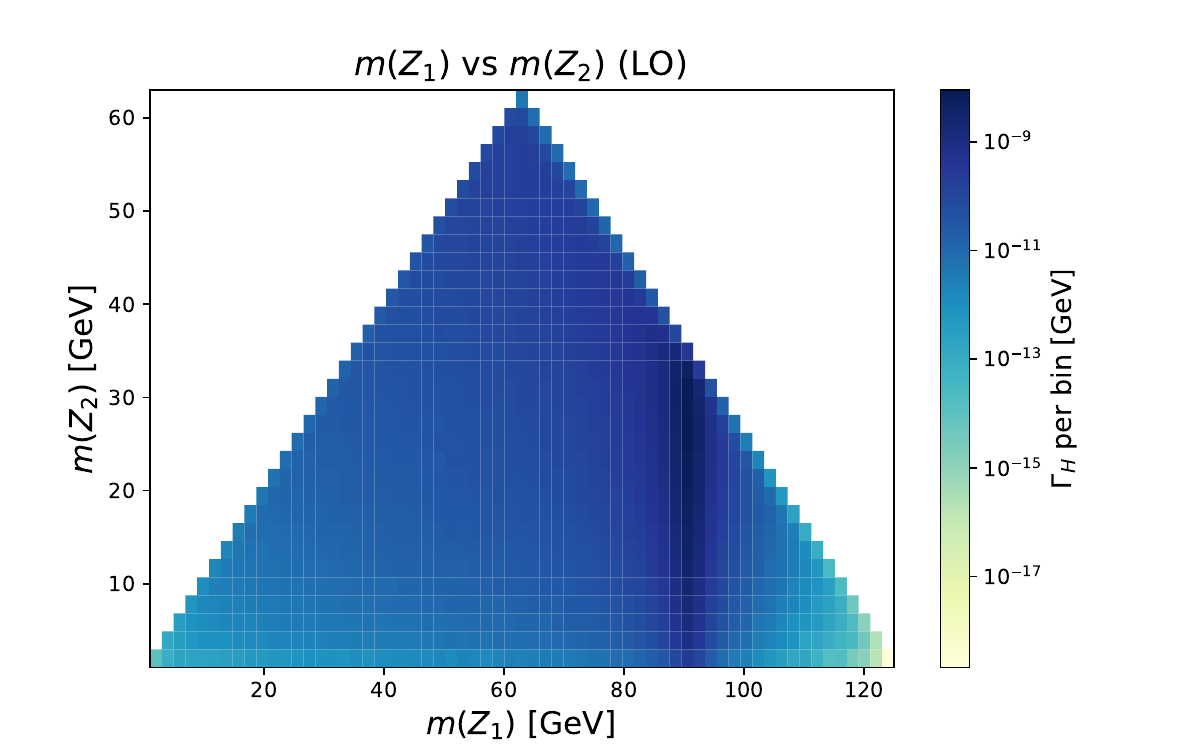}
         \caption{}
         \label{figGb}
     \end{subfigure}
        \caption{Partial decay width obtained using $H\to e^+e^-\mu^+\mu^-$ events simulated at LO, as a function of the invariant masses $m(e^+e^-)$  and $m(\mu^+ \mu^-)$ (a) and as a function of $m({Z}_{1})$ and $m({Z}_{2})$ (b).}
        \label{figG}
\end{figure}

Figure~\ref{figGa} shows the distribution of $\Gamma_H$, {\it i.e.} the  $H \rightarrow e^+ e^- \mu^+ \mu^- $ partial decay width, as a function of the masses of the two reconstructed $Z$, $m(e^+e^-)$  and $m(\mu^+ \mu^-)$. The regions that are more populated are those where one $Z$ is produced approximately on shell and the other $Z$ is highly off-shell.
One can therefore order the two bosons according to their invariant masses and introduce the labels $Z_1$ and $Z_2$ for the leading and trailing boson, respectively.
The quantity $\Gamma_H$ as a function of the invariant masses of $Z_1$ and $Z_2$, respectively denoted as $m(Z_1)$ and $m(Z_2)$, is shown in Fig.~\ref{figGb}. Here it is also clearly visible that the regions where both the $Z$ bosons are off-shell, at the top of the triangle, is suppressed w.r.t.~to the region where $Z_1$ is on-shell.

Comparing Fig.~\ref{figGa} with Fig.~\ref{figI3a}, it can be seen that in the most populated regions ${\cal I}_3$ typically takes values larger than 2, up to the maximum value allowed by QM. Thus, at the inclusive level, we expect that this leads to ${\cal I}_3 >$ 2. 
Figure~\ref{figI3a} is also helpful when investigating the possibility of isolating a phase-space region where we expect to find the largest correlations. One can conclude that it corresponds to the region obtained by putting a lower cut on $m(Z_2)$ $>$ 30 GeV. This selection removes close to all the events that would lead to ${\cal I}_3 <$ 2. However, such a region also corresponds to a suppressed portion of phase space, as can be seen in Fig.~\ref{figGb}.

\medskip 
As mentioned in Sec.~\ref{sec:QIintro}, we parametrise the $\rho$ density matrix using the helicity basis, and therefore the reference system that we use in order to measure angular distributions is constructed as follows~\cite{Aguilar-Saavedra:2022wam}:
\begin{itemize}
\item The $\hat z$ axis is taken in the direction of the $Z_1$ three-momentum in the $H$ rest frame.
\item The $\hat x$ axis is defined as $\hat x = \mathrm{sign}(\cos \theta) (\hat p - \cos \theta \hat z) / \sin \theta$, where $\hat p = (0, 0, 1)$ would be the direction of the beam in the laboratory frame and $\cos \theta = \hat z \cdot \hat p$.
\item The $\hat y$ axis is defined such that $\hat y = \hat z \times \hat x$.
\end{itemize}

Given the $(\hat x, \hat y, \hat z) $ directions, it is possible to define $(\theta_1, \phi_1)$ as the polar coordinates of the three-momentum of the negatively charged lepton $\ell^-_1$ from the $Z_1$ in the $Z_1$ rest frame. Similarly, $(\theta_2, \phi_2)$ correspond to the angles that give the direction of the $\ell^-_2$ momentum in the $Z_2$ rest frame.\footnote{Notice that unlike the previous section we do not use the convention $H\rightarrow V_a V_b \rightarrow f_1 \bar f_2 f_3 \bar f_4$ but $H\rightarrow Z_1 Z_2 \rightarrow \ell^-_1 \ell^+_1 \ell^-_2 \ell^+_2$, where $\ell^{\pm}_i$ originates from $Z_i$} Once the angles are defined, it is directly possible to numerically calculate the integrals in Eqs.~\eqref{eqn:coeff_defA}--\eqref{eqn:coeff_defC}, compute all the $A$ and $C$ coefficients and in turn the entries of the spin density matrix. In order to solve the integrals, we split them in real and imaginary components. We always verify that the integral of the imaginary components are compatible with zero within the numerical error. The $A$ and $C$ coefficients are therefore real numbers, unless otherwise specified. We remind the reader that the entries of the spin density matrix as a function of all the $A$ and $C$ coefficients are given in Appendix~\ref{sec:rhogeneral}. 

If no cuts are imposed on the momenta of the four leptons, the LO spin density matrix obtained through the tomographic procedure is: 
\begin{equation}
\hspace{-5mm}
\rho_{\text{LO}} = 
\begin{pmatrix}
\cdot & \cdot & \cdot & \cdot & \cdot & \cdot & \cdot & \cdot & \cdot \\
\cdot & \cdot & \cdot & \cdot & \cdot & \cdot & \cdot & \cdot & \cdot \\
\cdot & \cdot & ~~\blue{0.195(2)} & \cdot & \blue{-0.313(3)} & \cdot & ~~\blue{0.194(1)} & \cdot & \cdot \\
\cdot & \cdot & \cdot & \cdot & \cdot & \cdot & \cdot & \cdot & \cdot \\
\cdot & \cdot & \blue{-0.313(3)} & \cdot & ~~\blue{0.612(1)} & \cdot & \blue{-0.313(3)} & \cdot & \cdot \\
\cdot & \cdot & \cdot & \cdot & \cdot & \cdot & \cdot & \cdot & \cdot \\
\cdot & \cdot & ~~\blue{0.194(1)} & \cdot & \blue{-0.313(3)} & \cdot & ~~\blue{0.195(3)} & \cdot & \cdot \\
\cdot & \cdot & \cdot & \cdot & \cdot & \cdot & \cdot & \cdot & \cdot \\
\cdot & \cdot & \cdot & \cdot & \cdot & \cdot & \cdot & \cdot & \cdot \\
\end{pmatrix}\,, \label{eq:rhoLOinclusive}
\end{equation}
\smallskip
where we have written in parentheses the numerical error in the Monte Carlo simulation on the last digit, and denoted with a dot all the vanishing entries. 

The uncertainties have been evaluated by repeating the QT procedure several times on independent simulations. The elements of the spin density matrix are evaluated as the mean and standard deviation across the values calculated in each run.
The shape of the spin density matrix at LO follows the structure shown in Eq.~\eqref{rhovv} and the values obtained are in good agreement with the literature~\cite{Aguilar-Saavedra:2022wam}. We have also verified that all the relations from Eq.~\eqref{eq:xexplicit} to Eq.~\eqref{eq:coeff_LOrelation1ok} are satisfied.

\medskip

The numerical results have been obtained {\aNLO}\footnote{In particular have we used the version 3.5.5 importing the {\UFO} model \cite{Degrande:2011ua,Darme:2023jdn} {\tt  loop\_qcd\_qed\_sm\_Gmu} and setting the Complex-Mass-Scheme on, consistently with  what is discussed later in Sec.~\ref{sec:theoNLOEW} for the NLO EW case.} using as input parameters\footnote{The value reported here are the input parameters used for both LO and NLO EW simulations. Since, as we will explain later in Sec.~\ref{sec:theoNLOEW}, we will use the Complex-Mass-Scheme, they can be identified as the value of the pole mass and total width used in the simulations.  }
\begin{equation}
 M_Z = 91.188 \text{ GeV}, \quad M_W = 80.419 \text{ GeV}, \quad G_{\mu} = 1.16639 \times 10^{-5} \text{ GeV}^{-2} \ , \label{eq:inputLOandNLO}
\end{equation}
and the top quark and Higgs boson masses  set equal to
\begin{equation}
M_H = 125 \text{ GeV}, \quad m_t = 173.3 \text{ GeV} \ .   \label{eq:inputNLO}\\
\end{equation}
In fact, the LO computation does not depend on the quantities in Eq.~\eqref{eq:inputNLO}, which instead enter the computation of NLO EW correction that is discussed in Sec.~\ref{sec:NLOeffects}.

\begin{table}
\begin{center}
\small
\renewcommand{\arraystretch}{1.5}
\resizebox{\textwidth}{!}{
\begin{tabular}{ c | c | c | c | c ||}
\cline{2-5}
                       & inclusive & $m(Z_2) > $ 10 GeV & $m(Z_2) > $ 20 GeV &$m(Z_2) > $ 30 GeV \\ \hhline{ - = = = = }
\multicolumn{1}{|| c ||}{$C_{2,2,2,-2}$} & 0.581(5) & 0.619(4) & 0.713(4) & 0.775(3) \\ \hline
\multicolumn{1}{|| c ||}{$C_{2,1,2,-1}$} & -0.938(4) & -0.975(3) & -1.017(3) & -1.014(3) \\ \hline
\multicolumn{1}{|| c ||}{${\cal I}_3 \left(O_B^{O_A, U_\mathrm{fix}}\right)$} & 2.601(6) & 2.672(4) & 2.772(4) & 2.794(5) \\ 
\hline
\multicolumn{1}{|| c ||}{${\cal I}_3 \left(O_B^{(O_A)}\right)$} & 2.63 & 2.69 & 2.77 & 2.80 \\ \hline
\end{tabular}
}
\captionsetup{width=\textwidth}
\caption {Independent $C$ coefficients of the spin density matrix, extracted with a QT approach using a LO simulation. The table also includes the ${\cal I}_3$ values obtained with two different approaches. The variables are presented as a function of the minimum mass required for $Z_2$. The numerical error on the last digit is in parentheses. \label{tab:LO_coeff}}
\end{center}
\end{table}
\begin{figure}
    \centering
    \includegraphics[width=.75\linewidth]{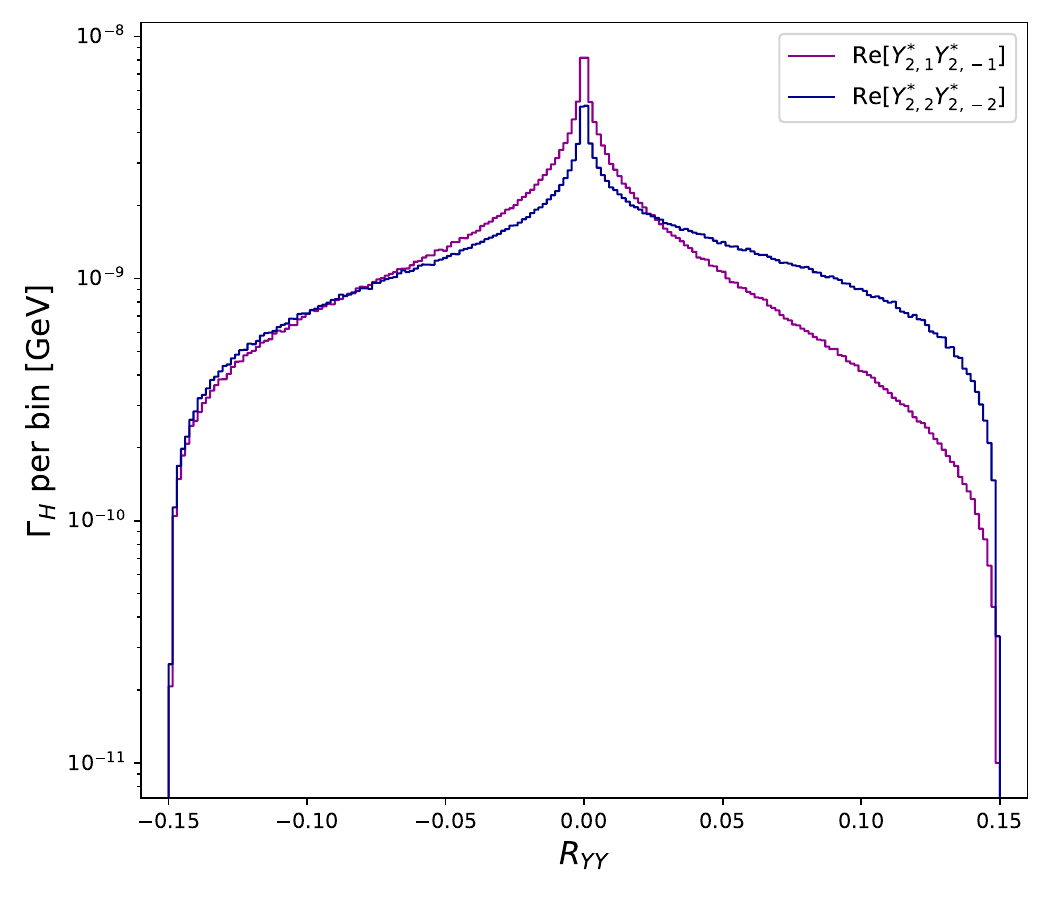}
    \caption{$\Gamma_H$ as a function of $R_{YY}$ for $R_{YY}=\Re\left[Y^*_{2,2} Y^*_{2,-2}\right]$ and $R_{YY}=\Re\left[Y^*_{2,1} Y^*_{2,-1}\right]$. The average of the distributions are respectively proportional to $C_{2,2,2,-2}$ and $C_{2,1,2,-1}$. See Eq.~\eqref{eq:numerical_INT}.}
    \label{fig:LOcoeff}
\end{figure}

The results for the only two independent and non-zero $C$ coefficients at LO (see Eqs.~\eqref{eq:xexplicit} and \eqref{eq:yexplicit}) are shown in Tab.~\ref{tab:LO_coeff}. There, we also show their values for three different cuts on the invariant mass of the trailing $Z$ boson. In Tab.~\ref{tab:LO_coeff} we also report the values of ${\cal I}_3$,  both evaluated as ${\cal I}_3 \left(O_B^{O_A, U_\mathrm{fix}}\right)$ (see Eq.~\eqref{eqn:I3JA}) and by optimising the random $U$ matrix according to the specific region of the phase-space, ${\cal I}_3 \left(O_B^{(O_A)}\right)$, as explained in Sec.~\ref{sec:IntroBell}.

As one can see, the coefficients $C_{2,2,2,-2}$ and $C_{2,1,2,-1}$ are significantly different from zero, with the deviation increasing with the mass cut on $m(Z_2)$, indicating that the two $Z$ are always entangled, following the condition defined in Eq.~\eqref{entanglement_vv}.
The value of ${\cal I}_3$, with both methods, is larger than the locality limit, and approaches the maximum value expected for QM when the lower bound on $m(Z_2)$ is increased, probing smaller values of $\beta$. Simultaneously, the difference between the ${\cal I}_3$ values obtained with both methods decreases and, regardless of the cuts imposed, is always very small, {\it i.e.}, at most 1\% of the  ${\cal I}_3 $ value itself.

Before moving to the NLO results we want to briefly describe how we practically calculate via numerical methods the integrals that appear in  Eqs.~\eqref{eqn:coeff_defA}--\eqref{eqn:coeff_defC}. In Fig.~\ref{fig:LOcoeff} we show the partial decay width $\Gamma_H$ at the differential level w.r.t.~two different quantities. In the $x$ axis of the plot we display the quantity $R_{YY}$, which for the blue line corresponds to $\Re[Y^*_{2,2}Y^*_{2,-2}]$, while for the purple line to $\Re[Y^*_{2,1}Y^*_{2,-1}]$. As can be easily seen from the definition of the spherical harmonics as well as from the plot, for both cases $|R_{YY}|\leq 15/(32 \pi)\simeq 0.15$.  In fact, the distributions in Fig.~\ref{fig:LOcoeff} correspond to the value of $\Gamma_H$ for different bins of $R_{YY}$. It is easy to understand that, since $\Gamma_H$ by definition is a real quantity, the sum of the values of $\Gamma_H$ in each bin multiplied by the average value of $R_{YY}$ in the corresponding bin, in the limit of zero bin-width,\footnote{For the calculation of the integrals we use bins of width 0.0025 for the $A$ coefficients and bins of width 0.0015 for the $C$ coefficients.} converges to the quantity 
\begin{eqnarray}
\int R_{YY} \frac{d\Gamma_H}{ dR_{YY}}d R_{YY}
&=& \Re\left[\int Y^*_{L_1,M_1}(\Omega_1) Y^*_{L_2,M_2}(\Omega_2) \frac{d\Gamma_H}{ d\Omega_1 d\Omega_2} d\Omega_1 d\Omega_2 \right] \nonumber\\
&=& \Gamma_H \frac{B_{L_1}^1 B_{L_2}^2}{(4\pi)^2} C_{L_1, M_1, L_2, M_2}\,, \label{eq:numerical_INT}
\end{eqnarray}
if $R_{YY}\equiv \Re\left[Y^*_{L_2,M_1}(\Omega_1) Y^*_{L_2,M_2}(\Omega_2)\right] $. This is precisely the quantity appearing in Eqs.~\eqref{eqn:coeff_defA}--\eqref{eqn:coeff_defC}, where the symbol $\sigma$ appears but is equivalent to $\Gamma_H$ used here.

It is important to note that the quantities in Eq.~\eqref{eq:numerical_INT} do not assume any intermediate vector boson;  the information required is the angular distribution of the negatively charged leptons in the reconstructed $Z$ frames, which are directly defined via the pair of leptons with opposite charges. Thus, they are observables that can be calculated and measured regardless of the reconstruction of the $\rho$ density matrix via QT and its interpretation in the context of QI. The same applies to the coefficient $C_{L_1, M_1, L_2, M_2}$ itself, that is just the r.h.s.~of  Eq.~\eqref{eq:numerical_INT} divided by a number. As we will discuss in detail in the next sections, the equivalence of $C_{L_1, M_1, L_2, M_2}$ as the term appearing in Eq.~\eqref{eq:rhoexp}, and its more general meaning as an observable derived from the momenta of the final-state leptons, critically relies on having the correct value of $B_{L_1}^1 B_{L_2}^2$. We will show that while this is trivial at LO in the SM, at NLO EW accuracy or including BSM effects the situation completely changes.

\FloatBarrier

\section{$ H (\rightarrow ZZ^*) \rightarrow e^+ e^- \mu^+ \mu^- $ at  NLO EW}
\label{sec:NLOeffects}

In this section we discuss the impact of NLO EW corrections on the analysis of the $H (\rightarrow ZZ^*) \rightarrow e^{+} e^{-} \mu^{+} \mu^{-}$ decay and the relevant quantities for QI already described in the previous section. In Sec.~\ref{sec:theoNLOEW} we discuss the theoretical framework employed for the calculation of NLO EW corrections, and its consequences for the QT approach. In Sec.~\ref{sec:resultsNLOEW} we present numerical results and we scrutinise the stability of QI variables under radiative corrections for the processes considered and in general for the $H (\rightarrow VV^*) \rightarrow f_1 \bar f_2 f_3 \bar f_4$ decays.

\subsection{Theoretical framework}
\label{sec:theoNLOEW}

In general, the NLO EW corrections consist of any contribution of $\mathcal O (\alpha_{\rm EW})$ w.r.t.~the LO prediction of the observable that is considered. They originate from two classes of contributions: virtual and real-emission corrections.  On the one hand, virtual corrections arise from  the interference of one-loop diagrams, which are both infrared (IR) and ultraviolet (UV) divergent, with tree-level diagrams for the same process considered at the LO. On the other hand, real-emission corrections arise from the squared amplitude for the same process considered at the LO plus a real emission of one photon. After the renormalisation of the one-loop amplitude, when virtual and real-emission contributions are combined, a finite and therefore physical prediction is achieved for inclusive, or more in general IR-safe, observables.\footnote{A vast literature on this topic is present and more details can be found, {\it e.g.}, in Refs.~\cite{Denner:1991kt, Denner:2019vbn, Frederix:2018nkq} and references therein.}

The calculation of the NLO EW corrections for the $H (\rightarrow ZZ^*) \rightarrow e^{+} e^{-} \mu^{+} \mu^{-}$ decay was performed for the first time in Ref.~\cite{Bredenstein:2006rh}. Nowadays, this calculation can be performed via public tools, such as {\aNLO}, which indeed has been used for obtaining the numerical results presented in this work. As already mentioned,  NLO EW corrections to the $C$ and $A$ coefficients have also been recently presented in Ref.~\cite{Grossi:2024jae}, where the {\sc \small  MoCaNLO} framework \cite{Denner:2021csi,Denner:2023ehn} has been used for the numerical simulations.

\begin{figure}[!t]
  \hspace{0.10\textwidth}
    \begin{subfigure}[t]{0.15\textwidth}
        \centering
\begin{tikzpicture}
  \begin{feynman}
    \centering
    \vertex (a1);
    \vertex[right=1.0cm of a1] (a2);
    \vertex[above right=1.0cm and 1.0cm of a2] (b1);
    \vertex[below right=1.0cm and 1.0cm of a2] (b2);
    \vertex[above right=0.5cm and 1.0cm of b1] (c1);
    \vertex[below right=0.5cm and 1.0cm of b1] (c2);
    \vertex[above right=0.5cm and 1.0cm of b2] (d1);
    \vertex[below right=0.5cm and 1.0cm of b2] (d2);
    \vertex[below right=0.5cm and 1.0cm of d2] (d3);
    \vertex[above right=1.0cm and 1.0cm of d2] (e1);
    
    \diagram* {
      (a1) -- [dashed, edge label=\(H\)] (a2),
      (a2) -- [boson, edge label=\(Z\)] (b1),
      (a2) -- [boson, edge label'=\(Z\)] (b2),
      (b1) -- [anti fermion,edge label=\(\mu^+\)] (c1),
      (b1) -- [fermion,edge label'=\(\mu^-\)] (c2),
      (b2) -- [anti fermion, edge label=\(e^+\)] (d1),
      (b2) -- [fermion, edge label'=\(e^-\)] (d2),
      (d2) -- [fermion] (d3),
      (d2) -- [photon, edge label'=\(\gamma\)] (e1),
      
    };
  \end{feynman}
\end{tikzpicture}
\captionsetup{width=2\textwidth, justification=raggedright }
\caption{Photon radiation in the final state. \label{fig:realph}}
\label{fig:Hlllla}
\end{subfigure}
\hspace{0.20\textwidth}
\begin{subfigure}[t]{0.15\textwidth}
        \centering
\begin{tikzpicture}
  \begin{feynman}
   \centering
    \vertex (a1);
    \vertex[right=1.0cm of a1] (a2);
    \vertex[above right=1.0cm and 1.0cm of a2] (b1);
    \vertex[below right=1.0cm and 1.0cm of a2] (b2);
    \vertex[below=1.0cm of b1] (b3);
    \vertex[above right=0.5cm and 1.5cm of b1] (c1);
    \vertex[above right=0.5cm and 1.5cm of b3] (c2);
    \vertex[below right=0.2cm and 1.0cm of b2] (d1);
    \vertex[above right=0.5cm and 1.5cm of d1] (e1);
    \vertex[below right=0.5cm and 1.5cm of d1] (e2);
    
    \diagram* {
      (a1) -- [dashed, edge label=\(H\)] (a2),
      (a2) -- [boson, edge label=\(W^+\)] (b1),
      (a2) -- [boson, edge label'=\(W^-\)] (b2),
      (b1) -- [fermion,edge label=\(\nu_{e}\)] (b3),
      (b3) -- [boson,edge label=\(W\)] (b2),
      (b1) -- [anti fermion,edge label=\(e^+\)] (c1),
      (b3) -- [fermion,edge label'=\(e^-\)] (c2),
      (b2) -- [photon, edge label'=\(\gamma\)] (d1),
      (d1) -- [fermion, edge label=\(\mu^-\)] (e1),
      (d1) -- [anti fermion, edge label'=\(\mu^+\)] (e2),
      
    };
  \end{feynman}
\end{tikzpicture}
\captionsetup{width=2\textwidth, justification=raggedright }
\caption{Loop involving $W$.}
\label{fig:Hllllww}
\end{subfigure}

\vspace{1cm}

 \begin{subfigure}[t]{0.15\textwidth}
        \centering
\begin{tikzpicture}
  \begin{feynman}

    \vertex (a1);
    \vertex[right=1.0cm of a1] (a2);
    \vertex[above right=1.5cm and 1.0cm of a2] (b1);
    \vertex[below right=1.5cm and 1.0cm of a2] (b2);
    \vertex[below=1.0 cm of b1] (b3);
    \vertex[below=1.0 cm of b3] (b4);
    \vertex[above right=0.5cm and 1.5cm of b1] (c1);
    \vertex[above right=0.5cm and 1.5cm of b3] (c2);
    \vertex[below right=0.5cm and 1.5cm of b4] (e1);
    \vertex[below right=0.5cm and 1.5cm of b2] (e2);
    
    \diagram* {
      (a1) -- [dashed, edge label=\(H\)] (a2),
      (a2) -- [boson, edge label=\(Z\)] (b1),
      (a2) -- [boson, edge label'=\(Z\)] (b2),
      (b1) -- [anti fermion,edge label=\(e^+\)] (b3),
      (b3) -- [photon,edge label=\(\gamma\)] (b4),
      (b4) -- [fermion,edge label=\(\mu^-\)] (b2),
      (b1) -- [fermion,edge label=\(e^-\)] (c1),
      (b3) -- [anti fermion,edge label'=\(e^+\)] (c2),
      (b4) -- [anti fermion, edge label=\(\mu^+\)] (e1),
      (b2) -- [fermion, edge label'=\(\mu^-\)] (e2),
      
    };
  \end{feynman}
\end{tikzpicture}
\captionsetup{width=2\textwidth, justification=raggedright }
\caption{Loop with photon-exchange between different-flavour leptons.}
\label{fig:tt_dilep}
\end{subfigure}
\hspace{0.12\textwidth}
 \begin{subfigure}[t]{0.15\textwidth}
        \centering
\begin{tikzpicture}
  \begin{feynman}
    \centering
    \vertex (a1);
    \vertex[right=1.0cm of a1] (a2);
    \vertex[above right=1.5cm and 1.0cm of a2] (b1);
    \vertex[below right=1.5cm and 1.0cm of a2] (b2);
    \vertex[above right=0.5cm and 1.5cm of b1] (c1);
    \vertex[right=1.0 cm of b2] (b3);
    \vertex[below right=0.5cm and 1.5cm of b3] (c2);
    \vertex[above right=2.0cm and 1.0cm of b3] (c3);
    \vertex[above right=0.5cm and 1.0cm of c3] (e1);
    \vertex[below right=0.5cm and 1.0cm of c3] (e2);
    
    \diagram* {
      (a1) -- [dashed, edge label=\(H\)] (a2),
      (a2) -- [boson, edge label=\(Z\)] (b1),
      (a2) -- [boson, edge label'=\(Z\)] (b2),
      (b1) -- [anti fermion,edge label=\(\mu^+\)] (c1),
      (b1) -- [fermion,edge label=\(\mu^-\)] (b2),
      (b2) -- [fermion,edge label=\(\mu^-\)] (b3),
      (b3) -- [fermion,edge label=\(\mu^-\)] (c2),
      (b3) -- [boson,edge label=\(Z\)] (c3),
      (c3) -- [anti fermion, edge label=\(e^+\)] (e1),
      (c3) -- [fermion, edge label'=\(e^-\)] (e2),
      
    };
  \end{feynman}
\end{tikzpicture}
\captionsetup{width=2\textwidth, justification=raggedright }
\caption{Lepton-pair not originating from a $Z$ boson decay.}
\label{fig:HZZNLO}
\end{subfigure}
\hspace{0.18\textwidth}
\begin{subfigure}[t]{0.15\textwidth}
        \centering
\begin{tikzpicture}
 \begin{feynman}
    \centering
    \vertex (a1);
    \vertex[right=1.0cm of a1] (a2);
    \vertex[above right=1.0cm and 1.0cm of a2] (b1);
    \vertex[below right=1.0cm and 1.0cm of a2] (b2);
    \vertex[above right=0.5cm and 1.0cm of b1] (c1);
    \vertex[below right=0.5cm and 1.0cm of b1] (c2);
    \vertex[below right=1.0cm and 1.0cm of b2] (b3);
    \vertex[right=1.0cm of b3] (b4);
    \vertex[above right=0.5cm and 1.0cm of b4] (d1);
    \vertex[below right=0.5cm and 1.0cm of b4] (d2);
    
    \diagram* {
      (a1) -- [dashed, edge label=\(H\)] (a2),
      (a2) -- [boson, edge label=\(Z\)] (b1),
      (a2) -- [boson, edge label'=\(Z\)] (b2),
      (b1) -- [anti fermion,edge label=\(\mu^+\)] (c1),
      (b1) -- [fermion,edge label'=\(\mu^-\)] (c2),
      (b2) -- [fermion,half left] (b3),
      (b2) -- [anti fermion,half right] (b3),
      (b3) -- [boson, edge label=\(Z\)] (b4),
      (b4) -- [anti fermion, edge label=\(e^+\)] (d1),
      (b4) -- [fermion, edge label'=\(e^-\)] (d2),
      
    };
  \end{feynman}
\end{tikzpicture}
\captionsetup{width=2\textwidth, justification=raggedright}
\caption{$HVV^*$ genuine topology.}
\label{fig:corrections_loops}
\end{subfigure}
\caption{Representative diagrams contributing to NLO EW corrections for  the $H \rightarrow e^{+} e^{-} \mu^{+} \mu^{-}$ process. Besides \ref{fig:corrections_loops} all the displayed loop diagrams modify the intermediate state compared to the LO picture. The diagram~\ref{fig:Hlllla} is associated to QED final-state radiation.}
\label{fig:corrections}
\end{figure}
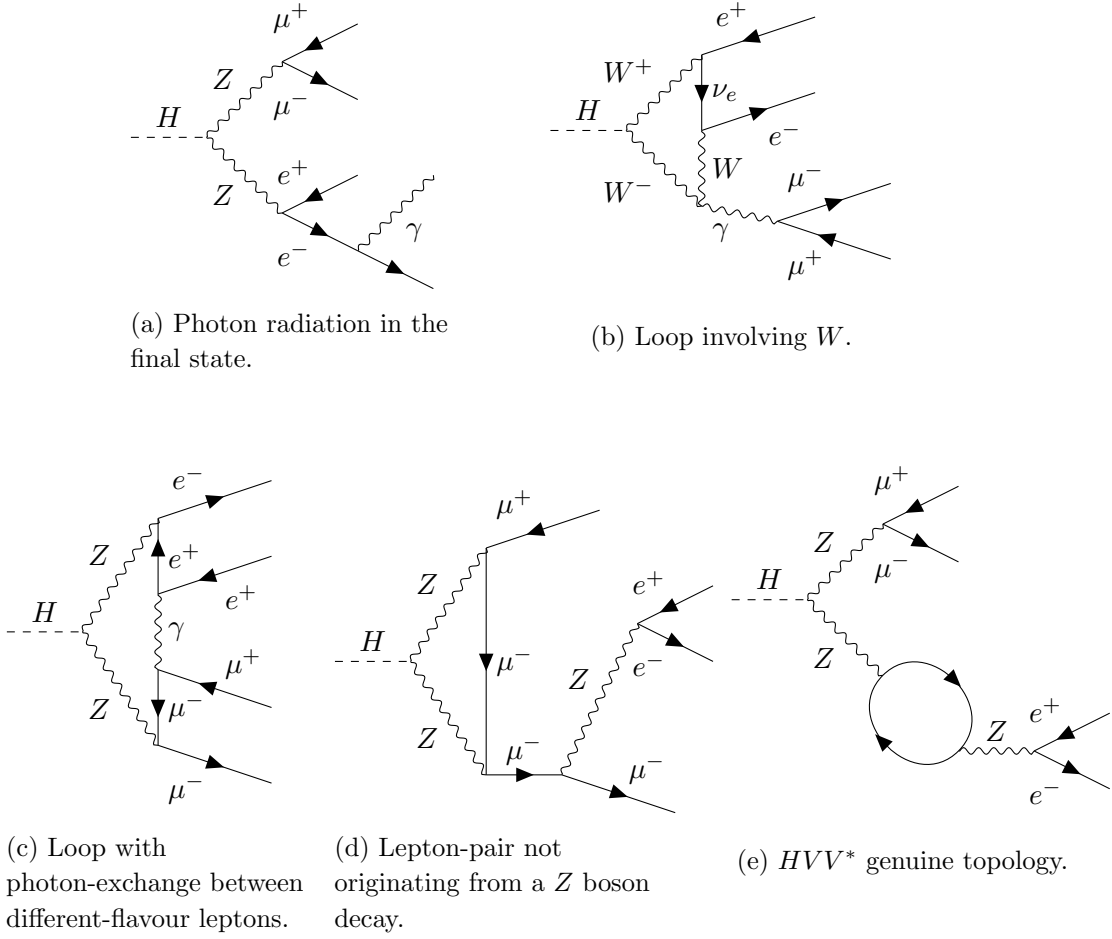

Representative diagrams for both the one-loop and real emission contributions to the NLO EW corrections for the $H (\rightarrow ZZ^*) \rightarrow e^{+} e^{-} \mu^{+} \mu^{-}$ decay are shown in Fig.~\ref{fig:corrections}. Looking at diagrams like those in Figs.~\ref{fig:Hllllww}, \ref{fig:tt_dilep} and \ref{fig:HZZNLO}, it is manifest that at NLO EW accuracy there are also contributions that do \textit{not} correspond to an intermediate $H \rightarrow ZZ^*$ decay. This is not a special feature of this process; in general, given a final state, higher-order corrections can involve a set of possible different intermediate states. Moreover, since at least one of the two $Z$ bosons must be off-shell in the $H (\rightarrow ZZ^*) \rightarrow e^{+} e^{-} \mu^{+} \mu^{-}$ decay, a theoretically consistent approximation for isolating only loop diagrams featuring a  $H \rightarrow ZZ^*$ topology (as, {\it e.g.}, the Double-Pole-Approximation \cite{Denner:2000bj}) is not possible, to the best of our knowledge. Simply selecting such diagrams would lead to gauge-dependent and therefore unphysical results. Moreover, in addition to the only possible spin structure that is present at LO (vector-vector), the  diagrams contributing to NLO EW corrections can in principle feature additional spin structures that involve also scalars and tensors.
  
From the previous argument it is therefore obvious that the QT approach, which relies on the presence of an intermediate $H \rightarrow ZZ^*$ decay whose spin density matrix can be studied via the angular distribution of the fermions emerging from the $Z$ decays, may be formally inconsistent. Even considering the gauge-dependent contributions with $H (\rightarrow ZZ^*) \rightarrow e^{+} e^{-} \mu^{+} \mu^{-}$ topology, such as the one in Fig.~\ref{fig:corrections_loops}, the LO picture of the QT approach discussed in Sec.~\ref{QMT} would be affected. Indeed, due to the NLO EW corrections to the decay, the spin-analysing power $\alpha$ would depend on the invariant mass of the corresponding fermion pair, and would be different for the two different pairs. 

Nevertheless, since NLO EW corrections typically induce relative effects at the percent level,\footnote{Notable exceptions are the EW Sudakov logarithms, which however are not relevant in this context, or final-state radiation (FSR) of photons from light fermions which instead is relevant in this study. The impact of the latter in fact can be strongly reduce via large-cone recombination. We will discuss in detail such aspect in Sec.~\ref{sec:resultsNLOEW}.} one may simply argue that, although formally inaccurate, the NLO EW corrections may be calculated and interpreted via the QT approach based on the LO picture. If EW corrections are really small, they may be simply ignored. Indeed, percent effects are  anyway negligible in this context. First, because they are anyway much smaller than experimental accuracy achievable in the near future. Second, because they do not alter any qualitative conclusion regarding the condition of a quantum state ({\it e.g.}~entangled {\it vs.} separable).

The problem is that it has been shown in Ref.~\cite{Grossi:2024jae} that, rather than small, NLO EW for some of the $C$ coefficients are giant, of the order of $-90\%$. Thus, trying to understand if the QT approach is reliable for the $H (\rightarrow ZZ^*) \rightarrow e^{+} e^{-} \mu^{+} \mu^{-}$ process and more in general for the $H (\rightarrow VV^*) \rightarrow f_1 \bar f_2 f_3 \bar f_4$ decay is imperative. This is precisely the core subject of this paper and will be scrutinised, based on the numerical results obtained, in Sec.~\ref{sec:resultsNLOEW}.
Before moving on to this discussion, in the following we describe the details of the settings and input parameters for the calculation of the NLO EW corrections that we have performed.

\medskip

The renormalisation of the EW sector is performed in the so-called  $G_\mu$-scheme, with the input parameters already listed in Eqs.~\eqref{eq:inputLOandNLO} and \eqref{eq:inputNLO}. In order to dynamically treat on-shell and off-shell configurations for the two $Z$ bosons, we adopt the Complex-Mass-Scheme \cite{Denner:2006ic}, with the implementation described in Ref.~\cite{Frederix:2018nkq}. Since the leptons in the final state are treated as massless, to achieve IR-finiteness at the differential level they have to be recombined with photons into dressed leptons. In particular, we  cluster  leptons with photons into dressed leptons if
\begin{equation}
\Delta \hat R(\ell,\gamma) <\Delta  R\,,
\label{eq:Rphoton}
\end{equation}
where $\Delta  R$ is a positive value that has to be set in the calculation and $\Delta \hat R(\ell,\gamma)$ is defined as $\Delta \hat R (\ell,\gamma) \equiv \sqrt{(\Delta \phi)^2+(\Delta \eta)^2}$. The quantities $\Delta \phi$ and $\Delta \eta$ are respectively the azimuthal angle between $\ell$ and $\gamma$ and the difference between their pseudorapidities.

\subsection{Numerical results and interpretation}
\label{sec:resultsNLOEW}

\begin{figure}[t!]
     \centering
     \begin{subfigure}[b]{0.49\textwidth}
         \centering
         \includegraphics[width=\textwidth]{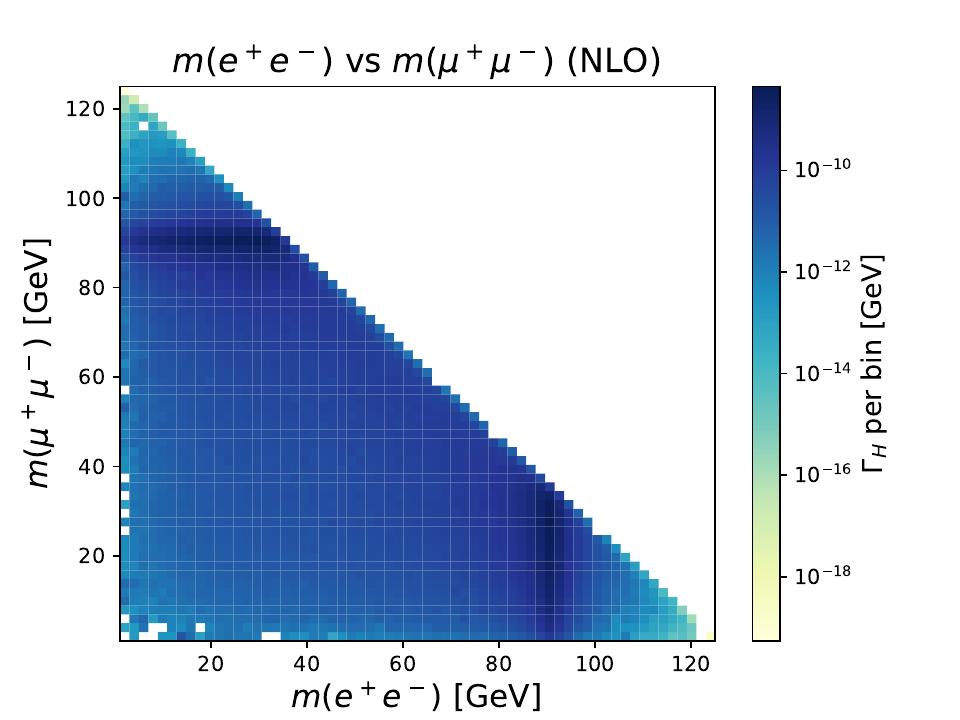}
         \caption{}
         \label{fig:GammaNLO}
     \end{subfigure}
     \hfill
     \begin{subfigure}[b]{0.49\textwidth}
         \centering
         \includegraphics[width=\textwidth]{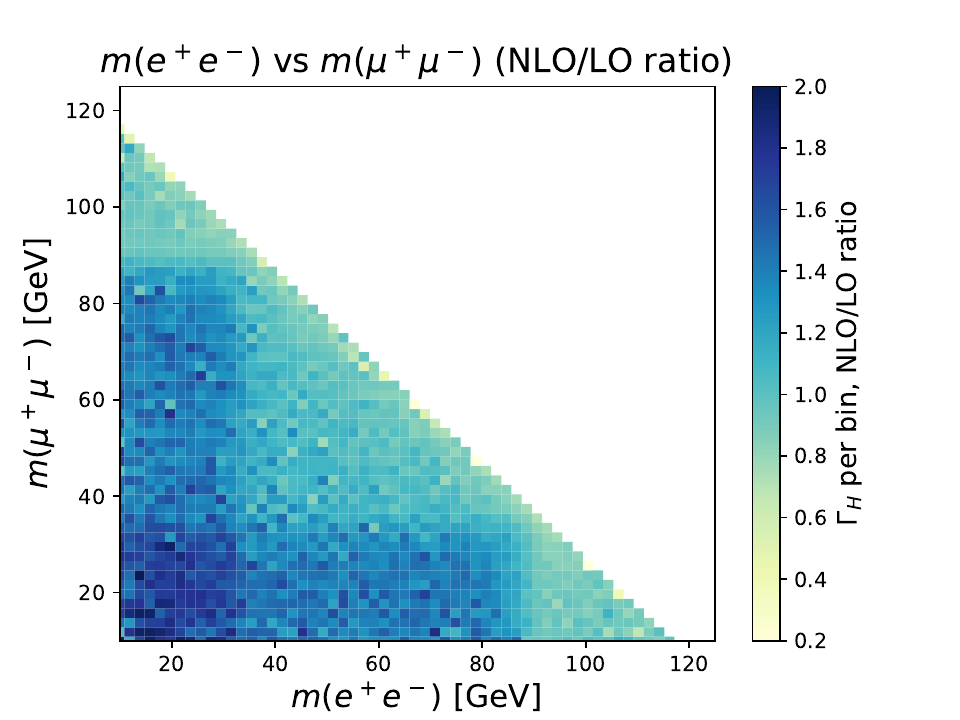}
         \caption{}
         \label{fig:GammaNLOLO}
     \end{subfigure}
    \captionsetup{width=.9\textwidth}
    \caption{(a) Partial decay width as a function of the invariant masses $m(e^+e^-)$  and $m(\mu^+ \mu^-)$ at NLO. The ratio with respect to the LO is shown in (b).}
    \label{fig:Gamma}
\end{figure}

\begin{figure}[t!]
     \centering
     \begin{subfigure}[b]{0.49\textwidth}
         \centering
         \includegraphics[width=\textwidth]{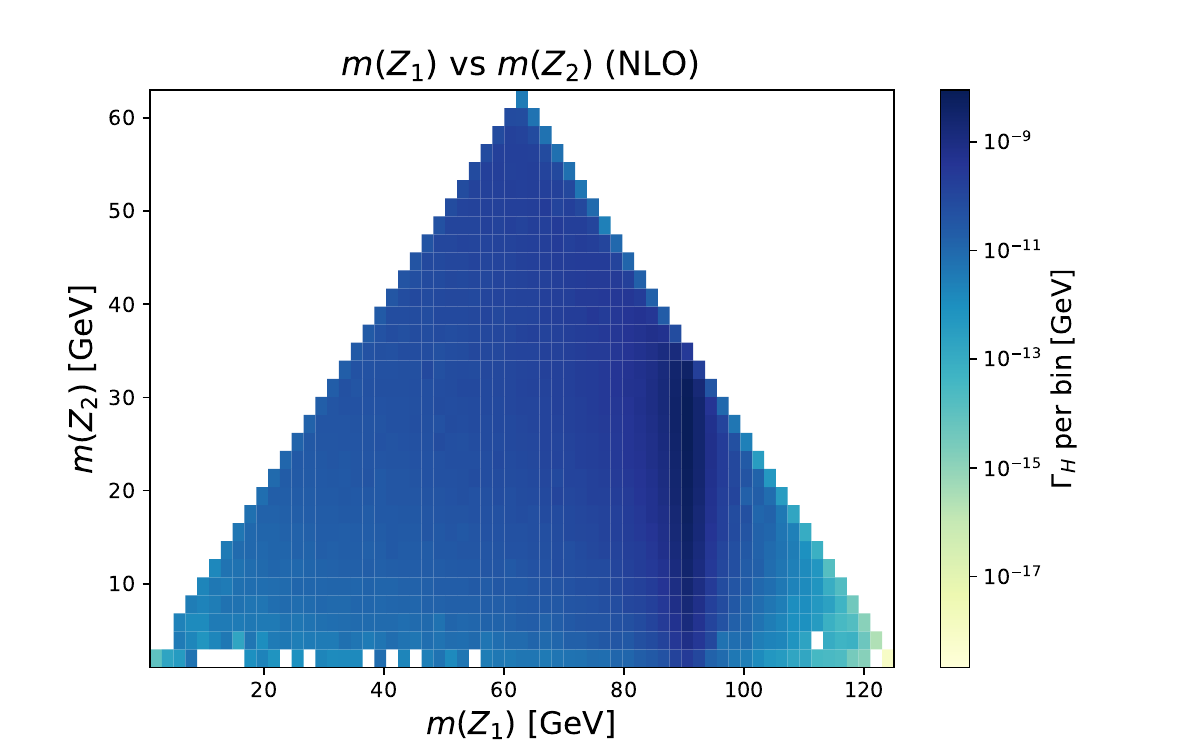}
         \caption{}
         \label{fig:GammaNLO_Z1Z2}
     \end{subfigure}
     \hfill
     \begin{subfigure}[b]{0.49\textwidth}
         \centering
         \includegraphics[width=\textwidth]{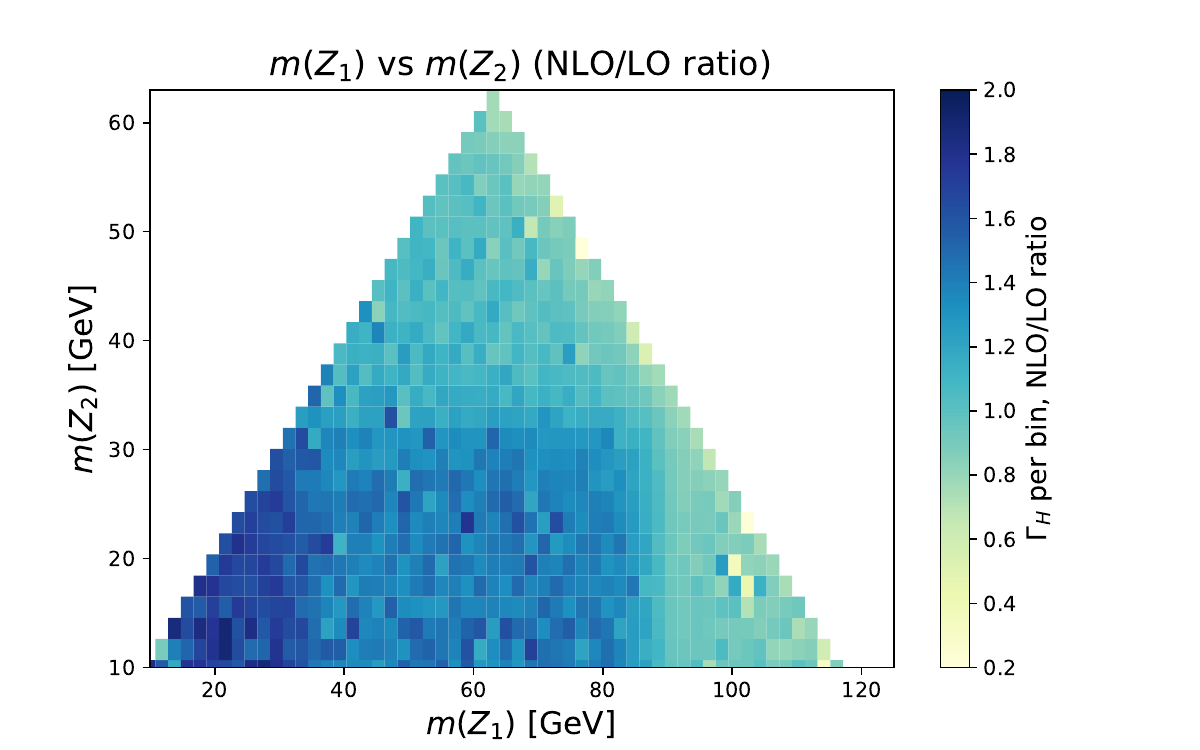}
         \caption{}
         \label{fig:GammaNLOLO_Z1Z2}
     \end{subfigure}
    \captionsetup{width=.9\textwidth}
    \caption{Same as Fig.~\ref{fig:Gamma} but as a function of  $m(Z_1)$ and $m(Z_2)$. \label{fig:Gamma_Z1Z2}}
\end{figure}

Before discussing the impact of NLO EW corrections on $C$ and $A$ coefficients and on the $\rho$ matrix, we discuss the NLO EW prediction for the partial decay width $\Gamma_H$ at the differential level,  w.r.t.~the mass invariants of the lepton pairs, as done at LO in Fig.~\ref{figG}. In Fig.~\ref{fig:Gamma} we consider the case in which the two invariant masses are $m(e^+e^-)$ and $m(\mu^+\mu^-)$, as in Fig.~\ref{figGa}, while in Fig.~\ref{fig:Gamma_Z1Z2} they are $m(Z_1)$ and $m(Z_2)$, as in Fig.~\ref{figGb}. In both figures we show in the left plot the predictions at NLO EW accuracy, and in the right plot the corresponding $K$-factor, {\it i.e.}, the ratio between the predictions at NLO EW and LO accuracies. The results have been obtained setting $\Delta R=0.1$.

As can be seen from the plots, the overall event distribution is similar to the one at LO, but it is also clear that the impact of the NLO EW corrections is not uniformly distributed across the plane. It is well known that the invariant mass of two leptons emerging from a $Z$ decay receives very large and positive corrections for invariant masses smaller than the mass of the $Z$, $m(Z)<M_Z$, especially close to the value $M_Z$, while negative and milder corrections for  $m(Z)>M_Z$. This effect is due to the radiative return induced by the large fraction of events populating the region $m(Z)\simeq M_Z$, which after the emission of a photon, if it is not recombined, migrates to values of $m(Z)<M_Z$ where at LO the rate is much smaller. The net effect is a relatively large correction. In the case of the Higgs decay there are two $Z$ bosons, therefore this effect happens once if $m(Z_1)< M_Z $ and twice if $m(Z_1)< (m_H- M_Z) $. Indeed, in the former case only $Z_1$ could be associated to a pair of leptons $\ell^+\ell^-$ such that $m(\ell^+\ell^-\gamma)\simeq M_Z$, while in the latter case such condition could be satisfied by both $Z_1$ and $Z_2$.

The previous argument clearly explains the shape of the $K-$factor in the $(m(Z_1),m(Z_2))$ plane (Fig.~\ref{fig:GammaNLOLO_Z1Z2}), and in turn also the one observed in the $(m(e^+e^-),m(\mu^+\mu^-))$ plane (Fig.~\ref{fig:GammaNLOLO}). In view of the following discussion, we note that the region where the impact of NLO EW corrections on $\Gamma_H$ is smaller corresponds to $m(Z_2) > 30$ GeV.

\medskip

We move now to the discussion of NLO EW corrections on $C$ and $A$ coefficients and their impact on the $\rho$ matrix.
The only inputs required for the reconstruction and QT approach described in Sec.~\ref{QMT}, and already implemented at LO in Sec.~\ref{sec:LO}, are the momenta of the four final-state leptons.\footnote{We consider the reference frame of the Higgs boson and not of the four final-state leptons. One should notice that at the experimental level the requirement $m(4\ell)\simeq m_H$ is imposed, such that the two definitions are equivalent. Both choices have been studied in Ref.~\cite{Grossi:2024jae} and we did not explore again this aspect. }   Consequently, the same method used to extract the $C$ and $A$ coefficients of the spin density matrix can also be applied to NLO simulations, using dressed leptons. As already mentioned in Sec.~\ref{sec:theoNLOEW}, this approach makes some important approximations, indeed it tries to restrict the complex structure depicted in Fig.~\ref{fig:corrections} to a simple structure of two intermediate $Z$ bosons decaying to leptons. While the angular distributions of the leptons are properly defined at NLO EW accuracy, the extraction of the $C$ and $A$ coefficients relies on a LO picture: the $\Gamma$ matrix at LO, see Eq.~\eqref{pmatrix}, parameterised by the spin-analysing power $\alpha$, which is also formally defined only at LO.

\begin{table}[t!]
\begin{center}
\renewcommand{\arraystretch}{1.5}
\begin{tabular}{ c | c | c | c ||}
\cline{2-4}
                       & LO & NLO & NLO$\,$/$\,$LO \\ \hhline{ - = = = }
\multicolumn{1}{|| c ||}{$A^1_{2,0}$} & $-$0.592(1) & $-$0.509(2) & 0.860(2) \\ 
\multicolumn{1}{|| c ||}{$A^2_{2,0}$} & $-$0.591(1) & $-$0.565(2) & 0.956(2) \\ \hline  \hline
\multicolumn{1}{|| c ||}{$C_{2,1,2,-1}$} & $-$0.937(2) & $-$0.943(4) & 1.006(3) \\ 
\multicolumn{1}{|| c ||}{$-C_{1,1,1,-1}$} & $-$0.94(1) & $-$0.16(2) & 0.17(2) \\ \hline \hline
\multicolumn{1}{|| c ||}{$A^1_{2,0}/\sqrt{2} + 1$} & 0.5817(7) & 0.640(1) & 1.101(2) \\ 
\multicolumn{1}{|| c ||}{$C_{2,2,2,-2}$} & 0.581(3) & 0.568(4) & 0.977(6) \\ 
\multicolumn{1}{|| c ||}{$-C_{1,0,1,0}$} & 0.59(1) & 0.03(2) & 0.06(4) \\ \hline \hline
\multicolumn{1}{|| c ||}{$C_{2,0,2,0}$} & 1.418(3) & 1.400(5) & 0.987(3) \\ 
\multicolumn{1}{|| c ||}{$C_{1,0,1,0} + 2$} & 1.41(1) & 1.97(2) & 1.39(1) \\ \hline
\end{tabular}
\captionsetup{width=.95\textwidth}
\caption {Non-vanishing $A$ and $C$ coefficients of the spin density matrix calculated using a LO simulation or including NLO EW corrections, the third column reports the ratio between the two results. The different blocks of rows highlight the equalities between the coefficients that should be present at LO. No phase-space restrictions are applied. The error on the last digit is in parentheses.}
\label{tab:nonzero_coeff}
\end{center}
\vspace{-5mm}
\end{table}

In Tab.~\ref{tab:nonzero_coeff} we show all the non-vanishing $C$ and $A$ coefficients at LO and NLO EW accuracy. Each horizontal block of the table displays coefficients (or relations) that have the same value at LO, but not at NLO EW, as we will discuss. The first interesting aspect to notice is that the same non-zero coefficients found at LO are still the only non-vanishing coefficients at NLO, as this is imposed from the geometry of the system. Another interesting observation is that the effects of the NLO corrections are uneven across the various coefficients. While some coefficients are only mildly impacted, with corrections ranging from 1$\%$ to 14$\%$, the $C_{L_1,M_1,L_2,M_2}$ coefficients with $L_1=L_2=1$ ($C_{1,1,1,-1}$ and $C_{1,0,1,0}$) are significantly modified at NLO EW, with deviations exceeding 90$\%$, in agreement with the findings of Ref.~\cite{Grossi:2024jae}.

\begin{figure}[t!]
     \centering
     \begin{subfigure}[b]{0.49\textwidth}
         \centering
         \includegraphics[width=\textwidth]{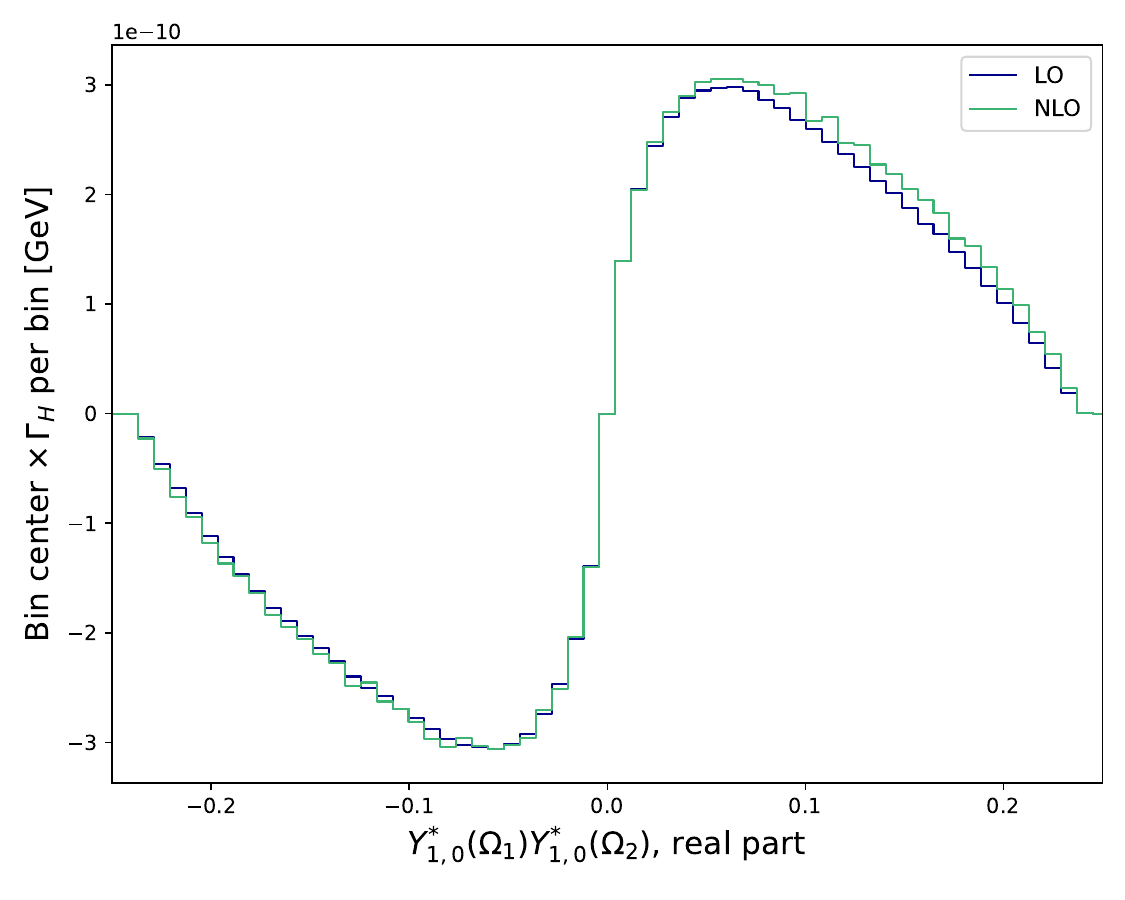}
         \caption{}
         \label{figa}
     \end{subfigure}
     \begin{subfigure}[b]{0.49\textwidth}
         \centering
         \includegraphics[width=\textwidth]{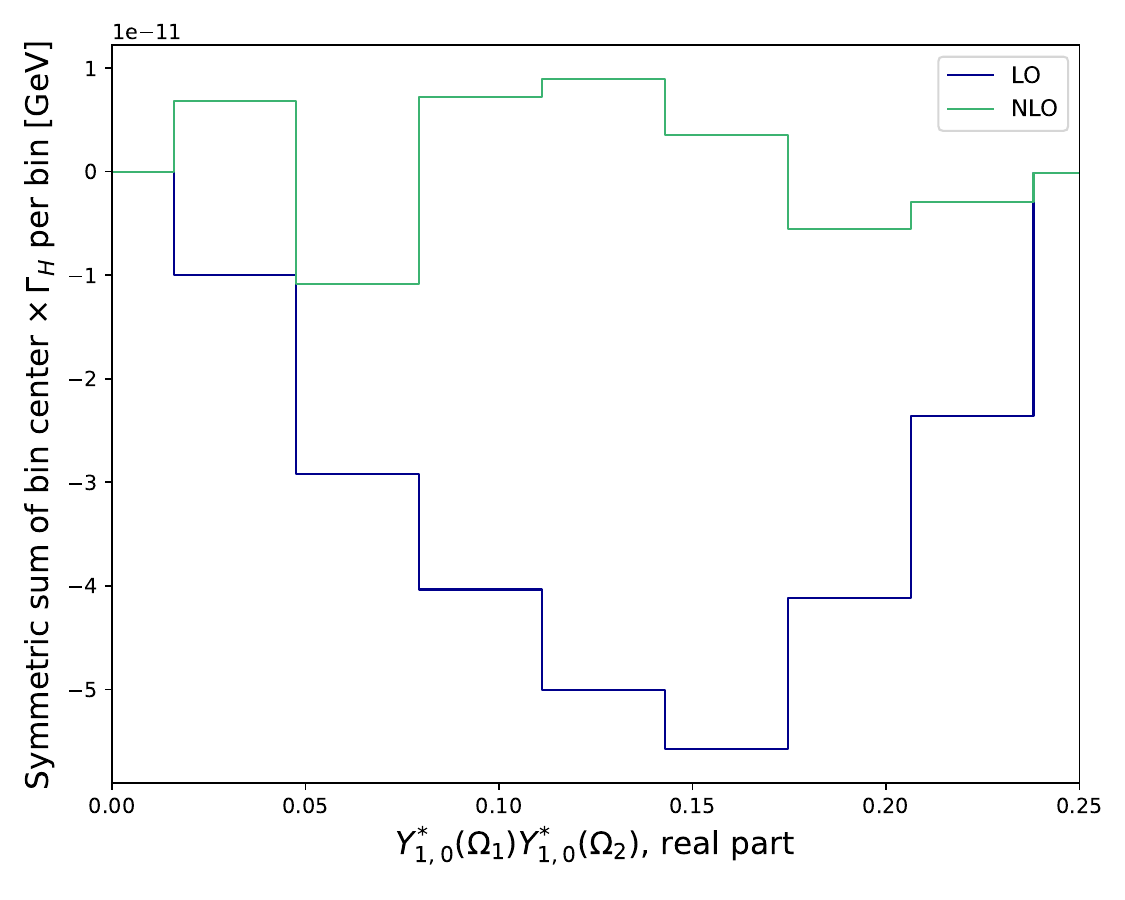}
         \caption{}
         \label{figb}
     \end{subfigure}
        \caption{Left Plot: $R_{YY} \times \Gamma_H(R_{YY})$ for $R_{YY} = \Re\left[Y^*_{1,0}Y^*_{1,0}\right]$, the spherical harmonic relevant for $C_{1,0,1,0}$, at LO and NLO. Right Plot: prediction for $R_{YY} \times \left(\Gamma_H(R_{YY})-\Gamma_H(-R_{YY})\right)$, see main text for more details. }
        \label{fig:c1010_NLO}
\end{figure}

\begin{figure}[t!]
     \centering
     \begin{subfigure}[b]{0.49\textwidth}
         \centering
         \includegraphics[width=\textwidth]{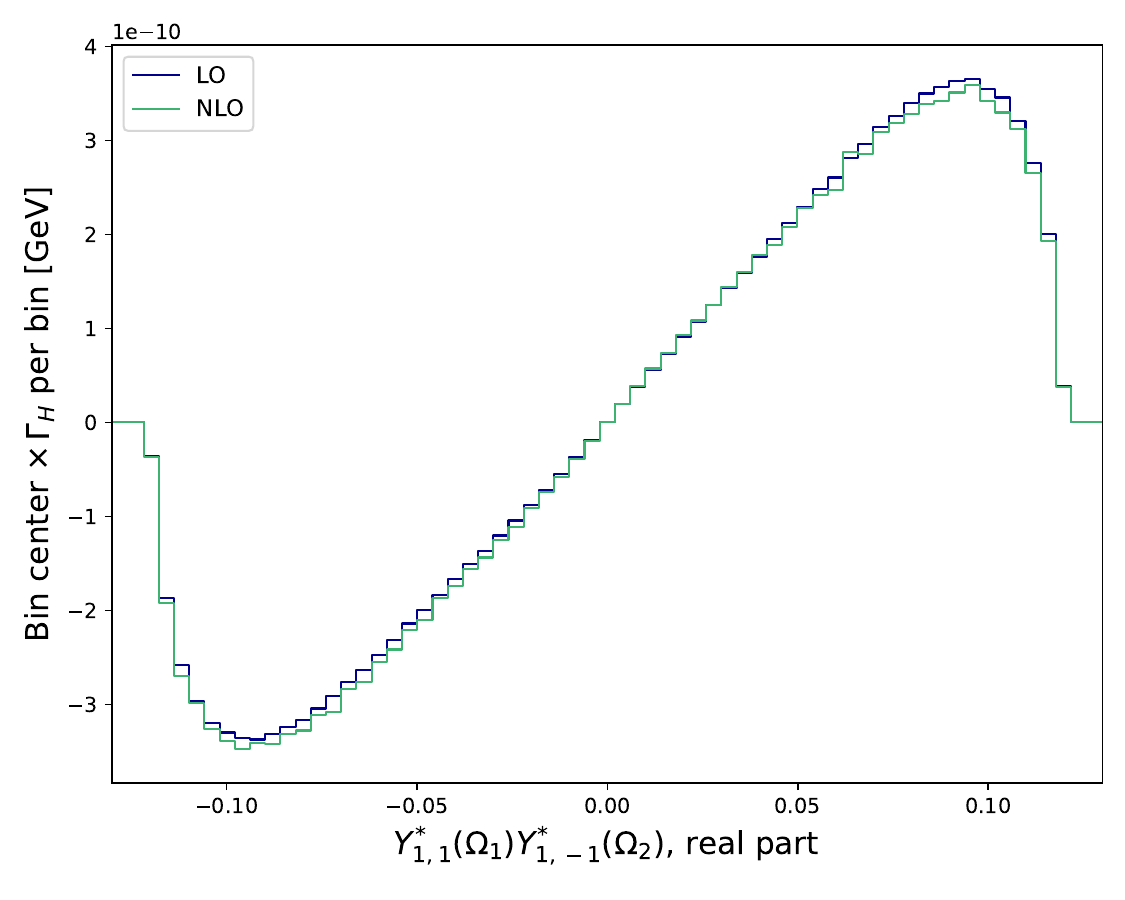}
         \caption{}
         \label{figa}
     \end{subfigure}
     \begin{subfigure}[b]{0.49\textwidth}
         \centering
         \includegraphics[width=\textwidth]{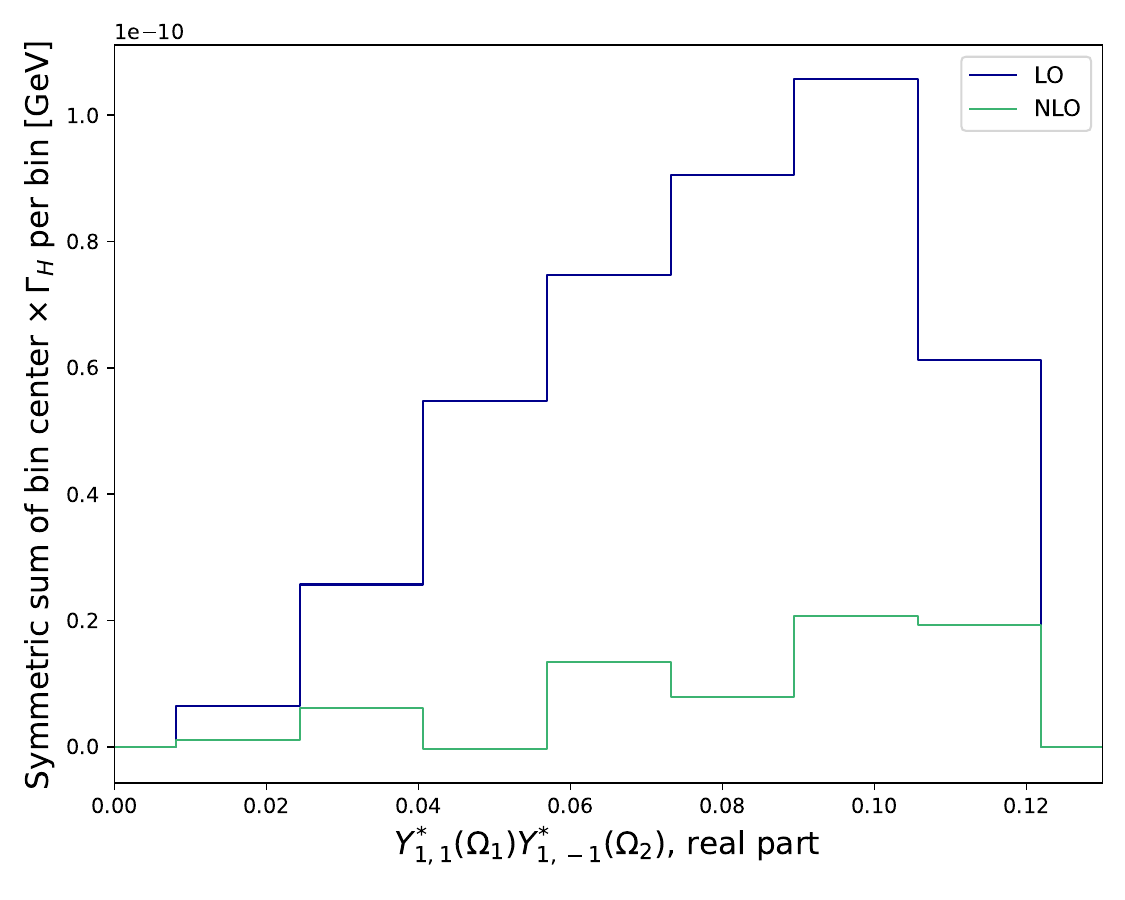}
         \caption{}
         \label{fig:c1111_NLO:figb}
     \end{subfigure}
        \caption{Same as Fig.~\ref{fig:c1010_NLO} for $C_{1,1,1,-1}$.}
        \label{fig:c1111_NLO}
\end{figure}

In order to better visualise the significant variations caused by NLO EW corrections on such coefficients, which we will denote in the following discussion as $C_{1,M,1,-M}$, two distributions are shown in Figs.~\ref{fig:c1010_NLO} and \ref{fig:c1111_NLO}, for $C_{1,0,1,0}$ and $C_{1,1,1,-1}$ respectively. While in Fig.~\ref{fig:LOcoeff} we have plotted $\Gamma_H$ for different bins of $R_{YY}$, for two different $R_{YY}$ functions, in Figs.~\ref{fig:c1010_NLO} (Figs.~\ref{fig:c1111_NLO}) we bin in $R_{YY}=\Re\left[Y^*_{1,0}Y^*_{1,0}\right]$ ($R_{YY}=\Re\left[Y^*_{1,1}Y^*_{1,-1}\right]$) and show both the LO (blue) and NLO (green) predictions. In the left panels, we display $R_{YY} \times \Gamma_H(R_{YY})$, while in the right panels, for $R_{YY}>0$ only, we plot $R_{YY} \times \left(\Gamma_H(R_{YY})-\Gamma_H(-R_{YY})\right)$. In both cases, summing the values across all bins yields a term proportional to the corresponding $C_{1,M,1,-M}$ coefficient, namely the contribution appearing in the second line of Eq.~\eqref{eq:numerical_INT}.

Consistently with the results shown in Tab.~\ref{tab:nonzero_coeff}, the right plots of both Figs.~\ref{fig:c1010_NLO} and \ref{fig:c1111_NLO} show that, while the prediction is negative at LO, it almost goes to zero at NLO. On the other hand, this effect is not clearly visible on the left plots of the same figures. The comparison between the left and right plots shows that the values of the $C_{1,M,1,-M}$ coefficients originate from large cancellations in the integral over the spherical harmonics. For $C_{1,M,1,-M}$, the NLO EW corrections lead to a symmetrisation of the distributions in the left plots, and consequently to much smaller predictions for $C_{1,M,1,-M}$. As it will be clear from the discussion later in this section, this symmetrisation is accidental.

Another surprising effect of the NLO EW corrections is the asymmetry between the two $A$ coefficients. In particular, NLO EW corrections have a much larger impact on $A^{1}_{2,0}$ ($\sim -15 \%$ {\it vs.} on $A^{2}_{2,0}$~$\sim -5 \%$), which is related to the $Z$ boson with leading mass. Before discussing the origin of these large effects on $A^{1}_{2,0}$ and the $C_{1,M,1,-M}$ coefficients, we analyse their implications, especially for the interpretation of the $\rho$ spin density matrix.

\medskip

The uneven impact of the NLO corrections on the various $A$ and $C$ coefficients has significant consequences, in particular it breaks the relations in Eqs.~\eqref{eq:veccoeff}--\eqref{eq:coeff_LOrelation1ok}. This is clearly visible in Tab.~\ref{tab:nonzero_coeff}, where a comparison within the same blocks shows that these relations hold at LO, but are largely broken when including NLO EW corrections, especially those involving $C_{1,M,1,-M}$ coefficients. 
As a result, several of the cancellations that occur at LO, leading to a simplified form of the spin density matrix, are no longer present at NLO, resulting in a different structure for $\rho$:
\begin{multline}
\label{NLOrho}
\rho_{\text{NLO}} = \\[3mm]
{\small
\begin{pmatrix}
\teal{~0.099(4)} & \cdot& \cdot& \cdot& \cdot& \cdot& \cdot& \cdot& \cdot\\
 \cdot& 0.004(2) & \cdot& \teal{0.131(4)} & \cdot& \cdot& \cdot& \cdot& \cdot\\
 \cdot& \cdot& \blue{0.111(4)} & \cdot& \blue{-0.183(4)} & \cdot& \blue{0.189(1)} & \cdot& \cdot\\
 \cdot& \teal{0.131(4)} & \cdot& -0.009(2) & \cdot& \cdot& \cdot& \cdot& \cdot\\
 \cdot& \cdot& \blue{-0.183(4)} & \cdot& \blue{0.591(1)} & \cdot& \blue{-0.183(4)} & \cdot& \cdot\\
 \cdot& \cdot& \cdot& \cdot& \cdot& -0.009(2) & \cdot& \teal{0.131(4)} & \cdot\\
 \cdot& \cdot& \blue{0.189(1)} & \cdot& \blue{-0.183(4)} & \cdot& \blue{0.110(3)} & \cdot& \cdot\\
 \cdot& \cdot& \cdot& \cdot& \cdot& \teal{0.131(4)} & \cdot& 0.004(2) & \cdot\\
 \cdot& \cdot& \cdot& \cdot& \cdot& \cdot& \cdot& \cdot& \teal{0.099(3)~} \\
\end{pmatrix} }.
\end{multline}
\smallskip
In this matrix, entries that are already present at LO are highlighted in blue, while the new NLO entries which are at least of order 0.01 are shown in green.\footnote{We will see later in the text that this distinction is not only quantitative but also qualitative.} Some of these new entries are comparable to, or even larger than, the LO entries. Looking at the basic requirements for a spin density matrix it is visible that $\rho_{\text{NLO}}$ is ill defined, as it violates the condition:
\begin{equation}
    |\rho_{ij}|^2 \leq \rho_{ii} \rho_{jj}, \quad i \neq j \label{eq:rhoNLObad}
\end{equation}
and also leads to some negative eigenvalues. Thus, at this point it is already clear that the naive application of the LO QT approach onto NLO EW accurate results is not feasible for the $H\to e^+ e^- \mu^+\mu^-$ decay. Not only NLO EW corrections are large, but they disrupt basics property of the $\rho$ density matrix. From this point the next step can be taken towards two different directions. First, declare the QT approach inconsistent with radiative corrections.  Second, try to understand the origin of this situation and find possible alternative methods or QI variables for probing the quantum correlations of two-qutrit system, which is nonetheless still present within the $H\to e^+ e^- \mu^+\mu^-$ decay. In the following, we pursue the latter direction and focus in particular on understanding whether this situation is specific to the $H \to e^+ e^- \mu^+ \mu^-$ decay or represents a more general issue for the QT approach. We anticipate and spoil the surprise: yes, it is peculiar to this decay, and more precisely, it is a feature of the four-lepton final state.

We have performed several tests in order to investigate the origin of these large corrections. In the following, we report all the tests carried out and show explicit results for some of them.
First of all, we simply tried to modify the renormalisation scheme, using instead of the $G_{\mu}$ scheme the so-called $\alpha(M_Z)$ scheme, where the input parameters are $\alpha_{\rm EW}, M_W$ and $M_Z$. While, in general, the NLO/LO ratio for the $C$ and $A$ coefficients changes at the 1\% level, consistently with what is typically observed in similar comparisons for other observables, the situation is different for the $C_{1,M,1,-M}$ coefficients. For $C_{1,1,1,-1}$ we obtain in the $\alpha(M_Z)$ scheme NLO/LO = 0.02, so a $-14\%$ difference relative to the LO w.r.t.~the $G_{\mu}$ result in Tab.~\ref{tab:nonzero_coeff}. For $C_{1,0,1,0}$ we find in the $\alpha(M_Z)$ scheme NLO/LO = $-0.06$, so a $-9\%$ difference relative to the LO w.r.t.~the $G_{\mu}$ result. This first test indicates a large sensitivity to the input parameters and the renormalisation condition.

Next, we checked if the origin of this pattern is related to the presence of an off-shell $Z$ boson. We have increased the mass of the Higgs boson to 250 GeV, such that both $Z$ bosons can be on-shell, while keeping all the interactions and independent parameters fixed to the SM values.  Also in this case we see very large corrections only on the $C_{1,M,1,-M}$ coefficients, although very different from the SM case with $m_H=125$ GeV, {\it i.e.}, of the order of 55$\%$.

These results raise a question about the validity of the approximations made in the derivation of the spin density matrix, in particular on the lack of corrections applied to the $\Gamma$ matrix of Eq.~\eqref{pmatrix} and especially the spin-analysing power $\alpha$. 
Indeed, even ignoring the theoretical problems introduced in Sec.~\ref{sec:theoNLOEW}, one can see that the determination of the $C$ and $A$ coefficients critically depends on the value of the $B$ (see Eqs.~\eqref{eqn:coeff_defC} and \eqref{eq:Bcoeffs}, and the discussion around Eqs.~\eqref{eq:G111-1} and \eqref{eq:G1010}). In particular, the $C_{1,M,1,-M}$ coefficients are the only ones that depend directly on the spin-analysing power $\alpha$. Even more important, they are normalised by $\alpha^{2}$, which for charged leptons is accidentally a very small number ($\alpha^2\simeq 0.048$), see also Tab.~\ref{tab:alphavalue}.

The smallness of the parameter $\alpha$ for charged leptons in $Z$ decays originates from the fact that $\sin^2\theta_W\simeq 1/4$. This condition is accidental, and $\alpha$ would be exactly equal to zero for $\sin^2\theta_W=1/4$. For this reason, the $\alpha$ parameter is extremely sensitive to the value of $\sin^2\theta_W$. As an example, if $M_W$ is increased by a $0.1\%$, $\sin^2\theta_W$ decreases by $-0.7\%$ and $\alpha$ increases by $5.5 \% $. It is therefore plausible that NLO EW corrections induce effects that in first approximation are equivalent to changing the effective value of $\alpha$ by a few tens of percents. However, performing QT and using the LO value of $\alpha$, as we are doing, these effects are erroneously propagated to the extraction of the $\rho$ matrix for the $H\to Z Z^*$ system. In other words, $\rho$ is not receiving large corrections, the $\Gamma$ matrix, and especially $\alpha$ within it, is receiving large corrections. However, we cannot account for them, and the effects propagate to $\rho$. Indeed, as already mentioned  in Sec.~\ref{sec:theoNLOEW}, to the best of our knowledge the $\Gamma$ matrix for an off-shell $Z$ boson at NLO EW is not a gauge-invariant quantity. Moreover, it would depend on the virtuality of the $Z$ boson. 

Nevertheless, we tested whether it was possible to obtain the structure of the $\rho_{\text{NLO}}$ matrix by simply assuming that all effects arise from the $\Gamma$ matrix, {\it i.e.}, in reality $\rho_{\text{NLO}}=\rho_{\text{LO}}$ and the effects we see are just a propagation of the error in the value of $\alpha$ on $\rho_{\text{NLO}}$. We find that if we assume that all the effects on the $\Gamma$ matrix can be obtained by rescaling $\alpha$ by a factor $\sim$ 0.39,  all the green and blue entries of $\rho_{\text{NLO}}$ in Eq.~\eqref{NLOrho} can be obtained with a high level of accuracy from $\rho_{\text{LO}}$. In Appendix \ref{app:wrongalpha} we show how we have derived this value. It is interesting to see that a rescaling of a factor 0.39 on $\alpha$, so $-61\%$ corrections, can be induced by reducing $M_W$ by 1.1\%, or equivalently increasing $\sin^2\theta_W$ by 7.7$\%$. In other words, it is not so unrealistic that the typical percent effects induced by EW corrections translate to several tens of percents on the effective $\alpha$ at NLO EW, and in turn via QT to the extraction of $\rho_{\text{NLO}}$. 

It is important to note that some of the effects induced by the NLO EW corrections on $\rho_{\text{NLO}}$ or the $C$ and $A$ coefficients, still,  cannot be explained via the previous argument based on an effective $\alpha$ at NLO EW and the assumption of minimal of NLO EW corrections for $\rho$ ($\rho_{\text{NLO}}\simeq\rho_{\text{LO}}$). These effects include: the different impact of the NLO EW corrections on $C_{1,1,1,-1}$ and $C_{1,0,1,0}$, the small corrections present for all the other $C$ and $A$ coefficients, the different value of $A^1_{2,0}$ and $A^2_{2,0}$ at NLO EW accuracy and, due to these three aforementioned features,  the non-vanishing entries of the matrix in Eq.~\eqref{NLOrho} that are in black. Thus, this conjecture of the effective $\alpha$ value at NLO EW should not be used to correct $\rho_{\text{NLO}}$, but rather considered as a helpful argument in order to understand the origin of such large corrections for the $C_{1,M,1,-M}$ coefficients and how is possible that $\rho_{\text{NLO}}$ leads to negative eigenvalues and  Eq.~\eqref{eq:rhoNLObad}. 

To further support the idea that these large corrections originate from the small value of $\alpha$, we have performed additional tests. First, we have tried to change $M_Z$ in order to increase/decrease the value of $|\alpha|$. We see that as $|\alpha|$ increases, the impact of NLO EW corrections largely decreases. On the contrary, as $|\alpha|$ decreases, the impact of NLO EW corrections grows substantially and diverges in the limit $\sin^2\theta_W\to 1/4$. All these effects can be again largely explained via the aforementioned conjecture of the effective $\alpha$ value at NLO EW.

As a second step, instead of varying the SM input parameters, we considered similar processes in the SM where $\alpha$ is different: the generic $H\rightarrow V V^* \to 4f$. The purpose of this test is twofold. On one hand, we continue to check if the large NLO EW corrections to $C_{1,M,1,-M}$ are due to the smallness of $\alpha$. On the other hand, we investigate if this problem is peculiar only for the $H\rightarrow Z Z^* \to 4\ell$ case.

We have calculated the NLO EW corrections for the $H\rightarrow q\bar{q}q'\bar{q'}$ process, where $(q,q')=(u,c)$ or $(q,q')=(c,s)$, and all quarks are considered massless.  As can be seen in Tab.~\ref{tab:alphavalue}, for leptons $\alpha^2\simeq 0.048 $, for up-type quarks $\alpha^2\simeq 0.489$ and for down-type quarks $\alpha^2\simeq 0.885$.  Thus, the size of the NLO EW corrections for the $C_{1,M,1,-M}$ is expected to be very different in each final state. Given the dependence of the EW corrections in general on the particle electric charge, further effects are present for the other coefficients as well, but we will focus the discussion of the comparison on $C_{1,M,1,-M}$.  

The results of our calculations are shown in Tab.~\ref{tab:nonzero_coeff_qq}, where we list the relative impact of NLO EW corrections for all the non-vanishing coefficients. It is manifest that their impact is reduced for final states with quarks. Especially, the impact on the $C_{1,M,1,-M}$ coefficients drastically changes. While the relative NLO EW corrections are of the order of 90$\%$ for charged leptons, they reduce to $\approx 20\%$ for up-type quarks and to only a few percents for down-type quarks. We have calculated the NLO EW corrections for the process $H\rightarrow WW^{*}\rightarrow e^{+}\nu_{e} \mu^{-}\bar{\nu_{\mu}}$, where $\alpha^2=1$, {\it i.e.}, even larger than the case of down-type quarks for  $ZZ^*$. We see corrections of at most a few percents for all $C$ coefficients, including the $C_{1,M,1,-M}$ .

\begin{table}[t!]
\begin{center}
\renewcommand{\arraystretch}{1.5}
\begin{tabular}{ c | c | c | c ||}
\cline{2-4}
           $\frac{\rm NLO}{\rm LO} -1~[\%]$             & $\ell^{+}\ell^{-}\ell'^{+}\ell'^{-}$ & $u\bar{u}c\bar{c} $ & $d\bar{d}s\bar{s}$\\ \hhline{ - = = = }
\multicolumn{1}{|| c ||}{$A^1_{2,0}$} & $-$14.0 & $-$5.2 & $-$1.1 \\ 
\multicolumn{1}{|| c ||}{$A^2_{2,0}$} & $-$4.4 & $-$1.0 & $< 1$ \\ \hline  \hline
\multicolumn{1}{|| c ||}{$C_{2,1,2,-1}$} & $< 1$ & $< 1$ & $< 1$ \\ 
\multicolumn{1}{|| c ||}{$C_{1,1,1,-1}$} & $-$83.0 & $-$17.8 & $-$3.1 \\ \hline \hline
\multicolumn{1}{|| c ||}{$C_{2,2,2,-2}$} &  $-$2.3 &  $-$1.3 & $< 1$ \\ 
\multicolumn{1}{|| c ||}{$C_{1,0,1,0}$} & $-$94.0 & $-$14.9 & $-$2.2 \\ \hline \hline
\multicolumn{1}{|| c ||}{$C_{2,0,2,0}$} & $-$1.3 & $< 1$ &  $< 1$ \\ \hline
\end{tabular}
\captionsetup{width=.95\textwidth}
\caption {{Percentage deviation from LO due to NLO corrections in all non-vanishing $A$ and $C$ coefficients of the spin density matrix. The columns refer to different final states, and different blocks of rows highlight the equalities between the coefficients that should be present at LO. No phase-space restrictions are applied.} }
\label{tab:nonzero_coeff_qq}
\end{center}
\vspace{-5mm}
\end{table}

From the results of these tests we can confidently state that the large NLO EW corrections on the $C_{1,M,1,-M}$ coefficients, in the case of $Z$ decaying to leptons, arise specifically from the smallness of the value of $\alpha$ and in turn to its sensitivity to radiative corrections. Keeping the LO value of this parameter for the extraction of $\rho_{\rm NLO}$ from NLO EW simulations leads to inconsistent results.
As already explained before in the text, the calculation of the $\alpha$ parameter at NLO EW is not theoretically well defined for an off-shell $Z$ boson (gauge dependence) and would be anyway dependent on its invariant mass.
The main consequence is that, without a tomographic approach including NLO EW corrections, it is not possible to estimate the correct $C_{1,M,1,-M}$ coefficients, and consequently determine the entire $\rho$ matrix for the process. Still, is important to note that this conclusion should not be generalised to other processes, including similar ones such as the $H(\rightarrow WW^{*})\rightarrow 4f$ decay and the $H\rightarrow q\bar{q}q\bar{q'}$ process with $q,q'=d,s,b$. Although with milder effects, the other $H\rightarrow q\bar{q}q'\bar{q'}$ final states may suffer the same problem.   

Before discussing how to deal with this peculiar situation for the evaluation of entanglement markers and Bell inequality violation, Sec.~\ref{sec:entangNLO}, we discuss other effects induced by NLO EW corrections that can be seen in Tab.~\ref{tab:nonzero_coeff}.

\medskip

It is very well known that the $m(\ell^+ \ell^-)$ distribution for a pair of leptons emerging from a $Z$ decay, in general not only in the Higgs decay case discussed here, receives large corrections, especially in the region $m(\ell^+ \ell^-)\simeq M_Z$. We discussed the consequences of this effect when we have commented Figs.~\ref{fig:Gamma} and \ref{fig:Gamma_Z1Z2}, and mentioned it is due to the radiation of photons from the leptons. Since in Tab.~\ref{tab:nonzero_coeff} we observe very different NLO corrections for the $A^1_{2,0}$ and $A^2_{2,0}$ coefficients, with a larger impact for the former, we investigate whether this difference, which cannot be explained via the argument of previous discussion regarding effective $\alpha$ at NLO, is related to final-state radiation of photons. Indeed, it is expected that the $Z$ with larger invariant mass, and therefore very often on-shell, should receive large corrections.

The component of the NLO EW corrections associated to the final-state-radiation (FSR) of photons is originating from Feynman diagrams as the one in Fig.~\ref{fig:realph}. When a photon is emitted collinear to one of the leptons (see Eq.~\eqref{eq:Rphoton}), it is recombined with it, forming a dressed lepton. However, those not satisfying Eq.~\eqref{eq:Rphoton} are not recombined, carrying away a fraction of the momentum of the lepton. These emissions of photons, which can be hard, will consequently impact the QT ({\it i.e.}~leading to a biased reconstruction of the $Z$ rest frame), and in turn the spin density matrix. 

In order to investigate the impact of FSR we have evaluated the NLO EW corrections for different values of $\Delta R$ in Eq.~\eqref{eq:Rphoton}, where  the nominal choice $\Delta R =0.1$ has been used for previous results. We consider two extreme cases: $\Delta R =0.01$ and $\Delta R = 1$.
The results for the non-vanishing coefficients of the spin density matrix at LO and NLO EW accuracy, for the two new different recombination radii, are shown in Tab.~\ref{table:nonzero_coeff_dR}.

\begin{table}[t!]
\begin{center}
\renewcommand{\arraystretch}{1.5}
\begin{tabular}{ c | c | c | c | c |}
\cline{2-5}
  & \shortstack{ \textcolor{white}{-} \\ NLO \\ ($\Delta R =0.01$)} &  \shortstack{NLO$\,$/$\,$LO \\ ($\Delta R =0.01$) } &  \shortstack{ NLO \\ ($\Delta R = 1 $) } &  \shortstack{ NLO$\,$/$\,$LO \\ ($\Delta R = 1 $) } \\ \hhline{ - = = = =  }
\multicolumn{1}{| c |}{$A^1_{2,0}$}  & $-0.432 $ & $0.731 (4)$ & $-0.578 $ & $0.977(2)$\\ 
\multicolumn{1}{| c |}{$A^2_{2,0}$}  & $-0.548$ & $0.927(3)$ & $-0.579 $ & $0.979 (2)$\\ \hline \hline
\multicolumn{1}{| c |}{$C_{2,1,2,-1}$} & $-0.956$ & $1.020(6)$  & $-0.933 $ & $0.996(3)$\\ 
\multicolumn{1}{| c |}{$-C_{1,1,1,-1}$}  & $-0.15$ & $0.15(2)$ & $-0.17$ & $0.18(2)$ \\ \hline \hline
\multicolumn{1}{| c |}{$A^1_{2,0}/\sqrt{2} + 1$} & $0.694 $ & $1.193(3)$  & $0.591 $ & $1.016(1)$  \\ 
\multicolumn{1}{| c |}{$C_{2,2,2,-2}$} & $0.554 $ & $0.954(7)$ & $0.572 $ & $0.985(5)$ \\ 
\multicolumn{1}{| c |}{$-C_{1,0,1,0}$} & $0.05 $ & $0.09(5)$ & $0.03 $ & $0.05(3)$ \\ \hline \hline
\multicolumn{1}{| c |}{$C_{2,0,2,0}$} & $1.393 $ & $0.982(5)$ & $1.418 $ & $1.000(3)$ \\
\multicolumn{1}{| c |}{$C_{1,0,1,0} + 2$} & $1.95 $ & $1.38(2)$ & $1.97$ & $1.39(1)$  \\  \hline 
\end{tabular}
\captionsetup{width=.95\textwidth}
\caption {{Non-vanishing $A$ and $C$ coefficients of the spin density matrix calculated at NLO EW and ratio with the LO predictions, for different values of $\Delta R$ . The different blocks of rows highlight the equalities between the couplings that should be present at LO. No phase space restrictions are applied. The error on the last digit is in parentheses. }}
\label{table:nonzero_coeff_dR}
\end{center}
\vspace{-5mm}
\end{table}

First of all, changing $\Delta R$ does not appear to have a significant impact on the NLO EW corrections to the coefficients $C_{1,M,1,-M}$. The impact relative to the LO remains substantial, regardless of the value of $\Delta R$, and with minimal differences w.r.t.~the nominal case. This observation supports the idea that the leading effects on these coefficients derive from a misalignment of the QT approach and not from FSR effects.

Next, we observe that for the case of a larger radius ($\Delta R = 1$) the impact of NLO EW corrections (NLO/LO$-1$) for all the other coefficients is small, ranging from 0 to 3$\%$. On the contrary, with a small radius ($\Delta R = 0.01$) the impact of NLO EW corrections ranges from 2$\%$ to 27$\%$, while in the nominal case ($\Delta R =0.1$) of Tab.~\ref{table:nonzero_coeff_dR} the effects were in the range from $1\%$ to 14$\%$. Thus, as we expected, a large (small) radius implies that a large (small) fraction of the radiated photons are recombined, thereby diminishing (increasing) the influence of the FSR component of NLO EW corrections.
 
A quantity that is very sensitive to $\Delta R$ is the difference between the $A$ coefficients, 
\begin{equation}
\Delta A \equiv A^1_{2,0} - A^2_{2,0}\,.
\end{equation} 
For large $\Delta R$, at NLO EW, $\Delta A\to 0$,  restoring the LO condition. This test seems to support the idea that FSR is the main cause for $\Delta A \ne 0$ observed in Tab.~\ref{tab:nonzero_coeff} at NLO EW. Moreover, the $A$ coefficient that is more sensible to the value of $\Delta R$ is  $A^1_{2,0}$, which is associated to the  $Z$ boson that can go on-shell, and therefore receive the largest effects from FSR.

In order to further test this idea one should isolate the purely QED component of the NLO EW corrections (NLO QED) and repeat the test. We notice that, besides colour factors and the different couplings ($\alpha_s$ {\it vs.}~$\alpha_{\rm EW}$), NLO QED corrections are equivalent to the NLO QCD corrections for the $H(\rightarrow ZZ^*) \rightarrow u\bar{u}c\bar{c}$ process, with gluons clustered to quarks similarly to the case of photons and fermions in NLO EW. In both cases the only diagrams that are present are real emissions of photons/gluons from the final-state fermions and loops involving the exchange of a photon/gluon between a pair of final-state fermions. The NLO QCD calculation can be performed automatically with {\aNLO} and, reducing by a factor of ten the value of $\alpha_s$, leads to a good estimate of the impact of the solely QED effects. The result of this test is consistent with the expectations. There is a large asymmetry of the NLO corrections on the $A$ coefficients, with $A^1_{2,0}$ receiving larger corrections. Also in this case we observe that the difference reduces drastically ($\Delta A\to 0$) when using a large $\Delta R$. 

\medskip

\begin{table}[t!]
\begin{center}
\renewcommand{\arraystretch}{1.5}
\begin{tabular}{ c | c | c | c |}
\cline{2-4}
                       & LO & NLO & NLO$\,$/$\,$LO \\ \hhline{ - = = = }
\multicolumn{1}{| c |}{$A^1_{2,0}$} & $-$0.318(1) & $-$0.241(2) & 0.758(5)  \\ 
\multicolumn{1}{| c |}{$A^2_{2,0}$} & $-$0.318(1) & $-$0.306(2) & 0.962(4)  \\ \hline \hline
\multicolumn{1}{| c |}{$C_{2,1,2,-1}$} & $-$1.014(3) & $-$1.021(5) & 1.007(4)\\ 
\multicolumn{1}{| c |}{$-C_{1,1,1,-1}$} & $-$1.01(2) & $-$0.68(2) & 0.67(2)  \\ \hline \hline
\multicolumn{1}{| c |}{$A^1_{2,0}/\sqrt{2} + 1$} & 0.775(1) & 0.830(1) & 1.070(1)\\ 
\multicolumn{1}{| c |}{$C_{2,2,2,-2}$} & 0.775(3) & 0.766(4) & 0.988(4)\\ 
\multicolumn{1}{| c |}{$-C_{1,0,1,0}$} & 0.77(2) & 0.49(3) & 0.63(3)  \\ \hline \hline
\multicolumn{1}{| c |}{$C_{2,0,2,0}$} & 1.225(4) & 1.240(6) & 1.012(4)  \\ 
\multicolumn{1}{| c |}{$C_{1,0,1,0} + 2$} & 1.23(2) & 1.51(3) & 1.23(2)\\ \hline
\end{tabular}
\captionsetup{width=.95\textwidth}
\caption {Same as Tab.~\ref{tab:nonzero_coeff} but applying the  $m(Z_2) > 30$~GeV cut.}
\label{table:nonzero_coeff_mZ30}
\end{center}
\vspace{-5mm}
\end{table}

Another interesting aspect to investigate is the interplay between the different regions of the phase space and the impact of the NLO EW corrections.
As observed in Figure~\ref{fig:GammaNLOLO_Z1Z2}, the region corresponding to the cut
\begin{equation}
m(Z_2)>30~\gev\,,\label{eq:cut30}
\end{equation}
 is associated to a smaller impact of NLO EW corrections. This region is also noteworthy from a QI perspective, as discussed in Sec.~\ref{sec:entan}, because it is expected to exhibit the largest entanglement between the two $Z$ bosons, with the state tending toward a pure Bell state.
Tab.~\ref{table:nonzero_coeff_mZ30} shows the values of the non-zero $A$ and $C$ coefficients of the spin density matrix within this region of the phase space, including the comparison between LO and NLO.

In this case, the effects of the NLO EW corrections are similar or slightly reduced w.r.t.~the inclusive case, with two notable exceptions. With the cut in Eq.~\eqref{eq:cut30}, the NLO EW corrections on the $C_{1,M,1,-M}$ coefficients are significantly reduced, while the corrections on $A^{1}_{2,0}$ are larger than in the inclusive case. 

A region of the phase-space that does not overlap with the one defined by the cut in Eq.~\eqref{eq:cut30} can be defined via the cut
\begin{equation}
85 ~\gev< m(Z_1) < 95~\gev\,,\label{eq:cut8595}
\end{equation}
where $Z_1$ is almost on-shell. This region is particularly interesting from an experimental perspective, as requirements on the lepton-pair mass are typically employed to accurately combine the leptons for the reconstruction of the $Z$ boson~\cite{ATLAS:2020wny,CMS:2023gjz}.
The results for the non-zero $A$ and $C$ coefficients  with the cut in Eq.~\eqref{eq:cut8595} are presented in Tab.~\ref{table:nonzero_coeff_mZ1}.

From the comparison between LO and NLO, it is evident that the effect of the NLO corrections on all coefficients excluding $C_{1,M,1,-M}$ w.r.t.~the inclusive case are reduced: NLO/LO$-1$ is at most at the level of only 2$\%$. For this reason, $\Delta A\simeq 0$ and also the LO relation between $A^{1}_{2,0}$ and $C_{2,2,2,-2}$ (Eq.~\eqref{eq:coeff_LOrelation1ok}) is basically restored, at variance with the  inclusive region. On the contrary, with the cut in Eq.~\eqref{eq:cut8595}, the effects of the NLO EW corrections on the $C_{1,M,1,-M}$ coefficients are even larger: $-130 \%$ for both $C_{1,0,1,0}$ and $C_{1,1,1,-1}$. We notice that in this case the pattern observed may be very well explained via the incorrect LO value of $\alpha$ for the QT. By following the argument of Appendix $\ref{app:wrongalpha}$, one can easily understand that with $1/(\kappa_1 \kappa_2) \simeq -1.3  $, and $\kappa_1, \kappa_2$ defined as in Eq.~\eqref{eq:kappadef2}, the structure observed in Tab.~\eqref{table:nonzero_coeff_mZ1} can be obtained. Indeed, all the effects that this parametrisation cannot account for are minimal with the cut of  Eq.~\eqref{eq:cut8595}. Thus, for this particular case one may correct the $C_{1,M,1,-M}$ setting them by hand to their original LO value, via the usage of the two effective $\alpha$ for the two different $Z$ bosons, and therefore finding $\rho_{\rm NLO}\simeq \rho_{\rm LO}$ within the percent level. However, via this procedure one may absorb into the effective $\alpha$ values corrections that would be in fact associated to the $H\to V V^*$ subprocess. We therefore do not regard this procedure as a solution to the problem discussed.

\begin{table}[t!]
\begin{center}
\renewcommand{\arraystretch}{1.5}
\begin{tabular}{ c | c | c | c |}
\cline{2-4}
                       & LO & NLO & NLO$\,$/$\,$LO \\ \hhline{ - = = = }
\multicolumn{1}{| c |}{$A^1_{2,0}$} & -0.5551(9) & -0.552(2) & 0.995(2) \\ 
\multicolumn{1}{| c |}{$A^2_{2,0}$} & -0.5551(9) & -0.548(1) & 0.987(2) \\ \hline \hline
\multicolumn{1}{| c |}{$C_{2,1,2,-1}$} & -0.956(2) & -0.949(4) & 0.993(3) \\ 
\multicolumn{1}{| c |}{$-C_{1,1,1,-1}$} & -0.95(1) & 0.30(2) & -0.31(2) \\ \hline \hline
\multicolumn{1}{| c |}{$A^1_{2,0}/\sqrt{2} + 1$} & 0.6075(6) & 0.610(1) & 1.003(1) \\
\multicolumn{1}{| c |}{$C_{2,2,2,-2}$} & 0.607(3) & 0.603(4) & 0.993(5) \\ 
\multicolumn{1}{| c |}{$-C_{1,0,1,0}$} & 0.61(1) & -0.18(2) & -0.30(4) \\ \hline \hline
\multicolumn{1}{| c |}{$C_{2,0,2,0}$} & 1.392(3) & 1.382(5) & 0.993(3) \\ 
\multicolumn{1}{| c |}{$C_{1,0,1,0} + 2$} & 1.39(1) & 2.18(2) & 1.56(1) \\ \hline

\end{tabular}
\captionsetup{width=.95\textwidth}
\caption {Same as Tab.~\ref{tab:nonzero_coeff} but applying the 85 GeV $< m(Z_1) < 95$~GeV cut.}
\label{table:nonzero_coeff_mZ1}
\end{center}
\vspace{-5mm}
\end{table}

By comparing the numbers in Tab.~\ref{table:nonzero_coeff_mZ30} and Tab.~\ref{table:nonzero_coeff_mZ1} against those in Tab.~\ref{tab:nonzero_coeff} we can see that the corresponding change of the impact of NLO EW corrections on $C_{1,M,1,-M}$ (NLO/LO with a cut minus NLO/LO with no cuts) is opposite. It is reduced in the former case and increased in the latter. The same applies for the impact the of NLO EW corrections on $\Delta A$, but in the opposite direction. It is increased in the former case and reduced in the latter.
The behaviour of the $C_{1,M,1,-M}$ coefficients is consistent with what has been  observed and commented in Figs.~\ref{fig:GammaNLOLO} and \ref{fig:GammaNLOLO_Z1Z2}, while the behaviour of  $\Delta A$ is opposite. A possible explanation for the decrease of the impact of NLO EW corrections on $\Delta A$  with the cut in Eq.~\eqref{eq:cut8595} is the following. Requiring a strict window on the $Z_{1}$ invariant mass means rejecting events with a hard photon radiation, which are precisely those that could affect the reconstruction of $Z_{1}$.

Similarly for the inclusive case, Eq.~\eqref{NLOrho}, also in the case of the cuts of Eqs.~\eqref{eq:cut30} and \eqref{eq:cut8595} the matrix $\rho_{\text{NLO}} $ is not semi-positive defined and therefore cannot directly be used in order to extract QI observables. This is precisely the reason why we did not put the matrices in the main text. Nevertheless, we report them in Appendix \ref{app:matr}, together with those at the inclusive level for different $\Delta R$ values.

\subsection{NLO EW corrections to entanglement markers and Bell inequality violation}
\label{sec:entangNLO}

 In this section we discuss the consequences of the findings of Sec.~\ref{sec:resultsNLOEW} on entanglement markers and Bell inequality violation.

An important consequence of the unreliability of the $C_{1,M,1,-M}$ values at NLO EW accuracy, and in turn of the general structure of the $\rho$ matrix, is that the definitions of ${\cal C(\rho)}$ in Sec.~\ref{sec:entan} or ${\cal I}_3$ in Sec.~\ref{sec:IntroBell} are no longer applicable. Indeed, they have been formulated on the basis of the knowledge of all the terms of the spin density matrix and specifically for the structure of the LO matrix.
A solution would be to build new markers, both for Bell inequality violation and for entanglement, that are robust against NLO EW corrections. In other words, quantities that are not sensitive to the values of the $C_{1,M,1,-M}$ coefficients.

The definition of ${\mathcal{C}_{UB}}$ is already independent of the coefficients that are highly sensitive to the EW corrections, since it only depends on the $A$ coefficients and for both LO and NLO predictions only $A^1_{2,0}$ and $A^2_{2,0}$ are non-vanishing (see Eq.~\eqref{eqn:cub}). As a consequence, we can use the same definition, and numerical results are listed in Tab.~\ref{tab:upb}.

\begin{table}[t!]
 \renewcommand{\arraystretch}{1.5}
    \begin{center}
    \begin{tabular}{ c | c | c | c ||}
    \cline{2-4}
                           & no cuts & $m(Z_2) > $ 30 GeV & 85 GeV $< m(Z_1) < 95$ GeV\\ \hhline{ - = = = }
    \multicolumn{1}{|| c ||}{LO ${\mathcal{C}_{UB}}$} & 1.10 & 1.27  & 1.13  \\ \hline \hline
    \multicolumn{1}{|| c ||}{NLO ${{\cal C}_{UB}}$} & 1.12 & 1.27 & 1.13 \\ \hline
    \end{tabular}
    \captionsetup{width=.95\textwidth}
    \caption { {Values obtained for the upper bound of the concurrence squared, $({\cal C(\rho)})^2$}.}
    \label{tab:upb}
    \end{center}
\end{table} 

Both at LO and NLO the upper bound on the squared concurrence results larger than 1 in all investigated phase space regions, reaching the maximum in the region $m(Z_2) > $ 30~GeV.
The lower bound on the squared concurrence, defined in Eq.~\eqref{eqn:clb}, instead does depend on $C_{1,M,1,-M}$. However, it can be modified in order to obtain a new lower bound independent from $C_{1,M,1,-M}$:
{
\bea 
({\cal C(\rho)})^2 & \geq &  {{\cal C}_{ LB}} \geq {{\cal C}^{L>1}_{LB}} \nonumber\\
&\equiv& \frac{2}{9} \max\nonumber
\Bigg[\Big(-2-2\sum_{L,M} |A^{a}_{L,M}|^2+\sum_{L,M}|A^{b}_{L,M}|^2+\sum_{L_a > 1,L_b > 1,M_a,M_b}|C_{L_a,M_a,L_b,M_b}|^2\Big),\nonumber\\
 &&\Big(-2+\sum_{L,M} |A^{a}_{L,M}|^2-2\sum_{L,M}|A^{b}_{L,M}|^2+\sum_{L_a > 1,L_b > 1,M_a,M_b}|C_{L_a,M_a,L_b,M_b}|^2\Big)\Bigg]\,.
 \label{eqn:clb_noc1}
 \eea
 }

 \begin{table}[t!]
 \renewcommand{\arraystretch}{1.5}
    \begin{center}
    \begin{tabular}{ c | c | c | c ||}
    \cline{2-4}
                           & no cuts & $m(Z_2) > $ 30 GeV & 85 GeV $< m(Z_1) < 95$ GeV\\ \hhline{ - = = = }
    \multicolumn{1}{|| c ||}{LO ${\mathcal{C}_{LB}}$} & 0.94 & 1.18  & 0.97  \\ \hline
    \multicolumn{1}{|| c ||}{LO ${{\cal C}^{L>1}_{LB}}$} & 0.47 & 0.59 & 0.49 \\ \hline \hline
    \multicolumn{1}{|| c ||}{NLO ${{\cal C}^{L>1}_{LB}}$} & 0.49 & 0.55 & 0.48 \\ \hline
    \end{tabular}
    \captionsetup{width=.95\textwidth}
    \caption {{Values obtained for the lower bound of the concurrence squared, $({\cal C(\rho)})^2$}. At LO two different definitions of the lower bound are compared, where the second row is defined to be stable against NLO EW corrections.  
}
    \label{tab:lowb}
    \end{center}
\end{table}

The definition in Eq.~\eqref{eqn:clb_noc1} of the lower bound leads to values that by definition are smaller than those obtained via Eq.~\eqref{eqn:clb}: regardless of the choice of either NLO or LO approximation, ${{\cal C}^{L>1}_{LB}} < {{\cal C}_{LB}}$. The corresponding results are shown in Tab.~\ref{tab:lowb} for the three different sets of cuts already chosen in Sec.~\ref{sec:resultsNLOEW}. 
At LO, both definitions of the lower bounds on $({\cal C(\rho)})^2$ confirm the presence of entanglement in all investigated regions. However, the definition ${\mathcal{C}_{LB}}^{L>1}$ leads to a value significantly closer to 0, leading to a more difficult observation at the experimental level of a clear signal of the presence of entanglement in this final state. 
The trend of the lower bound is the same using the two definitions, with the maximum value reached in the region $m(Z_2) > $ 30~GeV.
At NLO the values obtained with ${\mathcal{C}_{LB}}^{L>1}$ are very similar to the LO, confirming the stability of this definition against the NLO EW corrections. Clearly, based on the discussion of Sec.~\ref{sec:resultsNLOEW}, the same consideration would not be true for ${\mathcal{C}_{LB}}$ and we refrain from showing the corresponding prediction in the Tab.~\ref{tab:lowb}, since as already explained several times, this result would be unreliable.

\medskip

Also in the case of Bell inequalities it is possible to define an operator independent of $C_{1,M,1,-M}$, by exploiting the possibility of applying unitary transformations to the Bell operator. Indeed, the evaluation of ${\cal{I}}_3$ with the $O_B^{(O_A)}$ operator requires a maximisation performed extracting random unitary matrices, and to remove the dependence on $C_{1,M,1,-M}$ we imposed a restriction to the sampled phase space of the unitary matrices, in order to include only those that render ${\cal{I}}_3$ independent of the $C_{1,M,1,-M}$ values. Technically this can be done by parametrising ${\cal{I}}_3$ as a function of the $C_{1,M,1,-M}$ coefficients\footnote{Where we used the relation $C_{1,-1,1,1} = C_{1,1,1,-1}$.}
\be
{{\cal{I}}_3} = A \cdot C_{1,0,1,0} + B \cdot C_{1,-1,1,1} + D\,,
\ee
and discarding all random unitary matrices leading to $|A|$ or $|B|$ larger than 0.01. Then the value of ${\cal{I}}_3$ can be maximised inside this reduced set of unitary matrices, by maximising $D$.
\begin{table}[t!]
\renewcommand{\arraystretch}{1.5}
    \begin{center}
    \begin{tabular}{ c | c | c | c |}
    \cline{2-4}
                           & no cuts & $m(Z_2) > $ 30 GeV & 85 GeV $< m(Z_1) < 95$ GeV \\ \hhline{ - = = = }
    \multicolumn{4}{|c|}  {${\cal I}_{3}$, LO} \\ \hline
    \multicolumn{1}{| c |}{$O_B^{(O_{A},U_{\mathrm{fix}})}$} & 2.600 $\pm$ 0.003 & 2.794 $\pm$ 0.004 & 2.639 $\pm$ 0.003 \\ 
    \multicolumn{1}{ |c |}{$O_B^{(O_A)}$} & 2.63 & 2.79 & 2.65 \\ 
    \multicolumn{1}{| c |}{$O_B^{(O_A,C_{L>1})}$}& 2.63 & 2.79 & 2.65 \\ \hline \hline
    \multicolumn{4}{|c|}{${\cal I}_{3}$, NLO}  \\ \hline
    \multicolumn{1}{| c |}{$O_B^{(O_A,C_{L>1})}$}& 2.60 & 2.72  & 2.64 \\ \hline
    \end{tabular}
    \captionsetup{width=.95\textwidth}
   \caption{\(\mathcal{I}_3\) calculated using \(\rho_{\rm LO}\) and \(\rho_{\rm NLO}\) in different regions of the phase space. Three different operators are applied at LO, each obtained with a different optimisation of the unitary transformation to the initial Bell operator.} 
    \label{tab:I3NLO}
    \end{center}
\end{table}

\begin{figure}[h!]
    \centering
    \includegraphics[width=0.8 \textwidth]{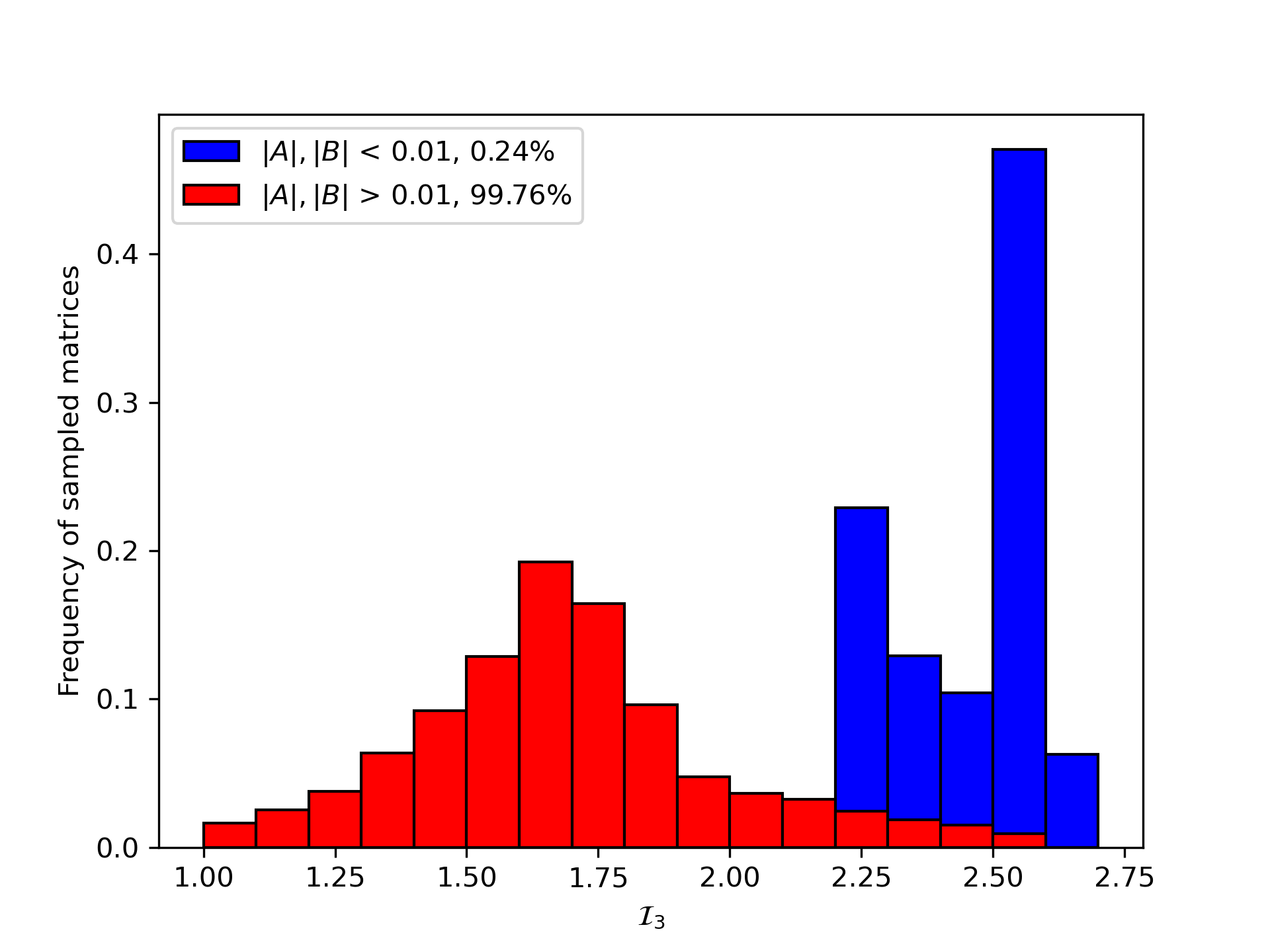}
    \caption{${\cal I}_3$ NLO values in the inclusive region obtained in the maximisation procedure $O_B^{(O_A)}$, run on $10^5$ three-dimensional unitary matrices. The blue histogram includes only results accepted using the $O_B^{(O_A,C_{L>1})}$ approach, while the discarded ${\cal I}_3$ values are included in the red histogram. Both histograms are normalised to unity and the acceptance rate is shown in the legend.}
    \label{fig:OBLlt1}
\end{figure}

Figure~\ref{fig:OBLlt1} shows, for the inclusive case, the spread of ${\cal{I}}_3$ values obtained during the maximisation only accepting matrices leading to no dependence on $C_{1,M,1,-M}$, which is denoted as $O_B^{(O_A,C_{L>1})}$. The figure also displays all the discarded ${\cal{I}}_3$, that showed a dependence on the $C_{1,M,1,-M}$ coefficients. It is clearly visible that, even if just 0.25$\%$ of the random matrices sampled are accepted to maximise ${\cal{I}}_3$, the actual values are all close to the maximum ${\cal{I}}_3$, while using all the random matrices leads to a larger spread in the ${\cal{I}}_3$ values. As a consequence, this method does not lead to a reduction of the maximum of ${\cal{I}}_3$, as shown in Tab.~\ref{tab:I3NLO}, and does not even require a significant increase in the computation time for the optimisation.
Regardless of the method employed, the Bell inequality is significantly violated at LO in all the investigated regions, with the highest value observed for $m(Z_2) > 30$~GeV. The three different definitions of ${\cal I}_{3}$ (see also discussion is Sec.~\ref{sec:IntroBell}) lead to very similar results, with a slightly larger ${\cal{I}}_3$ value obtained running the maximisation with random unitary matrices. 
Even if the NLO EW corrections generally reduce ${\cal I}_3$, its value remains significantly larger than 2 at both LO and NLO in all regions, with the highest value still observed in the $m(Z_2) > 30$~GeV region.

\section{New Physics effects and quantum-information observables}
\label{sec:NP}

Beyond their primary role in the characterisation of quantum properties at high energy, QI-inspired observables have recently attracted attention also due to their sensitivity to physics beyond the Standard Model (BSM) across a variety of final states~\cite{Aoude:2022imd,Fabbrichesi:2023jep,Bernal:2023ruk,Aoude:2023hxv,Maltoni:2024csn,Maltoni:2024tul,Sullivan:2024wzl}. 

A particularly striking example is the case of top-quark pair production near the threshold, where spin entanglement has been found to be stronger than what initially expected~\cite{ATLAS:2023fsd,CMS:2024pts}. This result is currently understood as a consequence of an enhancement of the cross section at threshold due to higher-order QCD effects and not due to BSM. The relevant aspect for our work is that  its detection has been possible thanks to quantum observables related to the entanglement~\cite{CMS:2025kzt} as the experimental invariant mass resolution alone is currently insufficient  to directly observe an excess or the narrow peak predicted by NRQCD and leaves open the possibility to study the problem with a spectral approach \cite{Nason:2025hix}. In the context of BSM, analogous effects in suitable QI observables are also expected when, instead of focusing on the phase-space region near the threshold, one considers an invariant mass window centred around the mass of a hypothetical heavy scalar or pseudoscalar particle decaying into a $t \bar t$ pair \cite{Maltoni:2024csn}.

In this section, we investigate how new interactions affecting the process \( H \to VV^* \) modify the quantum state of the final four-lepton system. We assume that no new light resonances are present, meaning that new physics effects arise only as deformations of Standard Model (SM) interactions, and that no BSM effects are present in the interactions among the vector bosons and the fermions. 

The sensitivity of the \( VV \) final state to the new physics (NP) scenario that we have just described has already been explored in previous studies, both in terms of anomalous couplings or via an effective field theory (EFT) approach. In particular Refs.~\cite{Fabbrichesi:2023jep,Bernal:2023ruk,Sullivan:2024wzl,Aoude:2023hxv} have investigated constraints on anomalous couplings using QI observables derived from the spin density matrix, such as entanglement and the asymmetric components of the spin correlation matrix. Notably, these constraints have been found to be competitive with, or complementary to, the best bounds obtained from global fits, see {\it e.g.}~Ref.~\cite{Giani:2023gfq}. 

However, all existing studies assume that NLO EW corrections to entanglement observables in the relevant kinematic regions remain below \( 1\% \), following the guideline provided in Ref.~\cite{Boselli:2015aha}. Given the competitiveness of these bounds and the insights provided by the NLO EW results discussed in the previous sections, it is valuable to conduct an analysis of the possible effects of EFT corrections on QI observables, comparing them against predictions that incorporate NLO accuracy for at least the SM.

This section is structured as follows. 
In Sec.~\ref{sec:anarho} we provide the analytical results at LO including BSM contributions for the $\rho$ density matrix of the $VV$ system at fixed invariant masses of the two $V$ bosons, decaying to the four fermions $f_1  \bar{f}_2 f_3 \bar{f}_4$. We stress that this approach focuses only on specific points in the phase space, and estimating the prediction for the spin density matrix obtained in the Higgs decay would require integration across the whole phase space. Nevertheless, the analytical formulas are very useful for understanding the impact of NP on the  $\rho$ density matrix.

In Sec.~\ref{sec:EFTnum} we instead provide numerical results for the $H\rightarrow e^{+}e^{-}\mu^{+}\mu^{-}$ process and we discuss the impact of BSM effects on the $A$ and $C$ coefficients and QI observables that are extracted from the four-lepton system, via the QT approach described in Sec.~\ref{sec:LO}. We also discuss the size of the impact of BSM effects in comparison with the effects of NLO EW corrections, which we calculate only for the SM component.\footnote{The calculation of EW corrections in the context of BSM in an EFT framework is developing in these last years. It has never been performed for QI observables (for the $H\to ZZ$ process see Ref.~\cite{Dawson:2024pft}) and it is clearly beyond the scope of this paper. Nevertheless, we acknowledge that given the size of the EW corrections found in the previous sections for the SM, also in the BSM context they   may lead to non-negligible results. }

\subsection{Analytical formulas for the $VV$ density matrix}
\label{sec:anarho}

In this section we provide analytical results at LO in the BSM context for the $\rho$ density matrix of the $VV$ system at fixed $\beta$, further specifying the dependence on the invariant masses of the two $V$ bosons: $m_a$ and $m_b$.
Rather than employing an explicit effective Lagrangian, as we will do in Sec.~\ref{sec:EFTnum}, we adopt a general Lorentz-invariant parameterisation of the $HVV$ vertex in order to systematically capture modifications to the spin density matrix from higher-dimensional operators. This approach accounts for potential modifications, including those from dimension-five or higher operators, through the parameters $a_1,\,a_2$ and $a_3$. Here, we establish a general theoretical framework for these effects, while the next section focuses on the explicit analysis of dimension-five operators.

The most general Lorentz-invariant vertex describing the coupling of the Higgs boson to two massive vector bosons \( V \) can be written as~\cite{Godbole:2007cn}:  
\ba 
\label{EFTvertex}
\mathcal{V}^{HV_aV_b}_{\mu\nu}=\frac{1}{v}\left[a_1 g_{\mu\nu}M_V^2+a_2\left(g_{\mu\nu} p_a \cdot p_b - (p_a)_{\nu} (p_b)_{\mu} \right) + a_3 \epsilon_{\mu\nu\alpha\beta} p_a^\alpha p_b^\beta \right]\,,
\ea
where \( p_a \) and \( p_b \) are the four-momenta of the massive vector bosons, which we consider as the qutrits ``$a$'' and ``$b$'', respectively.\footnote{In Sec.~\ref{sec:EFTnum} we will consider also cases where $V=\gamma$, but the $\gamma$ will be always off-shell and therefore still a qutrit. Moreover we will consider also the case where the two $V$ bosons are different: $Z$ and $\gamma$. We keep here the notation general, having these two possibilities in mind. \label{foot:VVdiff}} The parameter \( M_V \) represents the physical mass of the vector boson and has only normalisation purpose. In the case of the possible pairs of SM vector bosons stemming from the Higgs boson we conventionally use $M_Z$ for $ZZ,\gamma\gamma$ and $Z\gamma$, and $M_W$ for $WW$. The term
 \( \epsilon_{\mu\nu\alpha\beta} \) is the fully antisymmetric Levi-Civita tensor, and \( v \) is a scale conventionally taken as the vacuum expectation value (vev) of the Higgs field. The vertex in the above equation is also valid for massless vector bosons, except that the coupling 
$a_1$ is always zero for on-shell massless vector bosons (although these are not qutrits anymore). 

The coefficients \( a_1 \) and \( a_2 \) correspond to CP-even scalar couplings to the two qutrits, while \( a_3 \) is associated with the CP-odd scalar coupling. These coefficients are, in general, momentum-dependent form factors that can arise after having integrated out the heavy degrees of freedom; they are equivalent to the effect of higher-dimensional operators. More generally, if we assumed also the presence of lighter states, \( a_1 \), \( a_2 \), and \( a_3 \) could in principle be complex. In the case of an EFT approach, where the heavy degrees of freedom are integrated out, if the expansion is truncated at the lowest order, as done later in Sec.~\ref{sec:EFTnum}, the presence of an imaginary components in \( a_2 \) and \( a_3 \) would instead violate the hermicity of the Lagrangian. It is interesting to note that if we want to include in the parameterisation of Eq.~\eqref{EFTvertex} the loop effects of the SM itself, the \( a_1 \), \( a_2 \), and \( a_3 \) terms would have an imaginary part, which is related to the absorptive component of the loops featuring the light states (all the fermions besides the top and the photons). 

At the tree level in the SM, the only nonzero coupling is $a_1$ with the values $a_1=1$ for  $V_aV_b=ZZ$,  $a_1=2$ for $V_aV_b=WW$, while also $a_1$ is vanishing for $V_aV_b=\gamma\gamma$ and $V_aV_b=Z\gamma$.
Since the vertex in Eq.~(\ref{EFTvertex}) is constructed solely from the requirement of Lorentz invariance, it remains valid to all orders in perturbation theory, with \( a_1 \), \( a_2 \), and \( a_3 \) receiving potential loop-induced corrections. Notably, in this parametrization, each term independently satisfies the Ward identities.

\medskip

Using the vertex of $HVV$ in Eq.~(\ref{EFTvertex}), the amplitude for the process $H(p)\rightarrow V_a  V_b\rightarrow f_1  \bar{f}_2 f_3 \bar{f}_4$ reads:
\ba
\label{ampeft}
i{\cal M}_V&=& \frac{i}{v} \left(
a_1 g_{\mu\nu} M^2_V 
+ a_2(g_{\mu\nu}p_a \cdot p_b- (p_a)_\nu (p_b)_\mu) 
+  a_3\epsilon_{\mu\nu\alpha\beta} p_a^\alpha p_b^\beta \right)\,\nonumber\\
&&\,\times\left(\frac{
\bar u_1 \gamma^\mu
(c_L P_L+c_R P_R) v_2 \; \bar u_3 
\gamma^\nu (d_L P_L+d_R P_R) v_4  }{(m_a^2-M_{V_a}^2+iM_{V_a}\Gamma_{V_a})(m_b^2-M_{V_b}^2+iM_{V_b}\Gamma_{V_b})}\right)\,,
\ea
where we have introduced the two new symbols $d_L$ and $d_R$. In this section we use the symbols $c_L$ and $c_R$ only for the boson $V_a$;  for the boson $V_b$ we use the symbols $d_L$ and $d_R$ (see also footnote \ref{foot:VVdiff}). 

Squaring the amplitude and summing over the helicities of the final-state fermions leads to
\ba
\label{EFTampsq}
|\overline{{\cal M}}|^2&=&\sum_s {\cal M}_V^* {\cal M}_V=\frac{1}{D(m_a^2)D(m_b^2)}\left(|{\cal A}_1|^2+|{\cal A}_2|^2+|{\cal A}_3|^2+{\cal A}_{12}+{\cal A}_{13}+{\cal A}_{23}\right)\,,\nonumber\\
\ea
where
\ba
 D(m_{a(b)}^2)\equiv\left(m_{a(b)}^2-M_V^2\right)^2+\left(M_V \Gamma_V\right)^2\, ,
 \ea
 and each  ${\cal A}_{i}$  is the component of the amplitude, multiplied by $D(m_a^2)D(m_b^2)$, which is proportional to the coefficient $a_i$. The quantities  ${\cal A}_{ij}$ are simply defined as ${\cal A}_{ij}\equiv {\cal A}_{i}{\cal A}_{j}^*+{\cal A}_i^* {\cal A}_j=2\Re[{\cal A}_{i}{\cal A}_{j}^*]$.

Explicitly, the $|{\cal A}_{i}|^2$ terms in Eq.~\eqref{EFTampsq} read
\ba
\label{Eftampisq}
|{\cal A}_1|^2&=&16 |a_1|^2 M_V^4\left[\Pi_1(c_L^2 d_L^2+c_R^2 d_R^2)+\Pi_2(c_L^2 d_R^2+c_R^2 d_L^2)\right]\,,\\
|{\cal A}_2|^2&=&16 |a_2|^2 \left[H_1(c_L^2 d_L^2+c_R^2 d_R^2)+H_2(c_L^2 d_R^2+c_R^2 d_L^2)\right]\,,\\
|{\cal A}_3|^2&=&16 |a_3|^2 \left[H_3(c_L^2 d_L^2+c_R^2 d_R^2)+H_4(c_L^2 d_R^2+c_R^2 d_L^2)\right]\,,
\ea
while the ${\cal A}_{ij}$ terms read
 \ba
{\cal A}_{12}&=&8 M_V^2\Big[(a_1^* a_2+a_1 a_2^*)\left[(c_L^2 d_L^2+c_R^2 d_R^2)(\Sigma_1 K_1)+(c_L^2 d_R^2+c_R^2 d_L^2)(\Sigma_2 K_2)\right]\nonumber\\
&&+i (a_1^* a_2-a_1 a_2^*)\left[(c_L^2 d_L^2-c_R^2 d_R^2)\Pi_e\Sigma_3-(c_L^2 d_R^2-c_R^2 d_L^2)\Pi_e\Sigma_4\right]\Big]\,,
\\
{\cal A}_{13}&=&-8i M_V^2\Big[(a_1 a_3^* - a_1^* a_3)\left[(c_L^2 d_L^2-c_R^2 d_R^2)(\Sigma_3K_1)+(c_L^2 d_R^2-c_R^2 d_L^2)(\Sigma_4K_2)\right]\nonumber\\
&&-i (a_1^* a_3 + a_1 a_3^*)\left[(c_L^2 d_L^2+c_R^2 d_R^2)\Pi_e\Sigma_1-(c_L^2 d_R^2+c_R^2 d_L^2)\Pi_e\Sigma_2\right]\Big]\,,
\\
{\cal A}_{23}&=&-16i \Big[\Pi_0(a_2 a_3^* - a_2^* a_3)\left[(c_L^2 d_L^2-c_R^2 d_R^2)\Sigma_1 \Sigma_3+(c_L^2 d_R^2-c_R^2 d_L^2)\Sigma_2\Sigma_4\right]\nonumber\\
&&-i (a_2^* a_3+a_2 a_3^*)\left[(c_L^2 d_L^2+c_R^2 d_R^2)\Pi_e K_1-(c_L^2 d_R^2+c_R^2 d_L^2)\Pi_e K_2\right]\Big]\,, \label{Eftijampsq}
\ea 
with the momentum structures in the above equations defined in Tab.~\ref{tab:momentumstructrues}. 
\begin{table}[h!]
    \renewcommand{\arraystretch}{1.3}
    \centering
    \begin{tabular}{|c|c||c|c|}
        \hline
    Symbols & Definitions & Symbols & Definitions\\
    \hline
     $\Pi_0$ & $p_{12}p_{34}$  & $\Sigma_1$ & $p_{13}+p_{24}$ \\
     $\Pi_1$ & $p_{13}p_{24}$ &  $\Sigma_2$ & $p_{14}+p_{23}$\\
     $\Pi_2$ & $ p_{14}p_{23}$ &  $\Sigma_3$ & $p_{13}-p_{24}$\\
     $\Pi_\epsilon$ &$\epsilon_{\mu \nu \rho\sigma} p_1^\mu p_2^\nu p_3^\rho p_4^\sigma$ & $\Sigma_4$ & $p_{14}-p_{23}$\\
         \hline
     $H_1$ & $G_1+(\Pi_1-\Pi_2)^2$ & $G_1$ & $\Pi_0(\Pi_0+p_{13}^2+p_{24}^2-2\Pi_2)$\\
          $H_2$ & $G_2+(\Pi_1-\Pi_2)^2$ &  $G_2$ & $\Pi_0( \Pi_0+p_{14}^2+p_{23}^2-2\Pi_1)$\\
               $H_3$ & $G_3-(\Pi_1-\Pi_2)^2$&  $G_3$ & $\Pi_0(-\Pi_0+p_{13}^2+p_{24}^2+2\Pi_2)$\\
          $H_4$ & $G_4-(\Pi_1-\Pi_2)^2$ &  $G_4$ & $\Pi_0( -\Pi_0+p_{14}^2+p_{23}^2+2\Pi_1)$\\
        \hline
    $K_1$ & $\Pi_0+\Pi_1-\Pi_2$ & $K_2$ &$\Pi_0-\Pi_1+\Pi_2$\\
        \hline
    \end{tabular}
    \caption{Momentum structures contributing to Eqs.~\eqref{Eftampisq} -- \eqref{Eftijampsq}. In the table we use the notation $p_{ab}\equiv p_a\cdot p_b$.}
    \label{tab:momentumstructrues}
\end{table}

In order to compute the spin density matrix of the $VV$ system in the presence of a generic $HVV$ vertex, parametrised  as in Eq.~(\ref{EFTvertex}), we directly insert the squared amplitudes in Eqs.~\eqref{eqn:coeff_defA} and \eqref{eqn:coeff_defC}. The spin-analysing powers are defined as 
\ba
 \alpha_a\equiv\frac{c_R^2-c_L^2}{c_R^2+c_L^2}\,, \qquad \alpha_b\equiv\frac{d_R^2-d_L^2}{d_R^2+d_L^2}\,,
 \label{eq:alphatrick}
\ea
where the $d_{R/L}$ and $c_{R/L}$ values depend on the specific bosons and fermions considered.
In this way, starting from
the squared amplitude given in Eq.~\eqref{EFTampsq} we obtain the coefficients $A$ and $C$,  which by definition do not depend neither on $\alpha_a$ nor on $\alpha_b$. We determine the $A$ and $C$ coefficients independently for each momentum structure given in Tab.~\ref{tab:momentumstructrues}, and the corresponding values are provided in Tabs.~\ref{tab:tab1coffi}--\ref{tab:tab5coeff} of Appendix \ref{appendixAClist}. Substituting the value of each momentum structure in Eqs.~\eqref{Eftampisq} -- \eqref{Eftijampsq} and then plugging them into Eq.~\eqref{EFTampsq}, we obtain the value for each $A$ and $C$ coefficient.

\begin{table}[t!]
\renewcommand{\arraystretch}{1.3}
\centering
    \begin{tabular}{|l|l|}
        \hline
    Non-vanishing coefficients also in the SM & Vanishing coefficients in the SM\\  
    \hline
  $A^a_{2,0} = A^b_{2,0} $ & $A^a_{1,0} = -A^b_{1,0}$  \\
   $\frac{A^a_{2,0
}}{\sqrt{2}} + 1 = -C_{1,0,1,0}$& $\frac{A^b_{1,0}}{\sqrt{2}} = C_{2,0,1,0}$\\
  $  C_{2,0,2,0}=2+ C_{1,0,1,0}$ &  $C_{1,0,2,0}=-C_{2,0,1,0}$ \\
  $C_{1,-1,1,1}=-C_{2,-1,2,1}$   & $C_{1,-1,2,1}=-C_{2,-1,1,1}$\\
  \hline
   $C_{1,-1,1,1}=C_{1,1,1,-1}^*$ & $C_{1,-1,2,1}=C_{1,1,2,-1}^*$\\
   $C_{2,-1,2,1}=C_{2,1,2,-1}^*$&$C_{2,-1,1,1}=C_{2,1,1,-1}^*$\\
      $C_{2,2,2,-2} =C_{2,-2,2,2}^*$&\\
      \hline
    \end{tabular}
\caption{Non-zero $A$ and $C$ coefficients and relations among them at fixed $m_a$ and $m_b$ and assuming the parameterisation of Eq.~\eqref{EFTvertex} for the two qutrits $V_a V_b$ emerging from the $H\to V V^*$ decay.}
\label{EFT_coeffe}
\end{table}

Before showing the explicit formulas for (some of) the non-zero coefficients, we list them in Tab.~\ref{EFT_coeffe} together with their relations.
The left column lists coefficients that are also non-vanishing in the SM. We stress that they do not satisfy exactly the same relations as in the SM (Eqs.~\eqref{eq:veccoeff}--\eqref{eq:coeff_LOrelation1ok}), notably, the relation $C_{2,2,2,-2} = -C_{1,0,1,0}$ is not true. Moreover, some of these coefficients acquire non-zero imaginary contributions, which is not the case in the SM due to the presence of CP-conserving interactions only. The right column contains instead coefficients that are strictly zero in the SM but become non-vanishing due to the NP effects. It is important to note that these coefficients remain zero if all three parameters \( a_1 \), \( a_2 \), and \( a_3 \) are real, as is the case for the results in Sec.~\ref{sec:EFTnum}.  Additionally, for both columns, the relations among the coefficients in the lower rows arise due to cylindrical symmetry.

Using the information of Tab.~\ref{EFT_coeffe}, together with the formulas of Appendix \ref{sec:rhogeneral}, we can construct the density matrix at fixed $m_a$ and $m_b$, assuming the parameterisation of Eq.~\eqref{EFTvertex} for the two qutrits $VV$ emerging from the $H\to V V^*$ decay:
\begin{eqnarray}
\label{EFTrho}
&\rho_{\rm NP}(m_a, m_b)=\frac{1}{3} \times\\
&\left(
\begin{array}{ccccccccc}
\renewcommand{\arraystretch}{1.3}
 \cdot & \cdot & \cdot & \cdot & \cdot & \cdot & \cdot & \cdot & \cdot \\
 \cdot & \cdot & \cdot & \cdot & \cdot & \cdot & \cdot & \cdot & \cdot \\
 \cdot & \cdot & \blue{\left(1+\frac{1}{\sqrt{2}}A^2_{2,0}-\sqrt{\frac{3}{2}}A^2_{1,0}\right)}  & \cdot & \violet{\left(-C_{2,1,1,-1}+C_{2,1,2,-1}\right)} & \cdot & \blue{C_{2,2,2,-2}} & \cdot & \cdot \\
 \cdot & \cdot & \cdot & \cdot & \cdot & \cdot & \cdot & \cdot & \cdot \\
 \cdot & \cdot & \violet{\left( -C_{2,-1,1,1}+C_{2,-1,2,1}\right)}
   & \cdot &\blue{\left( 1-\sqrt{2}A^2_{2,0}\right)}& \cdot &\violet{\left(C_{2,1,1,-1}+C_{2,1,2,-1}\right)} & \cdot & \cdot \\
 \cdot & \cdot & \cdot & \cdot & \cdot & \cdot & \cdot & \cdot & \cdot \\
 \cdot & \cdot &\blue{C_{2,-2,2,2}} & \cdot &\violet{\left( C_{2,-1,1,1}+C_{2,-1,2,1}\right)} & \cdot &\blue{\left(1+\frac{1}{\sqrt{2}}A^2_{2,0}+\sqrt{\frac{3}{2}}A^2_{1,0}\right)} & \cdot & \cdot \\
 \cdot & \cdot & \cdot & \cdot & \cdot & \cdot & \cdot & \cdot & \cdot \\
 \cdot & \cdot & \cdot & \cdot & \cdot & \cdot & \cdot & \cdot & \cdot \\
\end{array}
\right)\,.\nonumber
\end{eqnarray}

We have highlighted entries that are equal or linearly dependent using different colours. The difference w.r.t.~Eq.~\eqref{beta_rho} is evident: while in the SM at fixed $\beta$ the $\rho$ matrix depends only on $\beta$, here, besides the dependence on separately $m_a$ and $m_b$, the dependence on $a_1, a_2$ and $a_3$ enters as well, leading to many more degrees of freedom, especially if the $A$ and $C$ coefficients are complex and even more if also $a_1, a_2$ and $a_3$ are complex. 

Assuming all three $a_i$ parameters as real, as we will in practice do for the numerical analysis of Sec.~\ref{sec:EFTnum}, all coefficients in the right column of Tab.~\ref{EFT_coeffe} become zero, while the remaining coefficients in the left column reduce to only three independent coefficients. These three coefficients can be fully expressed in terms of coefficients with only even $L$ values. The analytical expressions for these coefficients at fixed invariant masses of \(V_a\) and \(V_b\), denoted by \(m_a\) and \(m_b\), respectively, are given by:
\ba
A_{2,0}^{a,b}&=& \frac{64\sqrt{2}\,\pi^2}{9\langle|\overline{{\cal M}}|^2\rangle_{\Omega_{a,b}}}(c_L^2+c_R^2)(d_L^2+d_R^2) m_a^2 m_b^2(\b^2-1)\left(m_a^2 m_b^2(a_2^2+a_3^2)-M_V^4 a_1^2\right)  \nonumber\\
&=& \frac{\sqrt{2}}{\mathcal{N}}(\b^2-1)\left(m_a^2 m_b^2(a_2^2+a_3^2)-M_V^4 a_1^2\right)\,, \label{ACvalueEFT}
\ea
\ba
\nonumber\\
C_{2,-1,2,1}&=& -\frac{64\,\pi^2 }{3\langle|\overline{{\cal M}}|^2\rangle_{\Omega_{a,b}}}(c_L^2+c_R^2)(d_L^2+d_R^2) m_a^2 m_b^2\left(a_1 M_V^2\b+a_2 m_a m_b\right)\nonumber\\
&&\times\left(a_1 M_V^2+m_a m_b(a_2 \b -i a_3\sqrt{\b^2-1})\right) \label{ACvalueEFT2} \\
&=& -\frac{3 \sqrt{2}}{\mathcal{N}}\left(a_1 M_V^2\b+a_2 m_a m_b\right)\times\left(a_1 M_V^2+m_a m_b(a_2 \b -i a_3\sqrt{\b^2-1})\right)\,, \nonumber
\ea
\ba
\nonumber\\
C_{2,-2,2,2}&=& \frac{64\,\pi^2 }{3\langle|\overline{{\cal M}}|^2\rangle_{\Omega_{a,b}}}(c_L^2+c_R^2)(d_L^2+d_R^2) m_a^2 m_b^2\left(a_1 M_V^2+m_a m_b(a_2 \b -i a_3\sqrt{\b^2-1})\right)^2\nonumber\\
&=& \frac{3 \sqrt{2}}{\mathcal{N}}\left(a_1 M_V^2+m_a m_b(a_2 \b -i a_3\sqrt{\b^2-1})\right)^2\,,  \label{ACvalueEFT3}
\ea 
where the quantity $\langle|\overline{{\cal M}}|^2\rangle_{\Omega_{a,b}}$ is the squared matrix element integrated over the two solid angles  $\Omega_{a}$ and $\Omega_{b}$ and equal to
\ba
\label{crosssection}
\langle|\overline{{\cal M}}|^2\rangle_{\Omega_{a,b}}&=&\frac{64\pi^2}{9}(c_L^2+c_R^2)(d_L^2+d_R^2) m_a^2 m_b^2 \Big(  a_1^2 M_V^4(\b^2+2)+ a_2^2 m_a^2 m_b^2 (2\b^2+1)\nonumber\\
&&+2 a_3^2 m_a^2 m_b^2 (\beta^2-1) + 6a_1 a_2 m_a m_b M_V^2 \b\Big)\,,
\ea 
and $\mathcal{N}$ is a normalisation factor introduced for simplifying the equations and defined as 
\ba
\mathcal{N}&\equiv&  a_1^2 M_V^4(\b^2+2)+ a_2^2 m_a^2 m_b^2 (2\b^2+1)+2 a_3^2 m_a^2 m_b^2 (\beta^2-1) + 6a_1 a_2 m_a m_b M_V^2 \b\,.
\nonumber \\
\ea

The remaining non-zero coefficients are related to the expressions above, as summarised in the left column of Tab.~\ref{EFT_coeffe}. Clearly, the dependence on the couplings \(c_L\) and \(c_R\), and as well on \(d_L\) and \(d_R\), cancels between the numerator and denominator, since the $\rho$ matrix cannot depend on them.

It is interesting to note that, similarly to the SM case, for the given values of the \(A\) and \(C\) coefficients at fixed masses, the density matrix in Eq.~(\ref{ACvalueEFT}) simplifies to a representation corresponding to a pure state, defined as follows: 
\be  
\label{EFTstate}
|\psi\rangle=  a_L|0 \,0\rangle + a_+ |+ -\rangle + a_-|- +\rangle\,.
\ee
This state has a similar form to the SM LO state in Eq.~(\ref{state}), except that in the SM case, the relation  $a_+=a_-=a_T/\sqrt{2}$, holds, where both \(a_L\) and \(a_T\) are real. In contrast, in this scenario, these coefficients are given by:  
\ba
a_\pm &=&-\frac{a_1 M_V^2+ m_a m_b (a_2\beta \pm  i a_3 \sqrt{\beta^2-1})}{\sqrt{\mathcal{N}}}\,, \label{apmvaluesEFT}\\
a_L&=&\frac{a_1 M_V^2  \beta+a_2 m_a m_b }{\sqrt{\mathcal{N}}} \label{aLvaluesEFT}
\,.
\ea

Similarly to the SM case, without keeping $m_a$ and $m_b$ fixed, 
\be
\rho_{\rm NP}=\int \rho_{\rm NP}(m_a,m_b) w(m_a,m_b) dm_a dm_b\,.
\ee
 the system is not pure anymore, and the coefficients $C$ and $A$ receive contributions from different values of $m_a$ and $m_b$ with different weights $w(m_a,m_b) $. 
 
 \medskip
 
Before moving to the next section we want to highlight two limitations of the formulas presented in this section. First,  as we have already clarified several times, the spin of the $V$ bosons cannot be directly measured and therefore a QT approach has to be applied over the momenta of the four fermions. Having four fermions as $\mu^+ \mu^- e^+ e^-$, the possible intermediate $VV$ pairs are three: $ZZ$, $Z\gamma$ and $\gamma\gamma$. All of them can contribute, however a choice has to be made for the value of $\alpha_a$ and $\alpha_b$ in Eqs.~\eqref{eqn:coeff_defA} and \eqref{eqn:coeff_defC} and, especially,  it cannot be separately adapted for the three different $VV$ pairs. In other words, the condition in Eq.~\eqref{eq:alphatrick} can be true only for one of the three pairs. 

The second limitation, which again is not present in the $WW$ case or the aforementioned scenario for $\mu^+ \mu^- e^+ e^-$, is that in a general case not only the $ZZ$, $Z\gamma$ and $\gamma\gamma$ intermediate states are present, but they also interfere. This is precisely what happens at NLO EW accuracy where one-loop amplitudes, featuring $ZZ$, $Z\gamma$ and $\gamma\gamma$ intermediate states, contribute only via the interference with the tree-level amplitude, which instead features only the $ZZ$ intermediate state. The formulas in this section do not explicitly account for this effect.

\subsection{EFT analysis: numerical results via quantum tomography}
\label{sec:EFTnum}

In this section we provide numerical results and we discuss the impact of BSM effects on the $A$ and $C$ coefficients, and in turn on QI observables, extracted via the QT approach from the four-lepton system. We also discuss the size of BSM effects in comparison with the NLO EW corrections in the SM. Moreover, we investigate the sensitivity to NP that can be obtained via these quantities.

In order to assess the potential of $A$ and $C$ coefficients and QI observables in constraining the $a_i$ parameters introduced in the previous section in Eq.~\eqref{EFTvertex}, we perform  numerical simulations. As anticipated, we limit our study to the case of NP associated only to new heavy-states that do not interact with the fermions of the SM. 
Specifically, we employ the ``\textit{Higgs Characterisation model}"~\cite{Artoisenet:2013puc}, which provides an EFT parametrisation at the electroweak scale of a spin-0 neutral state ($X_0$) interacting with two SM vector bosons, taking into account both CP-even and CP-odd contributions. This model is written directly via the fields after the electroweak-symmetry breaking (EWSB), {\it i.e.}~$Z,\gamma$ and $W$, and includes only contributions of dimension four and five in the expansion in powers of $1/\Lambda$, where $\Lambda$ is the NP scale. The effective couplings employed here can be mapped to the SMEFT dimension-6 operators~\cite{Grzadkowski:2010es}, as shown in Appendix~\ref{appendixSMEFTOp}.

The effective Lagrangian for this model reads:
\begin{align}
\mathcal{L}_{X_0VV} &= 
\Bigg[ c_\phi \kappa_{\rm SM} \left[\frac{1}{2} g_{HZZ} Z^\mu Z_\mu + g_{HWW} W^+_\mu W^{-\mu} \right] \nonumber \\
&\quad - \frac{1}{4} \left[c_\phi \kappa_{H\gamma\gamma} g_{H\gamma\gamma} A_{\mu\nu} A^{\mu\nu} + s_\phi \kappa_{A\gamma\gamma} g_{A\gamma\gamma} A_{\mu\nu} \widetilde{A}^{\mu\nu} \right] \nonumber \\
&\quad - \frac{1}{2} \left[c_\phi \kappa_{HZ\gamma} g_{HZ\gamma} Z_{\mu\nu} A^{\mu\nu} + s_\phi \kappa_{AZ\gamma} g_{AZ\gamma} Z_{\mu\nu} \widetilde{A}^{\mu\nu} \right] \nonumber \\
&\quad - \frac{1}{4} \frac{1}{\Lambda} \left[c_\phi \kappa_{HZZ} Z_{\mu\nu} Z^{\mu\nu} + s_\phi \kappa_{AZZ} Z_{\mu\nu} \widetilde{Z}^{\mu\nu} \right] \nonumber \\
&\quad - \frac{1}{2} \frac{1}{\Lambda} \left[c_\phi \kappa_{HWW} W_{\mu\nu}^+ W^{-\mu\nu} + s_\phi \kappa_{AWW} W_{\mu\nu}^+ \widetilde{W}^{-\mu\nu} \right] \nonumber \\
&\quad - \frac{1}{\Lambda} c_\phi \left[ \kappa_{H\partial \gamma}  Z_\nu \partial_\mu A^{\mu\nu} + \kappa_{H\partial Z} Z_\nu \partial_\mu Z^{\mu\nu} + (\kappa_{H\partial W} W^+_\nu \partial_\mu W^{-\mu\nu} + \text{h.c.} \right)]  \Bigg] X_0 \, .
\label{langrangian}
\end{align}
The $V^{\mu\nu}$ terms are defined as $V^{\mu\nu}\equiv\partial^\mu V^\nu-\partial^\nu V^\mu$ and $\widetilde V^{\mu\nu} \equiv \frac{1}{2} \epsilon^{\mu\nu\rho\sigma}V_{\rho\sigma}$.
The terms $c_{\phi}\equiv\cos{\phi}$ and $s_{\phi}\equiv\sin{\phi}$ respectively parameterise the CP-even and CP-odd interactions via the CP-phase $\phi$. The coefficients $\kappa_{i}$ are real parameters (with the exception of $\kappa_{H\partial W}$) quantifying potential new physics contributions. The SM is recovered, {\it at LO}, by setting  $c_{\phi}=1$, $\kappa_{\rm SM} = 1$ and all other $\kappa_{i}$ = 0. As can be seen, the normalisation parameters $g_{ijk}$ are present only for the four-dimension operators and are defined as follows. 

In the case of $g_{HZZ}$ and $g_{HWW}$, the parameters are precisely set to the values that lead to the corresponding SM interactions after EWSB, when $c_{\phi}=1$ and $\kappa_{\rm SM} = 1$. In the case of $g_{H\gamma\gamma}$ and $g_{HZ\gamma}$,  the parameters are instead set to the values that lead to the corresponding UV-finite one-loop induced SM interactions, when $c_{\phi}=1$ and respectively $\kappa_{H\gamma\gamma} = 1$ and $\kappa_{HZ\gamma} = 1$, after having integrated out any mass dependence of the particles running in the loop. The cases of $g_{A\gamma\gamma}$ and $g_{AZ\gamma}$ are equivalent to those of $g_{H\gamma\gamma}$ and $g_{HZ\gamma}$ respectively, but assuming that the Higgs is a purely CP-odd pseudo-scalar with the same strength interactions of the SM purely CP-even scalar. The values of each normalisation parameter $g_{ijk}$ can be found in Tab. 2 of Ref.~\cite{Artoisenet:2013puc}. The dimension-5 operators are strictly speaking non-renormalisable, and therefore a normalisation procedure based on the same logic of the normalisation parameters $g_{ijk}$ is not possible, instead here enters the NP scale $\Lambda$ at denominator.

The Lagrangian includes interaction terms that modify the couplings of the Higgs boson not only to $ZZ$, but also to $WW,\, Z\gamma$ and $\gamma\gamma$. The relations among the couplings $a_i$ that have been introduced in the previous section and the $\kappa$ parameters in the effective Lagrangian for the different $VV$ pairs ($ZZ, WW, Z\gamma$ and $\gamma \gamma$) are summarised in Tab.~\ref{coupling_convers}. All the parameters appearing in Eq.~\eqref{langrangian} are real (except $\kappa_{H\partial W}$) and therefore, as anticipated in the previous section, all the $a_i$ parameters for the ``\textit{Higgs Characterisation model}" are also real (except $a_1$ for $WW$). For the numerical analysis that we are going to discuss, only the case of $WW$ is not relevant.

\begin{table}[t!]
\renewcommand{\arraystretch}{1.6}
\centering
\begin{tabular}{|c|l|c|l|}
\hline
 \(VV\) & & \(VV\) & \\ \hline
\(ZZ\) &
 \( a_1 = c_{\phi} \kappa_{\rm SM} + \frac{v}{\Lambda} \frac{\left(m_a^2 + m_b^2\right)}{2M_Z^2} c_{\phi} \kappa_{H\partial Z} \) & \(Z\gamma\) &
 \( a_1 = \frac{v}{\Lambda} \frac{m_{b}^2}{M_Z^2} c_{\phi} \kappa_{H\partial \gamma} \) \\ 
 & \( a_2 = \frac{1}{2} \frac{v}{\Lambda} c_{\phi} \kappa_{HZZ} \) &  & \( a_2 = v c_{\phi} \kappa_{HZ\gamma} g_{HZ\gamma} \)\\ 
 & \( a_3 = \frac{1}{2} \frac{v}{\Lambda} s_{\phi} \kappa_{AZZ} \) & &  \( a_3 = v s_{\phi} \kappa_{AZ\gamma} g_{AZ\gamma} \)\\ \hline
  \rule{0pt}{1.1\normalbaselineskip}
\(WW\) &
 \( a_1 = 2 c_{\phi} \kappa_{\rm SM} + \frac{v}{\Lambda} \frac{\left(\kappa_{H\partial W} m_a^2 + \kappa_{H\partial W}^* m_b^2\right)}{M_Z^2} c_{\phi} \) &
\(\gamma \gamma\) &
\( a_1 = 0 \) \\ 
 & \( a_2 = \frac{v}{\Lambda} c_{\phi} \kappa_{HWW} \) &  & \( a_2 = \frac{1}{2} v c_{\phi} \kappa_{H\gamma\gamma} g_{H\gamma\gamma} \) \\ 
 & \( a_3 = \frac{v}{\Lambda} s_{\phi} \kappa_{AWW} \) &  & \( a_3 = \frac{1}{2} v s_{\phi} \kappa_{A\gamma\gamma} g_{A\gamma\gamma} \) \\
 \hline
\end{tabular}
\caption{Parameters \(a_{1,2,3}\) from Eq.~\eqref{EFTvertex} expressed in terms of the parameters of Eq.~(\ref{langrangian}), for different $VV$ pairs.}
\label{coupling_convers}
\end{table}

\medskip

In the following analysis, we investigate multiple couplings by varying either a single $\kappa_i$ coefficient at the time or simultaneously with the CP phase ($\phi$). The simulation of the $H \rightarrow  e^+ e^- \mu^+ \mu^- $ process was performed at LO using {\aNLO}, incorporating the dedicated {\UFO} model for the modified Lagrangian.
The  $A$ and $C$ have been reconstructed from the four final-state lepton distributions, following the same procedure used for the SM at LO, as described in Sec.~\ref{sec:Lodetails}. Thus, the value assumed for the spin-analysing power $\alpha$ is the one of the $Z$ interacting with leptons, also in the presence of contributions involving the operators with one or more photon fields. Starting from the $A$ and $C$ coefficients, we have also reconstructed the observables sensitive to entanglement and the Bell operator, as outlined in Sec.~\ref{sec:entangNLO}.

Since we consider the amplitude squared, where all the NP interactions are present, in our predictions both linear and quadratic terms are present in the expansion $1/\Lambda$. Moreover, each contribution to our prediction is proportional to $\kappa_i \kappa_j$, where $\kappa_i$ and $\kappa_j$ is any of the $\kappa$ in Eq.~\eqref{langrangian}, besides those involving $W$ bosons.  However, the Lagrangian in Eq.~\eqref{langrangian} involves several free parameters, and we had to make choices in order to be definite in the discussion. To this purpose, we neglect the impact of $\kappa_{H\partial Z}$ by setting it equal to zero and fix $\kappa_{\rm SM}=1/ c_{\phi}$, such that $a_1=1$ for $ZZ$ and therefore we recover the SM prediction if any other $\kappa=0$. The advantage of this choice is that minimal variation of any of the $\kappa$ couplings will parametrise the departure from the SM prediction; from numerical results it will be manifest if the dominant effect is induced by linear or quadratic contributions in $1/\Lambda$. Special attention is necessary for the variation of $\phi$, and we will get back to this point in the discussion.

\medskip 

We start to examine the parameters affecting the $HZZ$ vertex: $\kappa_{HZZ}$, $\kappa_{AZZ}$ and $\phi$. The range of variation for these parameters around their SM value,  which is clearly equal to zero for all of them, has been chosen based on the limits imposed by recent experimental results. In particular, a study conducted by the CMS collaboration~\cite{CMS:2019ekd} has set constraints on the Higgs interaction couplings using a parameterisation of NP effects similar to the one adopted in Eq.~\eqref{EFTvertex}.

Specifically, the CMS analysis obtained the following constraints on these couplings:
\begin{equation}
\label{CMSpaprameter}
\frac{a^{\text{CMS}}_3}{a^{\text{CMS}}_1} \in [-1.13, 0.80]\,, \quad \frac{a^{\text{CMS}}_2}{a^{\text{CMS}}_1} \in [-0.12, 0.26]\,.
\end{equation}
The couplings $a^{\text{CMS}}_2$ and $a^{\text{CMS}}_3$, but {\it not}  $a^{\text{CMS}}_1$, that have been used in the CMS parametrisation, corresponds respectively to $a_2/2$ and $a_3/2$ in Eq.~(\ref{EFTvertex}). Keeping this in mind and using Tab.~\ref{coupling_convers}, we can determine reasonable ranges for varying the $\kappa_i$ coefficients. Setting $\Lambda=1~\tev$ they are
\begin{equation}
\label{kappavales}
\kappa_{HZZ} \in [-4.2, 2.0]\,, \quad \kappa_{AZZ} \in [-13.1, 18.5]\,.
\end{equation}

Since no intermediate photons are possible if only $\kappa_{HZZ}$ and $\kappa_{AZZ}$ are non-vanishing, for this case the formulas of Sec.~\ref{sec:anarho} can be exploited to understand which coefficients are non-vanishing and the relations among them.\footnote{Still we remind the reader that all the formulas in Sec.~\ref{sec:anarho} are for fixed $m_a$ and $m_b$, while here we integrate over the full phase space.}
Considering the relations in the left column of Tab.~\ref{EFT_coeffe}, only three $A$ and $C$ coefficients are independent. We focus on those with $L>1$, where NLO corrections are not very large.

From the fixed-mass analytical expressions for the coefficients in Eqs.~\eqref{ACvalueEFT2} and \eqref{ACvalueEFT3}, we observe that the $C$ coefficients can acquire nonzero imaginary components—an effect absent in the SM at both LO and NLO accuracy. This occurs when the coupling $\kappa_{AZZ}$ is activated and the phase $\phi$ is modified such that $s_\phi \neq 0$. As can be seen from Tab.~\ref{coupling_convers}, in this way $a_3$ becomes non-vanishing and in turn the $C$ coefficients acquire an imaginary component.

\begin{figure}[t!]
     \centering
     \begin{subfigure}[b]{0.49\textwidth}
         \centering
         \includegraphics[width=\textwidth]{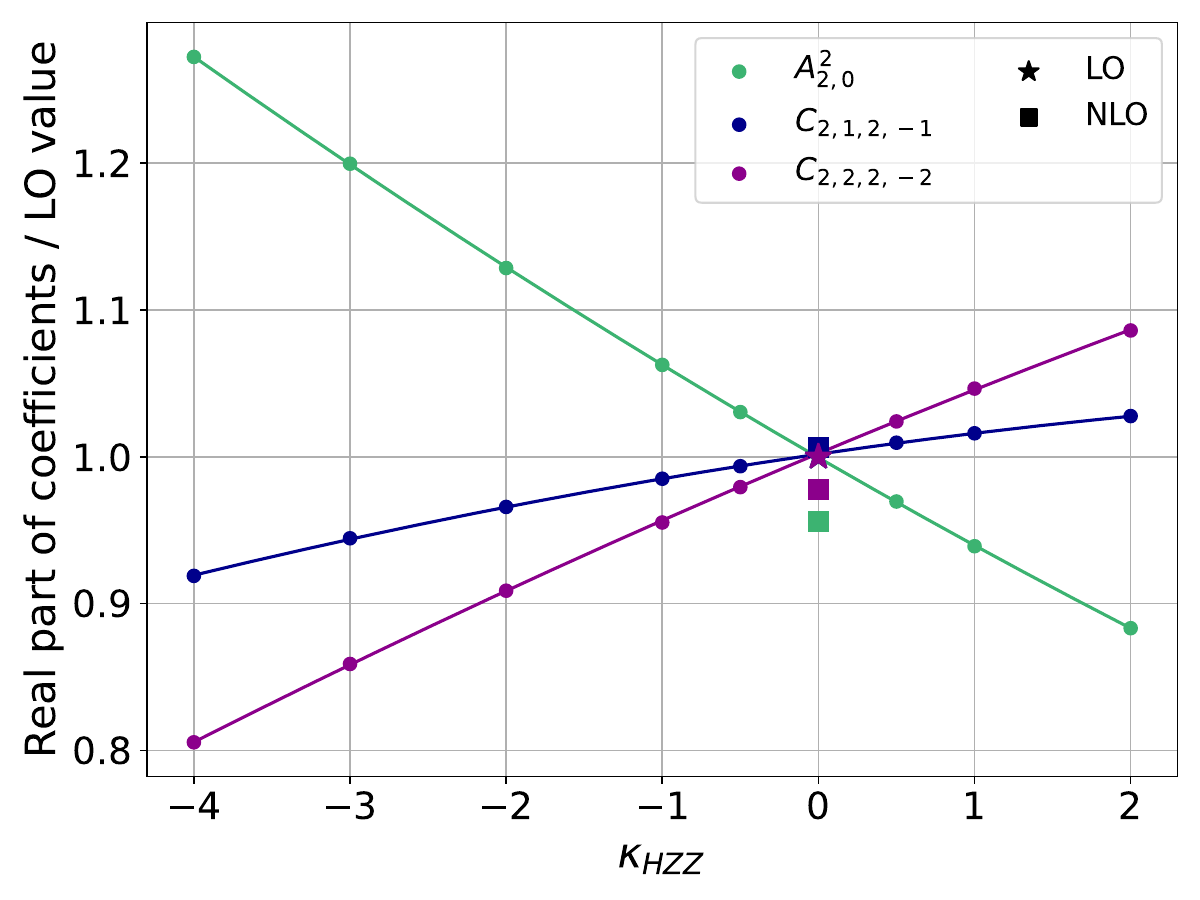}
         \caption{}
         \label{kHZZ_coeff}
     \end{subfigure}
     \begin{subfigure}[b]{0.49\textwidth}
         \centering
         \includegraphics[width=\textwidth]{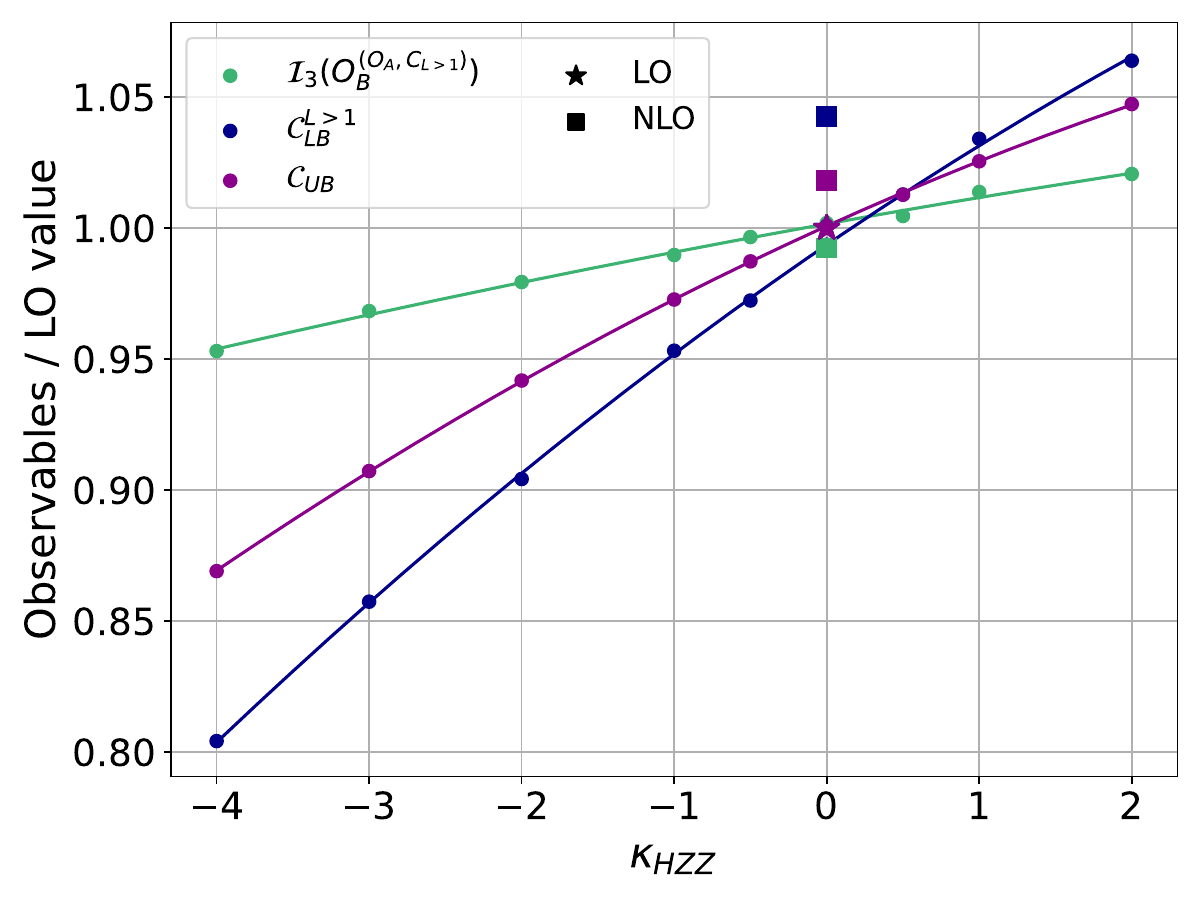}
         \caption{}
         \label{kHZZ_obs}
     \end{subfigure}
        \caption{Values of different  observables as functions of $\kappa_{HZZ}$. Each prediction is normalised over its LO  SM  value. The SM NLO EW predictions are also shown, for comparison.}
        \label{fig:kHZZ}
\end{figure}

The results for the $A$ and $C$ coefficients, as well as the observables ${\cal I}_3(O_B^{(O_A,C_{L>1})}),{\mathcal C}^{L>1}_{LB}$ and ${\mathcal C_{UB}}$, as a function of the variation of the $\kappa_{HZZ}$ coupling, are shown in Fig.~\ref{fig:kHZZ}.
The plots illustrate that all \( A \) and \( C \) coefficients exhibit a linear dependence on \( \kappa_{HZZ} \) within the explored range. For fixed invariant masses, this dependence can be derived from a linear expansion in powers of $1/\Lambda$ of Eq.~\eqref{ACvalueEFT2}--\eqref{ACvalueEFT3}. We notice that for all coefficients the common denominator contains both linear and quadratic terms in $1/\Lambda$, while the numerator's order depends on the specific coefficients. In the case of \( A_2^{2,0} \) it involves only $\Lambda$-independent (therefore of SM origin in this case) and $1/\Lambda^2$ terms, whereas for the \( C \) coefficients the numerator contains $\Lambda$-independent, $1/\Lambda$ and $1/\Lambda^2$ terms.

The coefficient that shows the largest sensitivity to \( \kappa_{HZZ} \) variations turns out to be  \( A^2_{2,0}\), reaching up to 30\% effects at the lower edge of the explored range. As can be seen from the comparison of Eq.~\eqref{ACvalueEFT} {\it vs.}~Eqs.~\eqref{ACvalueEFT2} and \eqref{ACvalueEFT3} the relative impact of $a_2$ on $A^2_{2,0}$ is anti-correlated to the one on the $C$ coefficients, when $a_1=1$ and $a_3=0$ as in our case. This suggests that simultaneous measurements of the $A$ and $C$ coefficients could impose more stringent constraints on $\kappa_{HZZ}$.

The observables examined in Fig.~\ref{kHZZ_obs} exhibit trends similar to those of the individual $A$ and $C$ coefficients. However, in particular, ${\cal I}_3 (O_B^{(O_A,C_{L>1})})$ and ${\mathcal C}^{L>1}_{LB}$  are more challenging to evaluate in real data, as they require the measurement of multiple coefficients and their correlations. An interesting aspect to note is that positive values of $\kappa_{HZZ}$ appear to enhance the quantum correlation between the two bosons.

\begin{figure}[t!]
     \centering
     \begin{subfigure}[b]{0.49\textwidth}
         \centering
         \includegraphics[width=\textwidth]{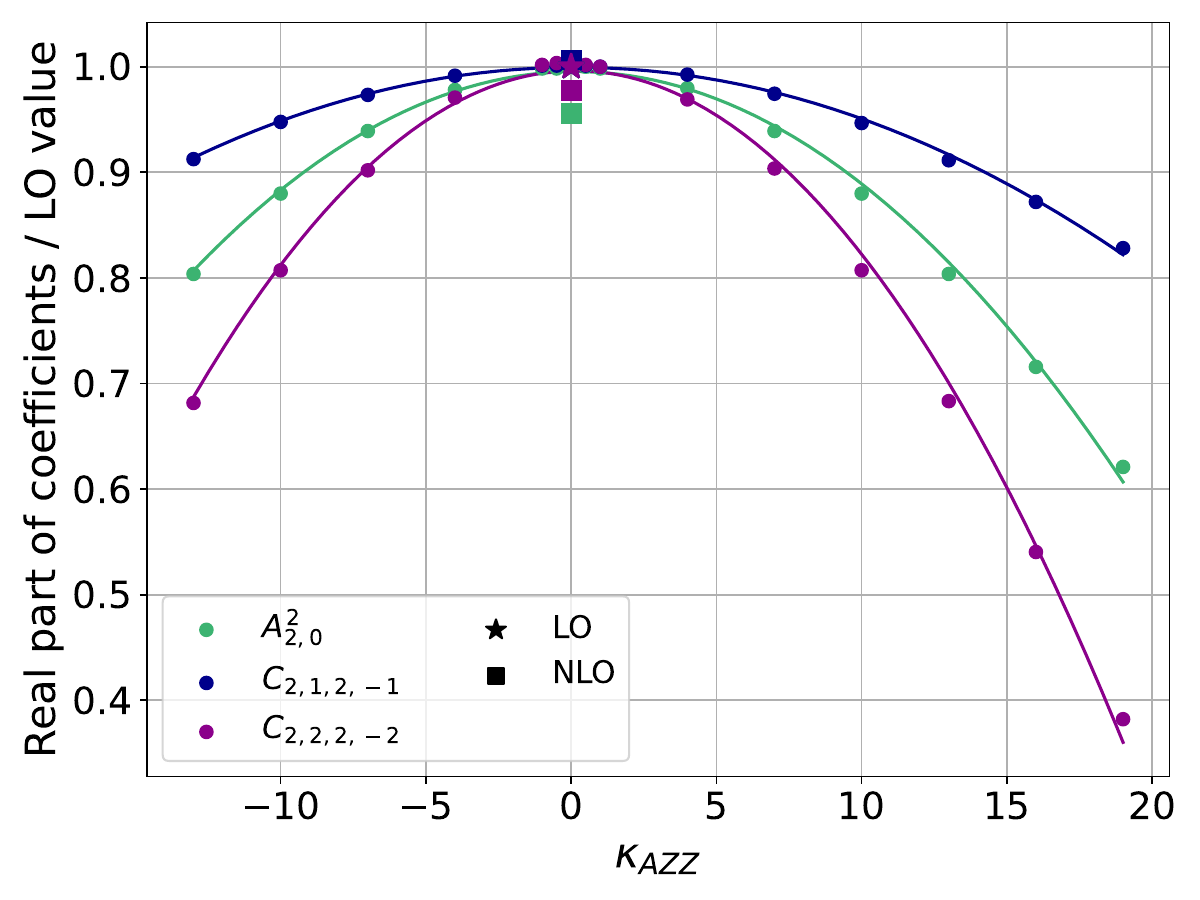}
         \caption{}
         \label{kAZZ_coeff}
     \end{subfigure}
     \begin{subfigure}[b]{0.49\textwidth}
         \centering
         \includegraphics[width=\textwidth]{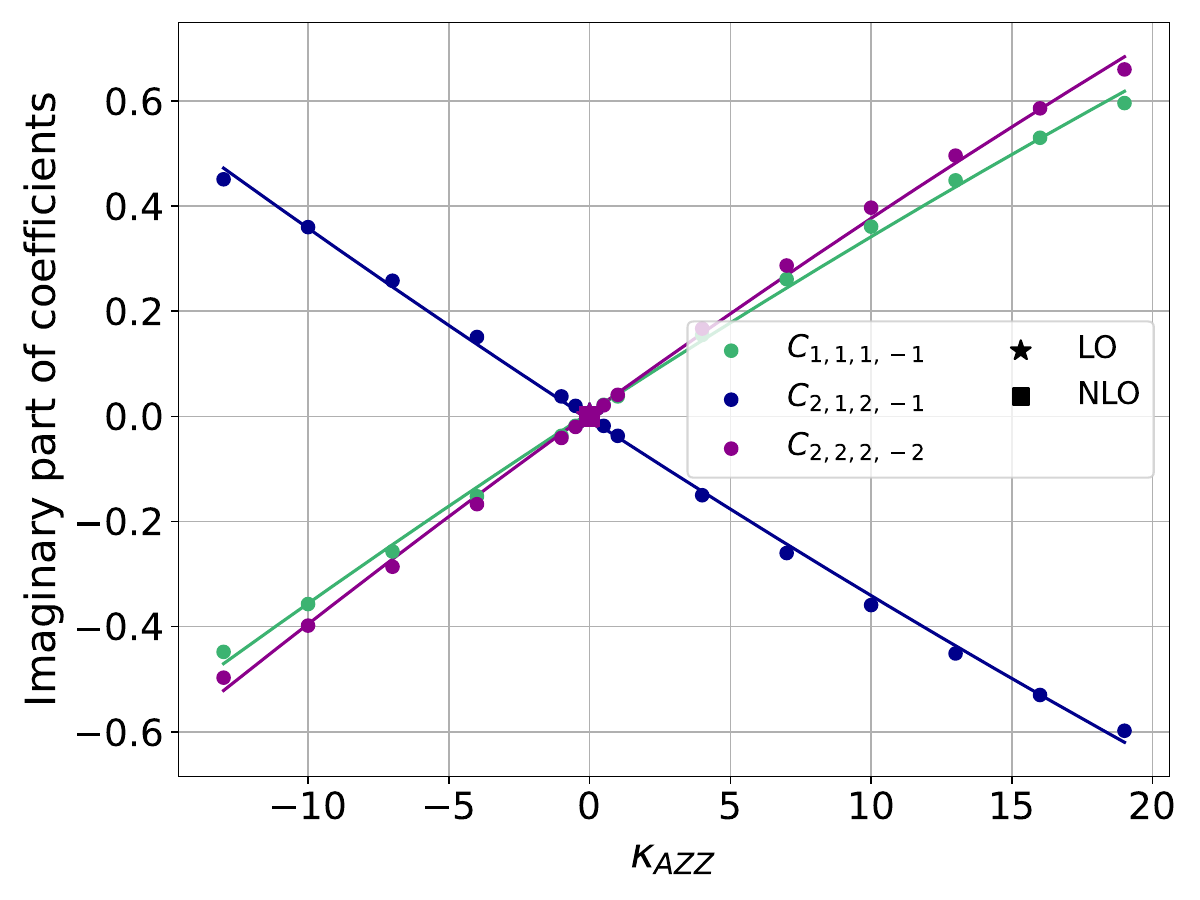}
         \caption{}
         \label{kAZZ_coeff_imag}
     \end{subfigure}
     \begin{subfigure}[b]{0.49\textwidth}
         \centering
         \includegraphics[width=\textwidth]{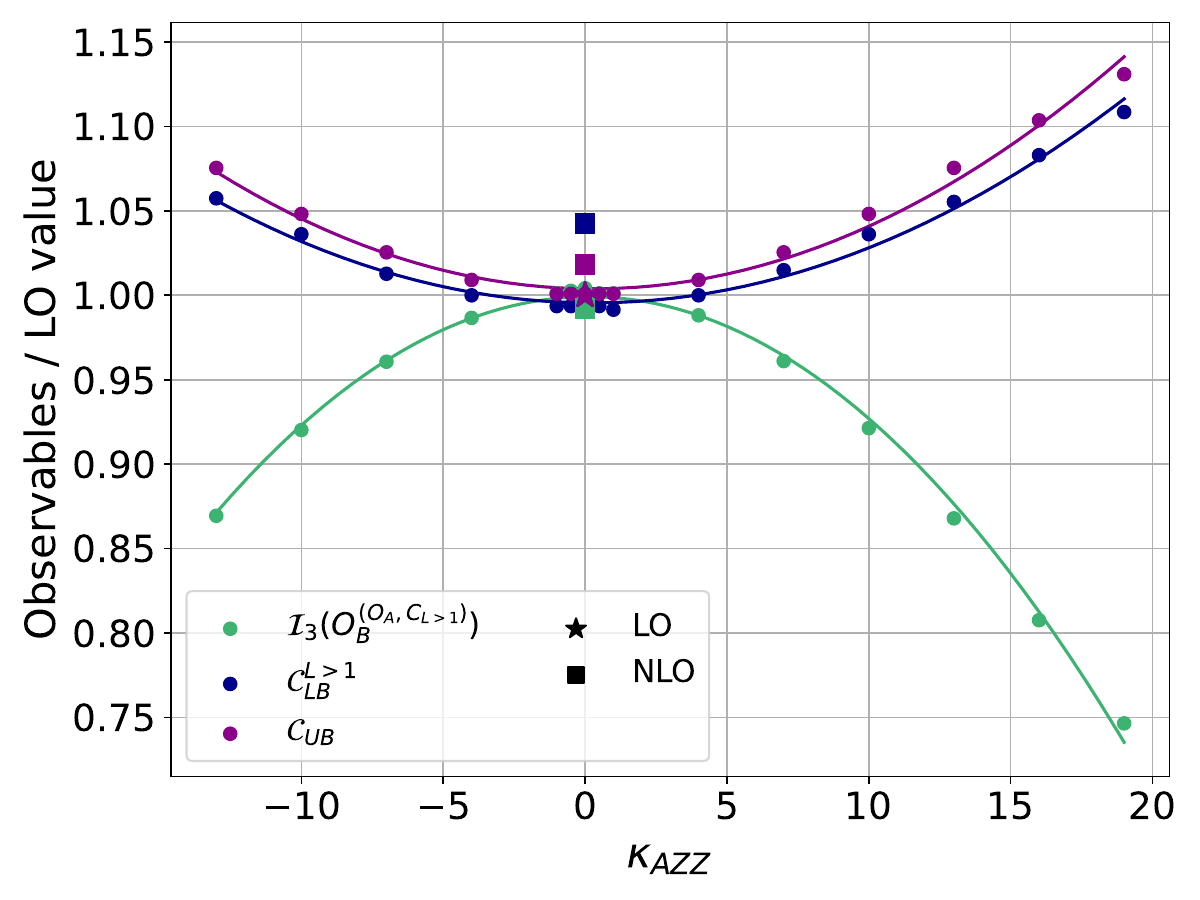}
         \caption{}
         \label{kAZZ_obs}
     \end{subfigure}
        \caption{Values of different  observables as functions of  $\kappa_{AZZ}$. Each prediction is normalised over its LO SM  value, besides those in the top-right plots showing the imaginary parts, which are zero in the SM. The SM NLO EW predictions are also shown, for comparison. }
        \label{fig:kAZZ}
\end{figure}

The results obtained setting ${\phi = \frac{\pi}{2}}$ and spanning on $\kappa_{AZZ}$ are shown in Fig.~\ref{fig:kAZZ}.\footnote{In practice, we have set ${\phi = \frac{\pi}{4}}$ and rescaled $\kappa_{AZZ}$ by a factor of $\sqrt{2}$ in order to keep numerically stable the value  $\kappa_{\rm SM}=1/ c_{\phi}$, which ensures the $a_1=1$  SM condition.}  In this case, the real part of the coefficients $A$ and $C$ displayed in Fig.~\ref{kAZZ_coeff}, and the observables displayed in Fig.~\ref{kAZZ_obs}, show a quadratic dependence on $\kappa_{AZZ}$. Also in this case the coefficients used to build the spin density matrix show a distinct sensitivity to NP effects, with $C_{2,2,2-2}$ reaching a variation of 60$\%$ compared to the SM value, at the right end of the investigated $\kappa_{AZZ}$ range. The quantities ${\cal I}_3 (O_B^{(O_A,C_{L>1})})$, ${\mathcal C_{LB}}^{L>1}$ and ${\mathcal C_{UB}}$ show some sensitivity to the $\kappa_{AZZ}$ variations, and $\mathcal{I}_3$ displays the most significant effects across the investigated range. Interestingly, the value ${\cal I}_3$ seems to decrease along the whole $\kappa_{AZZ}$ range, suggesting that the level of entanglement between the two bosons decreases, while the other two observables show an opposite trend. This is possible because ${\mathcal C_{LB}}^{L>1}$ and ${\mathcal C_{UB}}$ are just bounds for the concurrence and can not properly quantify the entanglement.

The reduction in entanglement in this case can be understood by examining Eqs.~\eqref{EFTstate}--\eqref{aLvaluesEFT}. The dependence of \( a_+ \) and \( a_- \) on \( \kappa_{HZZ} \) is identical, ensuring \( a_+ = a_- \), as in the SM at LO. However, their dependence on \( s_{\phi} \kappa_{AZZ} \), which enters via $a_3$, differs and leads to \( (a_+ - a_-) \propto s_{\phi} \kappa_{AZZ} \). Consequently, as \( |s_{\phi} \kappa_{AZZ}| \) increases, the system further deviates  from a maximally entangled state (\( a_+ = a_- = a_0 \)), explaining the overall decrease in \( {\cal I}_3 (O_B^{(O_A,C_{L>1})}) \) across the entire \( \kappa_{AZZ} \) range.

Figure~\ref{kAZZ_coeff_imag} illustrates the dependence of the imaginary component of the \( C \) coefficients on \( \kappa_{AZZ} \). All coefficients exhibit a linear dependence and similar sensitivity, receiving up to $\pm 60\%$ corrections at the edges of the investigated range. This figure also includes \( C_{1,1,1,-1} \) since the imaginary contributions to all coefficients from the SM, including NLO EW corrections, are zero. As expected from Tab.~\ref{EFT_coeffe}, \( C_{1,1,1,-1} \) contributes equally and oppositely to \( C_{2,1,2,-1} \).

\begin{figure}[t!]
     \centering
     \begin{subfigure}[b]{0.49\textwidth}
         \centering
         \includegraphics[width=\textwidth]{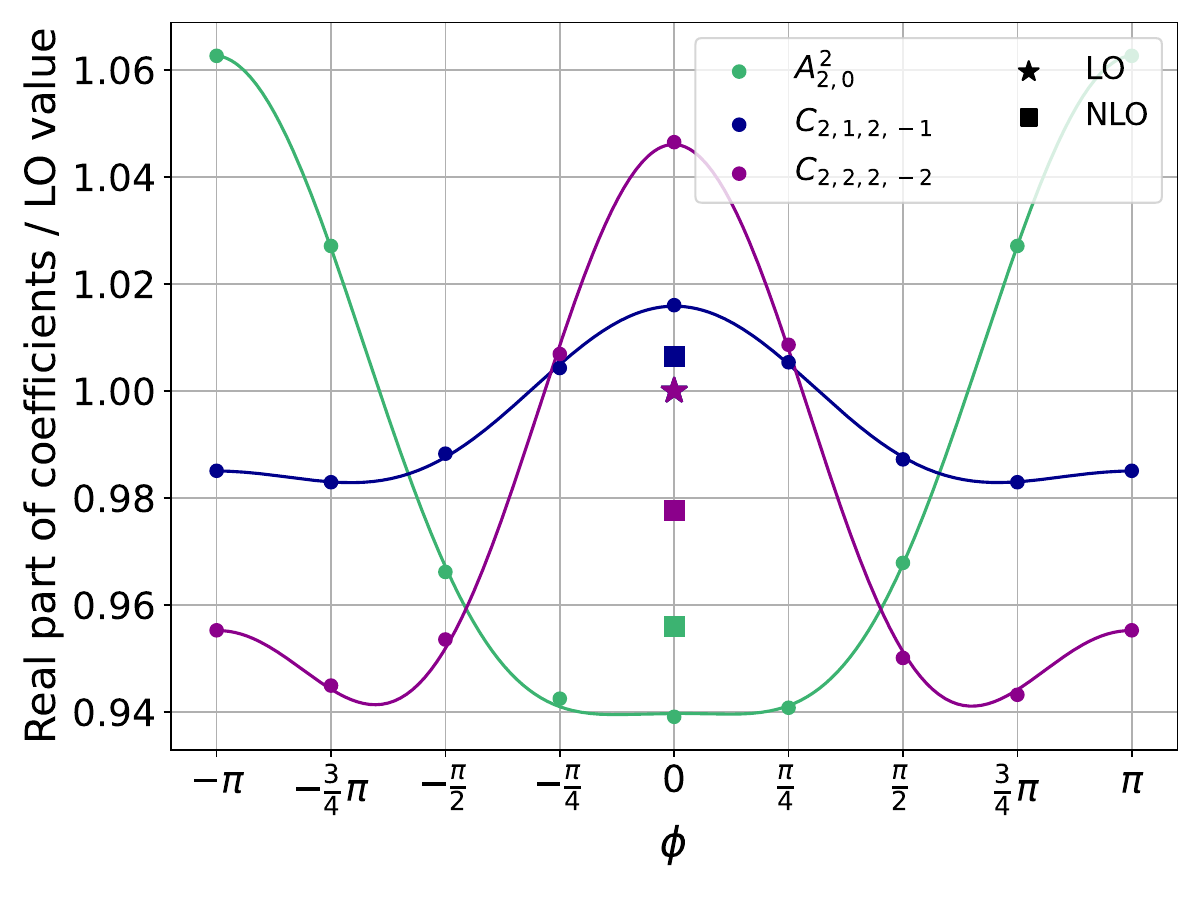}
         \caption{}
         \label{kXZZ_alpha_coeff}
     \end{subfigure}
     \begin{subfigure}[b]{0.49\textwidth}
         \centering
         \includegraphics[width=\textwidth]{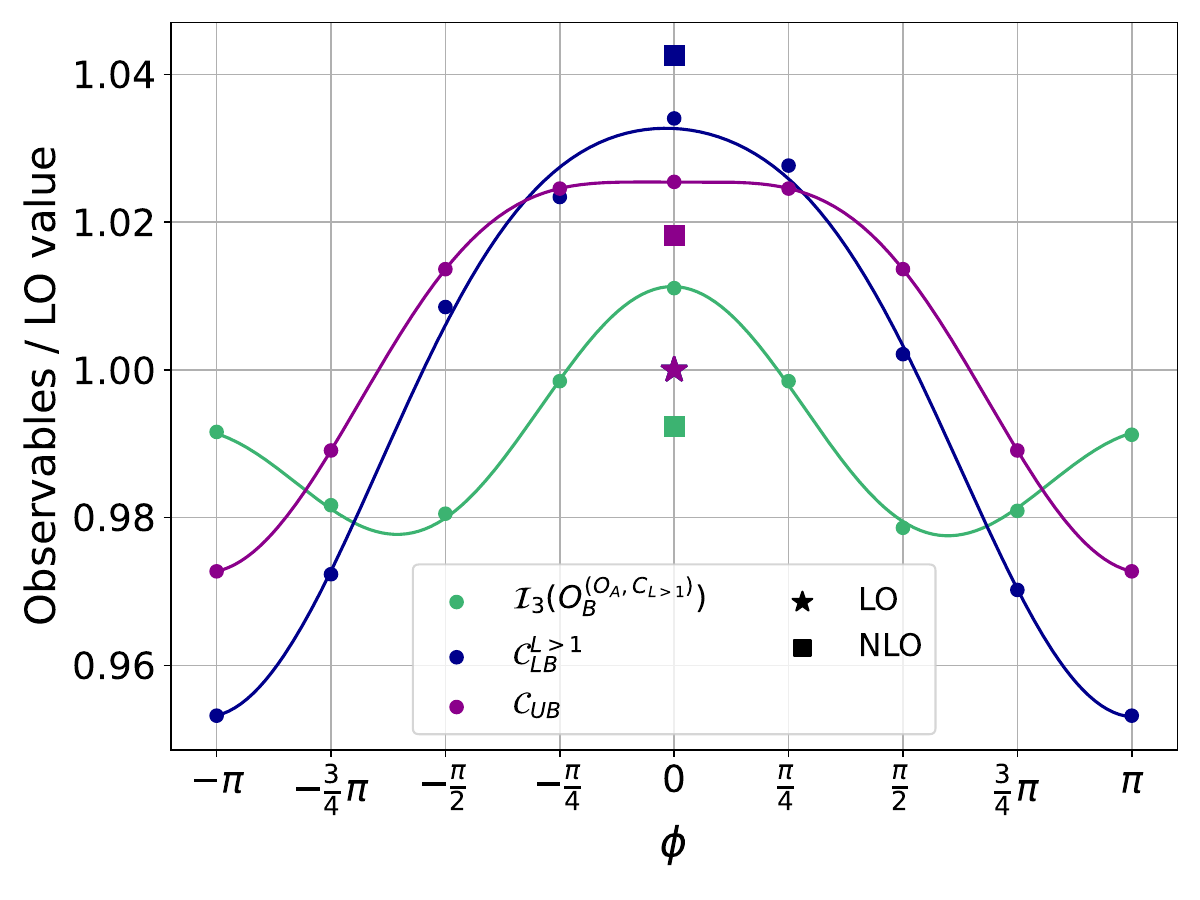}
         \caption{}
         \label{kXZZ_alpha_obs}
     \end{subfigure}
        \caption{Values of different  observables as functions of  $\phi$ for $\kappa_{HZZ} = 1$ and $\kappa_{AZZ} = 5$. Each prediction is normalised over its LO  SM  value. The SM NLO EW predictions are also shown, for comparison.
        }
        \label{fig:kHZZ_alpha}
\end{figure}

Figure~\ref{fig:kHZZ_alpha} shows the variations of the coefficients and QI-inspired observables as a function of the CP phase \( \phi \), while keeping \( \kappa_{HZZ} = 1 \) and \( \kappa_{AZZ} = 5 \), such that they lead to effects of similar size and are both within the allowed ranges of Eq.~\eqref{kappavales} at the same time.  All investigated variables exhibit a combination of sine and cosine dependence, with amplitudes determined by the coupling values. Since the coefficients are dominated by the SM contribution ($\kappa_{\rm SM}  c_{\phi}=1$), they show only a moderate dependence on the phase variations, with $C_{2,2,2,-2}$ displaying the greatest sensitivity. On the other hand, $C_{2,1,2,-1}$ seems quite flat with respect to the phase variation. In relation to the observables \( {\cal I}_3 (O_B^{(O_A,C_{L>1})}) \), ${\mathcal C_{LB}}^{L>1}$ and ${\mathcal C_{UB}}$ shown in  Fig.~\ref{kXZZ_alpha_obs}, the maximum value is observed at $\phi = 0$, where only the $\kappa_{HZZ}$ contributes to the decay width, in agreement with Fig.~\ref{kHZZ_obs}. The modification of the phase ($s_\phi \ne 0$), corresponding also to a non-vanishing contribution of the  pseudo-scalar component, reduces the observables. \( {\cal I}_3 (O_B^{(O_A,C_{L>1})}) \) is the observable where the oscillation as a function of $\phi$ shows the highest frequency, even if the magnitude of the variation is limited to $2\%$ with this choice of \( \kappa_{HZZ} = 1 \) and \( \kappa_{AZZ} = 5 \).
\medskip

\begin{figure}[t!]
    \centering
    \begin{subfigure}[b]{0.49\textwidth}
        \centering
        \includegraphics[width=\textwidth]{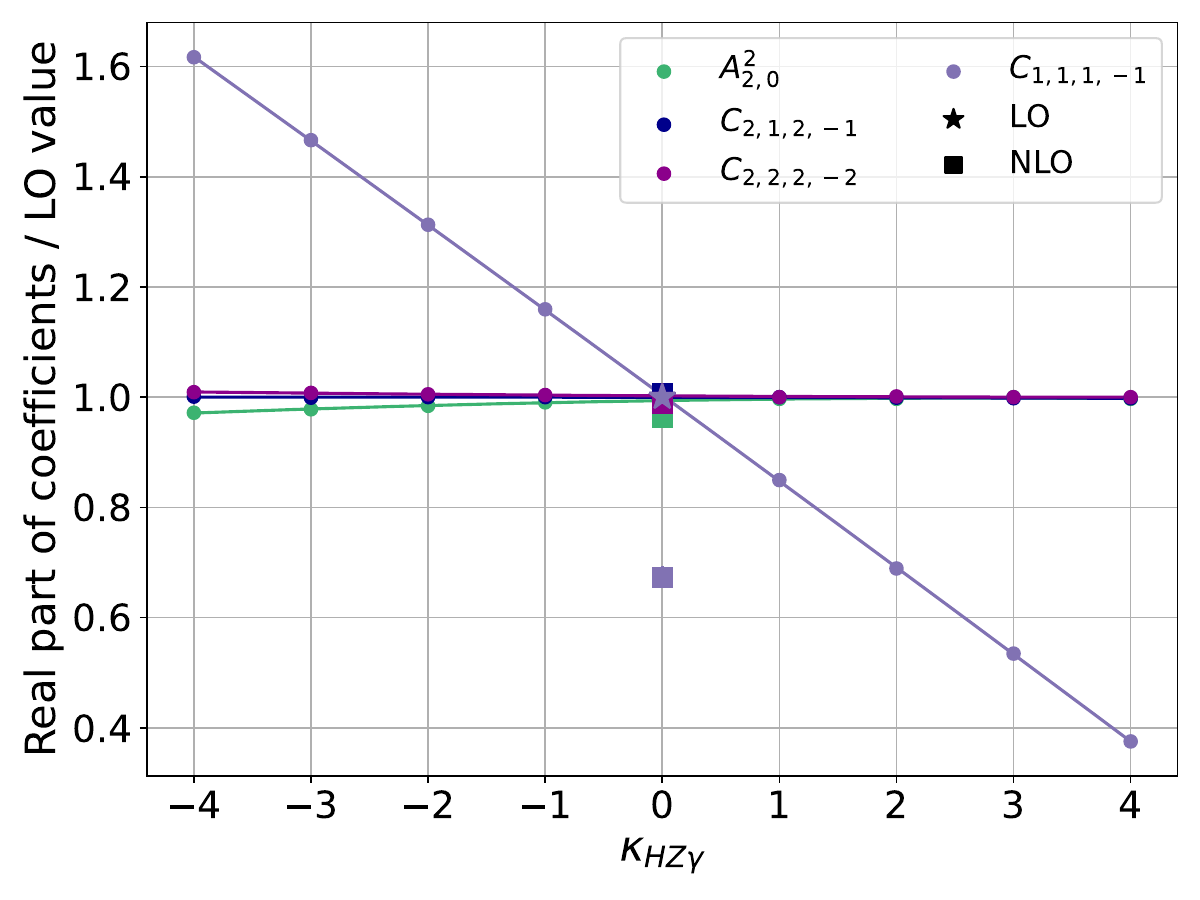}
        \caption{}
        \label{figa}
    \end{subfigure}
    \begin{subfigure}[b]{0.49\textwidth}
        \centering
        \includegraphics[width=\textwidth]{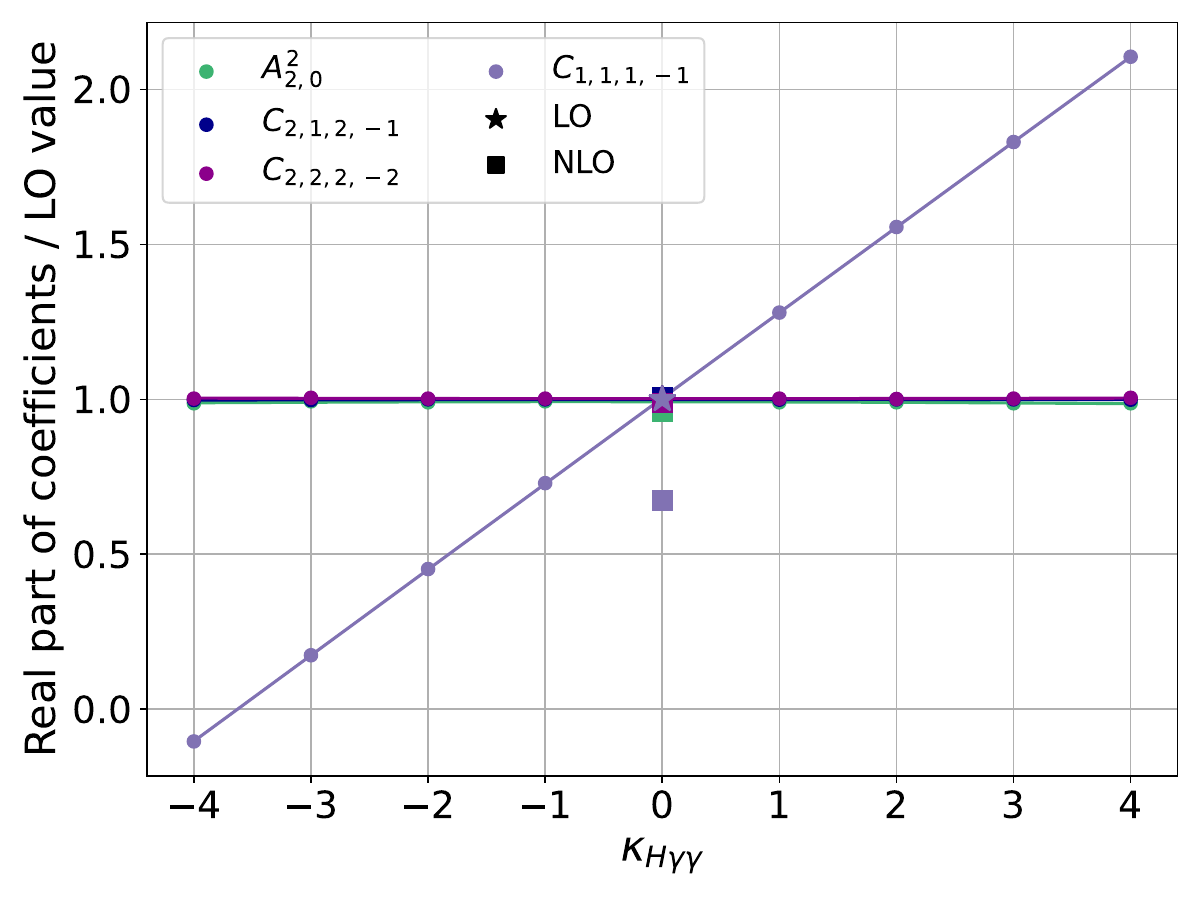}
        \caption{}
        \label{figb}
    \end{subfigure}
       \caption{Values of different  spin density matrix coefficients $A$ and $C$ as functions of $\kappa_{HZ\gamma}$ (left plot) and  $\kappa_{H\gamma\gamma}$ (right plot). Each prediction is normalised over its LO  SM  value. The SM NLO EW predictions are also shown, for comparison.}
       \label{fig:kHZg}
\vspace{-5mm}
\end{figure}

Finally, we investigated the sensitivity of a subset of the $A$ and $C$ coefficients to the variations of $\kappa_{HZ\gamma}$ and $\kappa_{H\gamma\gamma}$. We vary these coefficients while keeping $c_{\phi} \kappa_{\rm SM} = 1$.
This setup allows multiple types of diboson pairs to contribute to the four-lepton final state. While this scenario can be described by the effective Lagrangian shown in Eq.~\eqref{langrangian}, unlike the previous cases it cannot be directly derived from the analytical formulas given in Sec.~\ref{sec:anarho}. The problem is precisely what has been discussed at the end of Sec.~\ref{sec:anarho}: there are multiple possible intermediate $VV$ states, {\it i.e.}~$ZZ$, $Z\gamma$ and $\gamma\gamma$, but we are forced to assume a specific value of $\alpha$ to perform the QT. 
Also in this case, the $A$ and $C$ coefficients are reconstructed, as described in Section~\ref{sec:Lodetails},  employing the LO spin-analysing power of the leptons  originating from a  $Z$ decay. Thus, the $C_{1,M,1,-M}$ coefficients are not the true quantities entering the $\rho$ density matrix, since they depend on the mismatch between the value of $\alpha$ assumed and the one for $Z\gamma$ and $\gamma\gamma$ contributions. Moreover, in the simulations we impose a cut on $m(Z_2)$  in order to avoid divergences from the $\gamma\to \ell^+ \ell^-$  splitting.\footnote{This divergence is arising from the terms scaling as $\kappa_{H\gamma\gamma}^2$ or $\kappa_{A\gamma\gamma}^2$, while in the linear dependence on $\kappa_{H\gamma\gamma}$ or $\kappa_{A\gamma\gamma}$ it originates from the interference of the SM contribution, which does not involve photons, with diagrams featuring the $\gamma\to \ell^+ \ell^-$ splitting. This quantity, similar to the case of NLO EW corrections, leads to an integrable divergence.} We choose $m(Z_2) >30$ GeV, consistently with what has also been done in Sec.~\ref{sec:resultsNLOEW}. 

In this case, the concepts we have introduced in the last paragraph of Sec.~\ref{sec:LO} are particularly relevant for the  $C_{1,M,1,-M}$ coefficients: $C_{1,M,1,-M}$  can be calculated and measured regardless of the reconstruction of the $\rho$ density matrix via QT and its interpretation in the context of QI. In particular, $C_{1,M,1,-M}$ coefficients are affected by the spin-analysing power $\alpha$, and so they are expected to be sensitive to $\kappa_{HZ\gamma}$ and $\kappa_{H\gamma\gamma}$. In fact, an inconsistency similar to the one observed at NLO EW is present, and therefore a similar large sensitivity would not be surprising. Thus, we include $C_{1,1,1,-1}$ among the observables considered in our analysis.

Results, following the same style as the previous plots of this section, are shown in Fig.~\ref{fig:kHZg}. According to recent results based on the analysis of the $H\to Z\gamma$ decay~\cite{ATLAS:2023yqk}, a variation of $\kappa_{HZ\gamma}$ within a range $0 \lesssim \kappa_{HZ\gamma} \lesssim 2$ is still allowed at a $3\sigma$ confidence level. We decided to display a variation within a slightly larger interval, also for the case of $\kappa_{H \gamma\gamma}$, where instead the range is much more stringent ($0.7 \lesssim \kappa_{H \gamma\gamma} \lesssim 1.3$) if we assume a conservative $20\%$ accuracy on ${\rm Br}(H \to \gamma \gamma)$. 

While all the other $A$ and $C$ coefficients  show a minimal dependence on the variations of $\kappa_{HZ\gamma}$ and $\kappa_{H\gamma\gamma}$, the $C_{1,1,1,-1}$ coefficient  demonstrates, as expected, a very large sensitivity to the variations of these couplings. However, the NLO EW corrections in the SM are not negligible and can also lead to cancellations with the NP effects. This picture suggests that this observable could be used to derive independent limits on $\kappa_{HZ\gamma}$ and $\kappa_{H\gamma\gamma}$, but it  is essential to consider the NLO EW corrections in the simulation in order to extract reliable results. Especially, it is important to take into account EW corrections in order to avoid to interpret a possible deviation between data and LO simulations, which is clearly expected from NLO results, as a sign of NP effects.

As final comment we stress the fact that, according to  Tab.~\ref{EFT_coeffe}, the real parts of the $C$ coefficients satisfy the relation $C_{1,1,1,-1}=-C_{2,1,2,-1}$, which instead is manifestly false in Fig.~\ref{fig:kHZg}. The reason is precisely the fact that, since more $VV$ structures are present,  $C_{1,1,1,-1}$ also in this case cannot be interpreted as a coefficient of the density matrix, but only as the observable defined in Eq.~\eqref{eq:numerical_INT}. From the QI observable point of view this fact is unpleasant, but for the purpose of leveraging sensitivity to NP is actually convenient. There is not a problem on how to deal with large NLO EW corrections: they simply have to be taken into account for a reliable SM prediction. Rather, it is important to match NP contributions and predictions at NLO EW accuracy; NLO EW corrections already involve the contributions from terms equivalent to the case $\kappa_{HZ\gamma}=\kappa_{H\gamma\gamma}=1$, plus additional terms. We leave this study to future work, possibly with the inclusion of EW corrections to the NP contributions themselves.

\section{Conclusions and outlook}
\label{sec:conclusions}

In this paper we have explored the quantum properties of the Higgs boson decays into vector boson pairs ($H \to VV^*$), with a particular focus on the signature with two different-flavour lepton pairs ($H\to ZZ^*\to \mu^+ \mu^- e^+ e^-$). We have adopted an QI-inspired approach  to analyse the spin correlations of the two vector bosons, which are treated as a bipartite qutrit system.

First of all, we have reviewed the spin density matrix formalism of the $V V^*$ system using polarisation operators. We have also discussed in detail how, by employing QT techniques, the angular distributions of the decay products can be used to reconstruct the full quantum state of the intermediate vector boson pair. Since the spins of the vector bosons are not directly measurable at colliders, this reconstruction is crucial. 

Several QI observables are considered, such as measures of entanglement and markers for the violation of Bell-type inequalities. In particular, in order to quantify nonlocal correlations, we have focused on the upper and lower bounds of the concurrence and on the expectation values of Bell operators related to the CGLMP inequality. 

The core of the paper is the computation and the analysis of the NLO EW corrections. The results indicate that NLO EW effects can have a substantial impact on the extraction of the coefficients of the spin density matrix, and therefore on the evaluation of the QI observables. In some cases, these corrections are so large that the procedure to infer the parameters controlling the quantum correlations becomes unreliable.  We have explicitly shown that these effects are peculiar for the coefficients $C_{L_1,M_1,L_2,M_2}$ defined in the work and for the cases $L_1=L_2=1$. We have traced back the origin of this sensitivity to the smallness of the spin-analysing power $\alpha$ in the case of a $Z$ boson decaying into leptons, which is not protected by any symmetry and is subject to large radiative corrections. In this case, the precise knowledge of the $\alpha$ parameter is crucial for the extraction of coefficients of the form $C_{1,M,1,-M}$, and in turn for determining the structure of the $\rho$ density matrix of the $V V^*$ system. First, we have discussed the theoretical limitations in the definition of $\alpha$ at NLO EW accuracy. Second, we have explicitly  shown how this behaviour is special and in fact specific to the four-charged lepton final state.  We have shown that, thanks to a larger value of $\alpha$, the QT is not greatly affected by higher orders in $H\to ZZ^*\to q\bar q q' \bar q'$, especially for down-type quarks. Analogously, and consistently with our expectations, we have verified that QT allows to reconstruct the $\rho$ matrix reliably in the decays $H\to WW^*\to 4f$.

We have also investigated the impact of NLO EW corrections by varying the recombination radius $\Delta R$ of photons with leptons, and the impact of cuts on the invariant masses of the two lepton pairs. We find that the $A$ coefficients strongly depend on $\Delta R$, while the cuts on the invariant masses have a large impact on the size of NLO EW corrections to the $C_{1,M,1,-M}$ values: NLO EW corrections further increase when one on-shell $Z$ boson is selected.    

In order to overcome the limitations induced by higher-order corrections, we have explored modifications of the definition of QI observables. In particular, we have introduced modified definitions of $\mathcal{I}_3$ and the lower bound of the concurrence, ensuring that they do not depend on the $C_{1,M,1,-M}$ coefficients. This makes them robust against higher-order
corrections, even for the four-lepton signature. The only drawback of such approach is that the predicted value for the lower bound of the concurrence is smaller, leading to a more difficult observation of the entanglement. Thus, alternative approaches that would enable to account for EW corrections and at the same time not modify the definition of QI observables are desirable. In order to achieve such approaches, we notice that in addition to the problem related to the spin analysing power $\alpha$ and the $C_{1,M,1,-M}$ coefficients one would need to also consistently consider the multipartite entanglement (see {\it e.g.}~Refs.~\cite{Morales:2024jhj} and \cite{Horodecki:2025tpn}) from un-recombined real emissions of photons. 

Finally, in light of the improved predictions obtained in the SM, we have studied the sensitivity to new physics effects. To this aim, we have considered how modifications to the Higgs couplings, parameterised through an EFT approach framework, could affect the QI observables and the spin density matrix terms. Our analysis demonstrates that QI inspired observables are sensitive probes of potential new physics, and therefore can enhance the sensitivity of experimental measurements to deviations from the SM. We have provided both analytical formulas for amplitudes at fixed invariant masses of the two lepton pairs and numerical results for the full phase space. We have also discussed how, with BSM effects in the $HVV$ vertex, the final state $f\bar f f' \bar f'$ can also be reached through $\gamma \gamma$ and $Z\gamma$ intermediate states,  leading to complications in the QT approach for the extraction of the information on the $\rho$ density matrix. 

Future studies should focus on optimising the extraction of QI observables and exploring more realistic setups, as already done for the $ZZ$ production case in Ref.~\cite{Grossi:2024jae}. Simulations that incorporate the combination with the specific production mode and apply various selection cuts on the four-lepton final state would be highly beneficial. Given the size of NLO EW corrections for some of the observables considered, their calculation also for the BSM scenario would be desirable.

In conclusion, we have found that in specific cases, NLO EW corrections can induce large effects. By carefully studying their origin, we have reached the conclusion that this is a very peculiar case and will not be a typical pattern emerging in the study of QI observables in collider physics. Still, we acknowledge also in this context the paramount relevance of not ignoring effects from SM higher-order corrections before interpreting possible deviations from theory predictions.

\section*{Acknowledgments}

We are grateful to Chiara Arcangeletti, Shiuli Chatterjee, Baptiste Ravina and Juan Antonio Aguilar Saavedra for the helpful discussions.
D.P.~acknowledges financial support by the MUR through the PRIN2022 Grant 2022EZ3S3F and F.M.~and P.L.~through the PRIN2022 Grant 2022RXEZCJ, funded by the European Union – NextGenerationEU. F.F.~acknowledges financial support by the MSCA European Fellowship QUANTUMLHC - Exploring quantum observables at the LHC - program Horizon Europe G.A. 101107121 CUP J33C23001080006. F.M.~also acknowledges financial support by the project QIHEP-Exploring the foundations of quantum information in particle physics - financed through the PNRR, funded by the European Union – NextGenerationEU, in the context of the extended partnership PE00000023 NQSTI - CUP J33C24001210007.

\appendix

\section{Density matrix $\rho$ as a function of $A$ and $C$ coefficients}

\subsection{General expression}
\label{sec:rhogeneral}
{\small
\begin{align}
 \nonumber \rho_{1,1} &= \frac{1}{9} \left(\sqrt{\frac{3}{2}} A^1_{1,0}+\frac{A^1_{2,0}}{\sqrt{2}}+\sqrt{\frac{3}{2}} A^2_{1,0}+\frac{A^2_{2,0}}{\sqrt{2}}+\frac{3}{2} C_{1,0,1,0}+\frac{1}{2} \sqrt{3} C_{1,0,2,0}+\frac{1}{2} \sqrt{3} C_{2,0,1,0}+\frac{1}{2} C_{2,0,2,0}+1\right) \, , \\ \nonumber
\rho_{1,2} &= \frac{1}{9} \left(-\sqrt{\frac{3}{2}} A^2_{1,1}-\sqrt{\frac{3}{2}} A^2_{2,1}-\frac{3}{2} C_{1,0,1,1}-\frac{3}{2} C_{1,0,2,1}-\frac{1}{2} \sqrt{3} C_{2,0,1,1}-\frac{1}{2} \sqrt{3} C_{2,0,2,1}\right) \, , \\ \nonumber
\rho_{1,3} &= \frac{1}{9} \left(\sqrt{3} A^2_{2,2}+\frac{3 C_{1,0,2,2}}{\sqrt{2}}+\sqrt{\frac{3}{2}} C_{2,0,2,2}\right) \, , \\ \nonumber
\rho_{1,4} &= \frac{1}{9} \left(-\sqrt{\frac{3}{2}} A^1_{1,1}-\sqrt{\frac{3}{2}} A^1_{2,1}-\frac{3}{2} C_{1,1,1,0}-\frac{1}{2} \sqrt{3} C_{1,1,2,0}-\frac{3}{2} C_{2,1,1,0}-\frac{1}{2} \sqrt{3} C_{2,1,2,0}\right) \, , \\ \nonumber
\rho_{1,5} &= \frac{1}{9} \left(\frac{3}{2} C_{1,1,1,1}+\frac{3}{2} C_{1,1,2,1}+\frac{3}{2} C_{2,1,1,1}+\frac{3}{2} C_{2,1,2,1}\right) \, , \\ \nonumber
\rho_{1,6} &= \frac{1}{9} \left(-\frac{3 C_{1,1,2,2}}{\sqrt{2}}-\frac{3 C_{2,1,2,2}}{\sqrt{2}}\right) \, , \\ \nonumber
\rho_{1,7} &= \frac{1}{9} \left(\sqrt{3} A^1_{2,2}+\frac{3 C_{2,2,1,0}}{\sqrt{2}}+\sqrt{\frac{3}{2}} C_{2,2,2,0}\right) \, , \\ \nonumber
\rho_{1,8} &= \frac{1}{9} \left(-\frac{3 C_{2,2,1,1}}{\sqrt{2}}-\frac{3 C_{2,2,2,1}}{\sqrt{2}}\right) \, , \\ \nonumber
\rho_{1,9} &= \frac{1}{3} C_{2,2,2,2} \, , \\ \nonumber
\rho_{2,1} &= \frac{1}{9} \left(\sqrt{\frac{3}{2}} A^2_{1,-1}+\sqrt{\frac{3}{2}} A^2_{2,-1}+\frac{3}{2} C_{1,0,1,-1}+\frac{3}{2} C_{1,0,2,-1}+\frac{1}{2} \sqrt{3} C_{2,0,1,-1}+\frac{1}{2} \sqrt{3} C_{2,0,2,-1}\right) \, , \\ \nonumber
\rho_{2,2} &= \frac{1}{9} \left(\sqrt{\frac{3}{2}} A^1_{1,0}+\frac{A^1_{2,0}}{\sqrt{2}}-\sqrt{2} A^2_{2,0}-\sqrt{3} C_{1,0,2,0}-C_{2,0,2,0}+1\right) \, , \\ \nonumber
\rho_{2,3} &= \frac{1}{9} \left(-\sqrt{\frac{3}{2}} A^2_{1,1}+\sqrt{\frac{3}{2}} A^2_{2,1}-\frac{3}{2} C_{1,0,1,1}+\frac{3}{2} C_{1,0,2,1}-\frac{1}{2} \sqrt{3} C_{2,0,1,1}+\frac{1}{2} \sqrt{3} C_{2,0,2,1}\right) \, , \\ \nonumber
\rho_{2,4} &= \frac{1}{9} \left(-\frac{3}{2} C_{1,1,1,-1}-\frac{3}{2} C_{1,1,2,-1}-\frac{3}{2} C_{2,1,1,-1}-\frac{3}{2} C_{2,1,2,-1}\right) \, , \\ \nonumber
\rho_{2,5} &= \frac{1}{9} \left(-\sqrt{\frac{3}{2}} A^1_{1,1}-\sqrt{\frac{3}{2}} A^1_{2,1}+\sqrt{3} C_{1,1,2,0}+\sqrt{3} C_{2,1,2,0}\right) \, , \\ \nonumber
\rho_{2,6} &= \frac{1}{9} \left(\frac{3}{2} C_{1,1,1,1}-\frac{3}{2} C_{1,1,2,1}+\frac{3}{2} C_{2,1,1,1}-\frac{3}{2} C_{2,1,2,1}\right) \, , \\ \nonumber
\rho_{2,7} &= \frac{1}{9} \left(\frac{3 C_{2,2,1,-1}}{\sqrt{2}}+\frac{3 C_{2,2,2,-1}}{\sqrt{2}}\right) \, , \\ \nonumber
\rho_{2,8} &= \frac{1}{9} \left(\sqrt{3} A^1_{2,2}-\sqrt{6} C_{2,2,2,0}\right) \, , \\ \nonumber
\rho_{2,9} &= \frac{1}{9} \left(\frac{3 C_{2,2,2,1}}{\sqrt{2}}-\frac{3 C_{2,2,1,1}}{\sqrt{2}}\right) \, , \\ \nonumber
\end{align}
\begin{align}
\nonumber \rho_{3,1} &= \frac{1}{9} \left(\sqrt{3} A^2_{2,-2}+\frac{3 C_{1,0,2,-2}}{\sqrt{2}}+\sqrt{\frac{3}{2}} C_{2,0,2,-2}\right) \, , \\ \nonumber
\rho_{3,2} &= \frac{1}{9} \left(\sqrt{\frac{3}{2}} A^2_{1,-1}-\sqrt{\frac{3}{2}} A^2_{2,-1}+\frac{3}{2} C_{1,0,1,-1}-\frac{3}{2} C_{1,0,2,-1}+\frac{1}{2} \sqrt{3} C_{2,0,1,-1}-\frac{1}{2} \sqrt{3} C_{2,0,2,-1}\right) \, , \\ \nonumber
\rho_{3,3} &= \frac{1}{9} \left(\sqrt{\frac{3}{2}} A^1_{1,0}+\frac{A^1_{2,0}}{\sqrt{2}}-\sqrt{\frac{3}{2}} A^2_{1,0}+\frac{A^2_{2,0}}{\sqrt{2}}-\frac{3}{2} C_{1,0,1,0}+\frac{1}{2} \sqrt{3} C_{1,0,2,0}-\frac{1}{2} \sqrt{3} C_{2,0,1,0}+\frac{1}{2} C_{2,0,2,0}+1\right) \, , \\ \nonumber
\rho_{3,4} &= \frac{1}{9} \left(-\frac{3 C_{1,1,2,-2}}{\sqrt{2}}-\frac{3 C_{2,1,2,-2}}{\sqrt{2}}\right) \, , \\ \nonumber
\rho_{3,5} &= \frac{1}{9} \left(-\frac{3}{2} C_{1,1,1,-1}+\frac{3}{2} C_{1,1,2,-1}-\frac{3}{2} C_{2,1,1,-1}+\frac{3}{2} C_{2,1,2,-1}\right) \, , \\ \nonumber
\rho_{3,6} &= \frac{1}{9} \left(-\sqrt{\frac{3}{2}} A^1_{1,1}-\sqrt{\frac{3}{2}} A^1_{2,1}+\frac{3}{2} C_{1,1,1,0}-\frac{1}{2} \sqrt{3} C_{1,1,2,0}+\frac{3}{2} C_{2,1,1,0}-\frac{1}{2} \sqrt{3} C_{2,1,2,0}\right) \, , \\ \nonumber
\rho_{3,7} &= \frac{1}{3} C_{2,2,2,-2} \, , \\ \nonumber
\rho_{3,8} &= \frac{1}{9} \left(\frac{3 C_{2,2,1,-1}}{\sqrt{2}}-\frac{3 C_{2,2,2,-1}}{\sqrt{2}}\right) \, , \\ \nonumber
\rho_{3,9} &= \frac{1}{9} \left(\sqrt{3} A^1_{2,2}-\frac{3 C_{2,2,1,0}}{\sqrt{2}}+\sqrt{\frac{3}{2}} C_{2,2,2,0}\right) \, , \\ \nonumber
\rho_{4,1} &= \frac{1}{9} \left(\sqrt{\frac{3}{2}} A^1_{1,-1}+\sqrt{\frac{3}{2}} A^1_{2,-1}+\frac{3}{2} C_{1,-1,1,0}+\frac{1}{2} \sqrt{3} C_{1,-1,2,0}+\frac{3}{2} C_{2,-1,1,0}+\frac{1}{2} \sqrt{3} C_{2,-1,2,0}\right) \, , \\ \nonumber
\rho_{4,2} &= \frac{1}{9} \left(-\frac{3}{2} C_{1,-1,1,1}-\frac{3}{2} C_{1,-1,2,1}-\frac{3}{2} C_{2,-1,1,1}-\frac{3}{2} C_{2,-1,2,1}\right) \, , \\ \nonumber
\rho_{4,3} &= \frac{1}{9} \left(\frac{3 C_{1,-1,2,2}}{\sqrt{2}}+\frac{3 C_{2,-1,2,2}}{\sqrt{2}}\right) \, , \\ \nonumber
\rho_{4,4} &= \frac{1}{9} \left(-\sqrt{2} A^1_{2,0}+\sqrt{\frac{3}{2}} A^2_{1,0}+\frac{A^2_{2,0}}{\sqrt{2}}-\sqrt{3} C_{2,0,1,0}-C_{2,0,2,0}+1\right) \, , \\ \nonumber
\rho_{4,5} &= \frac{1}{9} \left(-\sqrt{\frac{3}{2}} A^2_{1,1}-\sqrt{\frac{3}{2}} A^2_{2,1}+\sqrt{3} C_{2,0,1,1}+\sqrt{3} C_{2,0,2,1}\right) \, , \\ \nonumber
\rho_{4,6} &= \frac{1}{9} \left(\sqrt{3} A^2_{2,2}-\sqrt{6} C_{2,0,2,2}\right) \, , \\ \nonumber
\rho_{4,7} &= \frac{1}{9} \left(-\sqrt{\frac{3}{2}} A^1_{1,1}+\sqrt{\frac{3}{2}} A^1_{2,1}-\frac{3}{2} C_{1,1,1,0}-\frac{1}{2} \sqrt{3} C_{1,1,2,0}+\frac{3}{2} C_{2,1,1,0}+\frac{1}{2} \sqrt{3} C_{2,1,2,0}\right) \, , \\ \nonumber
\rho_{4,8} &= \frac{1}{9} \left(\frac{3}{2} C_{1,1,1,1}+\frac{3}{2} C_{1,1,2,1}-\frac{3}{2} C_{2,1,1,1}-\frac{3}{2} C_{2,1,2,1}\right) \, , \\ \nonumber
\rho_{4,9} &= \frac{1}{9} \left(\frac{3 C_{2,1,2,2}}{\sqrt{2}}-\frac{3 C_{1,1,2,2}}{\sqrt{2}}\right) \, , \\ \nonumber
\rho_{5,1} &= \frac{1}{9} \left(\frac{3}{2} C_{1,-1,1,-1}+\frac{3}{2} C_{1,-1,2,-1}+\frac{3}{2} C_{2,-1,1,-1}+\frac{3}{2} C_{2,-1,2,-1}\right) \, , \\ \nonumber
\end{align}
\begin{align}
\nonumber \rho_{5,2} &= \frac{1}{9} \left(\sqrt{\frac{3}{2}} A^1_{1,-1}+\sqrt{\frac{3}{2}} A^1_{2,-1}-\sqrt{3} C_{1,-1,2,0}-\sqrt{3} C_{2,-1,2,0}\right) \, , \\ \nonumber
\rho_{5,3} &= \frac{1}{9} \left(-\frac{3}{2} C_{1,-1,1,1}+\frac{3}{2} C_{1,-1,2,1}-\frac{3}{2} C_{2,-1,1,1}+\frac{3}{2} C_{2,-1,2,1}\right) \, , \\ \nonumber
\rho_{5,4} &= \frac{1}{9} \left(\sqrt{\frac{3}{2}} A^2_{1,-1}+\sqrt{\frac{3}{2}} A^2_{2,-1}-\sqrt{3} C_{2,0,1,-1}-\sqrt{3} C_{2,0,2,-1}\right) \, , \\ \nonumber
\rho_{5,5} &= \frac{1}{9} \left(-\sqrt{2} A^1_{2,0}-\sqrt{2} A^2_{2,0}+2 C_{2,0,2,0}+1\right) \, , \\ \nonumber
\rho_{5,6} &= \frac{1}{9} \left(-\sqrt{\frac{3}{2}} A^2_{1,1}+\sqrt{\frac{3}{2}} A^2_{2,1}+\sqrt{3} C_{2,0,1,1}-\sqrt{3} C_{2,0,2,1}\right) \, , \\ \nonumber
\rho_{5,7} &= \frac{1}{9} \left(-\frac{3}{2} C_{1,1,1,-1}-\frac{3}{2} C_{1,1,2,-1}+\frac{3}{2} C_{2,1,1,-1}+\frac{3}{2} C_{2,1,2,-1}\right) \, , \\ \nonumber
\rho_{5,8} &= \frac{1}{9} \left(-\sqrt{\frac{3}{2}} A^1_{1,1}+\sqrt{\frac{3}{2}} A^1_{2,1}+\sqrt{3} C_{1,1,2,0}-\sqrt{3} C_{2,1,2,0}\right) \, , \\ \nonumber
\rho_{5,9} &= \frac{1}{9} \left(\frac{3}{2} C_{1,1,1,1}-\frac{3}{2} C_{1,1,2,1}-\frac{3}{2} C_{2,1,1,1}+\frac{3}{2} C_{2,1,2,1}\right) \, , \\ \nonumber
\rho_{6,1} &= \frac{1}{9} \left(\frac{3 C_{1,-1,2,-2}}{\sqrt{2}}+\frac{3 C_{2,-1,2,-2}}{\sqrt{2}}\right) \, , \\ \nonumber
\rho_{6,2} &= \frac{1}{9} \left(\frac{3}{2} C_{1,-1,1,-1}-\frac{3}{2} C_{1,-1,2,-1}+\frac{3}{2} C_{2,-1,1,-1}-\frac{3}{2} C_{2,-1,2,-1}\right) \, , \\ \nonumber
\rho_{6,3} &= \frac{1}{9} \left(\sqrt{\frac{3}{2}} A^1_{1,-1}+\sqrt{\frac{3}{2}} A^1_{2,-1}-\frac{3}{2} C_{1,-1,1,0}+\frac{1}{2} \sqrt{3} C_{1,-1,2,0}-\frac{3}{2} C_{2,-1,1,0}+\frac{1}{2} \sqrt{3} C_{2,-1,2,0}\right) \, , \\ \nonumber
\rho_{6,4} &= \frac{1}{9} \left(\sqrt{3} A^2_{2,-2}-\sqrt{6} C_{2,0,2,-2}\right) \, , \\ \nonumber
\rho_{6,5} &= \frac{1}{9} \left(\sqrt{\frac{3}{2}} A^2_{1,-1}-\sqrt{\frac{3}{2}} A^2_{2,-1}-\sqrt{3} C_{2,0,1,-1}+\sqrt{3} C_{2,0,2,-1}\right) \, , \\ \nonumber
\rho_{6,6} &= \frac{1}{9} \left(-\sqrt{2} A^1_{2,0}-\sqrt{\frac{3}{2}} A^2_{1,0}+\frac{A^2_{2,0}}{\sqrt{2}}+\sqrt{3} C_{2,0,1,0}-C_{2,0,2,0}+1\right) \, , \\ \nonumber
\rho_{6,7} &= \frac{1}{9} \left(\frac{3 C_{2,1,2,-2}}{\sqrt{2}}-\frac{3 C_{1,1,2,-2}}{\sqrt{2}}\right) \, , \\ \nonumber
\rho_{6,8} &= \frac{1}{9} \left(-\frac{3}{2} C_{1,1,1,-1}+\frac{3}{2} C_{1,1,2,-1}+\frac{3}{2} C_{2,1,1,-1}-\frac{3}{2} C_{2,1,2,-1}\right) \, , \\ \nonumber
\rho_{6,9} &= \frac{1}{9} \left(-\sqrt{\frac{3}{2}} A^1_{1,1}+\sqrt{\frac{3}{2}} A^1_{2,1}+\frac{3}{2} C_{1,1,1,0}-\frac{1}{2} \sqrt{3} C_{1,1,2,0}-\frac{3}{2} C_{2,1,1,0}+\frac{1}{2} \sqrt{3} C_{2,1,2,0}\right) \, , \\ \nonumber
\rho_{7,1} &= \frac{1}{9} \left(\sqrt{3} A^1_{2,-2}+\frac{3 C_{2,-2,1,0}}{\sqrt{2}}+\sqrt{\frac{3}{2}} C_{2,-2,2,0}\right) \, , \\ \nonumber
\rho_{7,2} &= \frac{1}{9} \left(-\frac{3 C_{2,-2,1,1}}{\sqrt{2}}-\frac{3 C_{2,-2,2,1}}{\sqrt{2}}\right) \, , \\ \nonumber
\rho_{7,3} &= \frac{1}{3} C_{2,-2,2,2} \, , \\ \nonumber
\end{align}
\begin{align}
\nonumber \rho_{7,4} &= \frac{1}{9} \left(\sqrt{\frac{3}{2}} A^1_{1,-1}-\sqrt{\frac{3}{2}} A^1_{2,-1}+\frac{3}{2} C_{1,-1,1,0}+\frac{1}{2} \sqrt{3} C_{1,-1,2,0}-\frac{3}{2} C_{2,-1,1,0}-\frac{1}{2} \sqrt{3} C_{2,-1,2,0}\right) \, , \\ \nonumber
\rho_{7,5} &= \frac{1}{9} \left(-\frac{3}{2} C_{1,-1,1,1}-\frac{3}{2} C_{1,-1,2,1}+\frac{3}{2} C_{2,-1,1,1}+\frac{3}{2} C_{2,-1,2,1}\right) \, , \\ \nonumber
\rho_{7,6} &= \frac{1}{9} \left(\frac{3 C_{1,-1,2,2}}{\sqrt{2}}-\frac{3 C_{2,-1,2,2}}{\sqrt{2}}\right) \, , \\ \nonumber
\rho_{7,7} &= \frac{1}{9} \left(-\sqrt{\frac{3}{2}} A^1_{1,0}+\frac{A^1_{2,0}}{\sqrt{2}}+\sqrt{\frac{3}{2}} A^2_{1,0}+\frac{A^2_{2,0}}{\sqrt{2}}-\frac{3}{2} C_{1,0,1,0}-\frac{1}{2} \sqrt{3} C_{1,0,2,0}+\frac{1}{2} \sqrt{3} C_{2,0,1,0}+\frac{1}{2} C_{2,0,2,0}+1\right) \, , \\ \nonumber
\rho_{7,8} &= \frac{1}{9} \left(-\sqrt{\frac{3}{2}} A^2_{1,1}-\sqrt{\frac{3}{2}} A^2_{2,1}+\frac{3}{2} C_{1,0,1,1}+\frac{3}{2} C_{1,0,2,1}-\frac{1}{2} \sqrt{3} C_{2,0,1,1}-\frac{1}{2} \sqrt{3} C_{2,0,2,1}\right) \, , \\ \nonumber
\rho_{7,9} &= \frac{1}{9} \left(\sqrt{3} A^2_{2,2}-\frac{3 C_{1,0,2,2}}{\sqrt{2}}+\sqrt{\frac{3}{2}} C_{2,0,2,2}\right) \, , \\ \nonumber
\rho_{8,1} &= \frac{1}{9} \left(\frac{3 C_{2,-2,1,-1}}{\sqrt{2}}+\frac{3 C_{2,-2,2,-1}}{\sqrt{2}}\right) \, , \\ \nonumber
\rho_{8,2} &= \frac{1}{9} \left(\sqrt{3} A^1_{2,-2}-\sqrt{6} C_{2,-2,2,0}\right) \, , \\ \nonumber
\rho_{8,3} &= \frac{1}{9} \left(\frac{3 C_{2,-2,2,1}}{\sqrt{2}}-\frac{3 C_{2,-2,1,1}}{\sqrt{2}}\right) \, , \\ \nonumber
\rho_{8,4} &= \frac{1}{9} \left(\frac{3}{2} C_{1,-1,1,-1}+\frac{3}{2} C_{1,-1,2,-1}-\frac{3}{2} C_{2,-1,1,-1}-\frac{3}{2} C_{2,-1,2,-1}\right) \, , \\ \nonumber
\rho_{8,5} &= \frac{1}{9} \left(\sqrt{\frac{3}{2}} A^1_{1,-1}-\sqrt{\frac{3}{2}} A^1_{2,-1}-\sqrt{3} C_{1,-1,2,0}+\sqrt{3} C_{2,-1,2,0}\right) \, , \\ \nonumber
\rho_{8,6} &= \frac{1}{9} \left(-\frac{3}{2} C_{1,-1,1,1}+\frac{3}{2} C_{1,-1,2,1}+\frac{3}{2} C_{2,-1,1,1}-\frac{3}{2} C_{2,-1,2,1}\right) \, , \\ \nonumber
\rho_{8,7} &= \frac{1}{9} \left(\sqrt{\frac{3}{2}} A^2_{1,-1}+\sqrt{\frac{3}{2}} A^2_{2,-1}-\frac{3}{2} C_{1,0,1,-1}-\frac{3}{2} C_{1,0,2,-1}+\frac{1}{2} \sqrt{3} C_{2,0,1,-1}+\frac{1}{2} \sqrt{3} C_{2,0,2,-1}\right) \, , \\ \nonumber
\rho_{8,8} &= \frac{1}{9} \left(-\sqrt{\frac{3}{2}} A^1_{1,0}+\frac{A^1_{2,0}}{\sqrt{2}}-\sqrt{2} A^2_{2,0}+\sqrt{3} C_{1,0,2,0}-C_{2,0,2,0}+1\right) \, , \\ \nonumber
\rho_{8,9} &= \frac{1}{9} \left(-\sqrt{\frac{3}{2}} A^2_{1,1}+\sqrt{\frac{3}{2}} A^2_{2,1}+\frac{3}{2} C_{1,0,1,1}-\frac{3}{2} C_{1,0,2,1}-\frac{1}{2} \sqrt{3} C_{2,0,1,1}+\frac{1}{2} \sqrt{3} C_{2,0,2,1}\right) \, , \\ \nonumber
\rho_{9,1} &= \frac{1}{3} C_{2,-2,2,-2} \, , \\ \nonumber
\rho_{9,2} &= \frac{1}{9} \left(\frac{3 C_{2,-2,1,-1}}{\sqrt{2}}-\frac{3 C_{2,-2,2,-1}}{\sqrt{2}}\right) \, , \\ \nonumber
\rho_{9,3} &= \frac{1}{9} \left(\sqrt{3} A^1_{2,-2}-\frac{3 C_{2,-2,1,0}}{\sqrt{2}}+\sqrt{\frac{3}{2}} C_{2,-2,2,0}\right) \, , \\ \nonumber
\rho_{9,4} &= \frac{1}{9} \left(\frac{3 C_{1,-1,2,-2}}{\sqrt{2}}-\frac{3 C_{2,-1,2,-2}}{\sqrt{2}}\right) \, , \\ \nonumber
\end{align}
\begin{align}
\nonumber \rho_{9,5} &= \frac{1}{9} \left(\frac{3}{2} C_{1,-1,1,-1}-\frac{3}{2} C_{1,-1,2,-1}-\frac{3}{2} C_{2,-1,1,-1}+\frac{3}{2} C_{2,-1,2,-1}\right) \, , \\ \nonumber
\rho_{9,6} &= \frac{1}{9} \left(\sqrt{\frac{3}{2}} A^1_{1,-1}-\sqrt{\frac{3}{2}} A^1_{2,-1}-\frac{3}{2} C_{1,-1,1,0}+\frac{1}{2} \sqrt{3} C_{1,-1,2,0}+\frac{3}{2} C_{2,-1,1,0}-\frac{1}{2} \sqrt{3} C_{2,-1,2,0}\right) \, , \\ \nonumber
\rho_{9,7} &= \frac{1}{9} \left(\sqrt{3} A^2_{2,-2}-\frac{3 C_{1,0,2,-2}}{\sqrt{2}}+\sqrt{\frac{3}{2}} C_{2,0,2,-2}\right) \, , \\ \nonumber
\rho_{9,8} &= \frac{1}{9} \left(\sqrt{\frac{3}{2}} A^2_{1,-1}-\sqrt{\frac{3}{2}} A^2_{2,-1}-\frac{3}{2} C_{1,0,1,-1}+\frac{3}{2} C_{1,0,2,-1}+\frac{1}{2} \sqrt{3} C_{2,0,1,-1}-\frac{1}{2} \sqrt{3} C_{2,0,2,-1}\right) \, , \\ \nonumber
\rho_{9,9} &= \frac{1}{9} \left(-\sqrt{\frac{3}{2}} A^1_{1,0}+\frac{A^1_{2,0}}{\sqrt{2}}-\sqrt{\frac{3}{2}} A^2_{1,0}+\frac{A^2_{2,0}}{\sqrt{2}}+\frac{3}{2} C_{1,0,1,0}-\frac{1}{2} \sqrt{3} C_{1,0,2,0}-\frac{1}{2} \sqrt{3} C_{2,0,1,0}+\frac{1}{2} C_{2,0,2,0}+1\right)\,.\\
\end{align}

\subsection{Non vanishing coefficients with cylindrical, up-down symmetry and with even parity}
\label{sec:rhocylandupdown}
\begin{align}
\teal{\rho_{1,1} }&= \frac{1}{9} \left(\frac{A^1_{2,0}}{\sqrt{2}}+\frac{A^2_{2,0}}{\sqrt{2}}+\frac{3}{2} C_{1,0,1,0}+\frac{1}{2} C_{2,0,2,0}+1\right) \, , \nonumber \\ 
\teal{\rho_{2,2} }&= \frac{1}{9} \left(\frac{A^1_{2,0}}{\sqrt{2}}-\sqrt{2} A^2_{2,0}-C_{2,0,2,0}+1\right) \, , \nonumber \\ 
\teal{\rho_{2,4} }&= \frac{1}{9} \left(-\frac{3}{2} C_{1,1,1,-1}-\frac{3}{2} C_{2,1,2,-1}\right) \, , \nonumber \\ 
\blue{\rho_{3,3} }&= \frac{1}{9} \left(\frac{A^1_{2,0}}{\sqrt{2}}+\frac{A^2_{2,0}}{\sqrt{2}}-\frac{3}{2} C_{1,0,1,0}+\frac{1}{2} C_{2,0,2,0}+1\right) \, , \nonumber \\ 
\blue{\rho_{3,5} }&= \frac{1}{9} \left(-\frac{3}{2} C_{1,1,1,-1}+\frac{3}{2} C_{2,1,2,-1}\right) \, , \nonumber \\ 
\blue{\rho_{3,7} }&= \frac{1}{3} C_{2,2,2,-2}  \, , \nonumber \\
\teal{\rho_{4,2} }&= \frac{1}{9} \left(-\frac{3}{2} C_{1,-1,1,1}-\frac{3}{2} C_{2,-1,2,1}\right) \nonumber \\
\teal{\rho_{4,4} }&= \frac{1}{9} \left(-\sqrt{2} A^1_{2,0}+\frac{A^2_{2,0}}{\sqrt{2}}-C_{2,0,2,0}+1\right) \nonumber \\
\blue{\rho_{5,3} }&= \frac{1}{9} \left(-\frac{3}{2} C_{1,-1,1,1}+\frac{3}{2} C_{2,-1,2,1}\right) \, , \nonumber \\ 
\blue{\rho_{5,5} }&= \frac{1}{9} \left(-\sqrt{2} A^1_{2,0}-\sqrt{2} A^2_{2,0}+2 C_{2,0,2,0}+1\right) \nonumber \\
\blue{\rho_{5,7} }&= \frac{1}{9} \left(-\frac{3}{2} C_{1,1,1,-1}+\frac{3}{2} C_{2,1,2,-1}\right) \nonumber \\
\teal{\rho_{6,6} }&= \frac{1}{9} \left(-\sqrt{2} A^1_{2,0}+\frac{A^2_{2,0}}{\sqrt{2}}-C_{2,0,2,0}+1\right) \, , \nonumber \\ 
\teal{\rho_{6,8} }&= \frac{1}{9} \left(-\frac{3}{2} C_{1,1,1,-1}-\frac{3}{2} C_{2,1,2,-1}\right) \, , \nonumber \\ 
\blue{\rho_{7,3} }&= \frac{1}{3} C_{2,-2,2,2}  \, , \nonumber \\
\blue{\rho_{7,5} }&= \frac{1}{9} \left(-\frac{3}{2} C_{1,-1,1,1}+\frac{3}{2} C_{2,-1,2,1}\right) \, , \nonumber \\ 
\blue{\rho_{7,7} }&= \frac{1}{9} \left(\frac{A^1_{2,0}}{\sqrt{2}}+\frac{A^2_{2,0}}{\sqrt{2}}-\frac{3}{2} C_{1,0,1,0}+\frac{1}{2} C_{2,0,2,0}+1\right) \, , \nonumber \\ 
\teal{\rho_{8,6} }&= \frac{1}{9} \left(-\frac{3}{2} C_{1,-1,1,1}-\frac{3}{2} C_{2,-1,2,1}\right) \, , \nonumber \\ 
\teal{\rho_{8,8} }&= \frac{1}{9} \left(\frac{A^1_{2,0}}{\sqrt{2}}-\sqrt{2} A^2_{2,0}-C_{2,0,2,0}+1\right) \, , \nonumber \\ 
\teal{\rho_{9,9} }&= \frac{1}{9} \left(\frac{A^1_{2,0}}{\sqrt{2}}+\frac{A^2_{2,0}}{\sqrt{2}}+\frac{3}{2} C_{1,0,1,0}+\frac{1}{2} C_{2,0,2,0}+1\right)\,. \nonumber
\end{align}
}

\clearpage

\section{Comparing NLO EW effects with  quantum tomography at LO using wrong $\alpha$}
\label{app:wrongalpha}

First of all we discuss which effects emerges if we try to simulate at LO the $H \rightarrow ZZ^* \rightarrow e^{+} e^{-} \mu^{+} \mu^{-}$ decay and perform the QT procedure assuming a  value for $\alpha$ that is different from the one used in the simulation. This is different than what has been discussed in the main text, where we have explained that NLO results can be very well approximated by assuming that NLO EW corrections on $\rho$ are small and are induced by the usage of incorrect $\alpha$ in QT. We will get to this point at the end of this Appendix.

\medskip 

If we want to perform the QT procedure assuming a value for $\alpha$ that is different from the one used in the simulation, we can actually allow for two separate values of $\alpha$ for $Z_1$ and $Z_2$, respectively denoted as $\alpha_1$ and $\alpha_2$ and defined as:
\begin{equation}
\alpha_1=\kappa_1 \alpha\,, \qquad \alpha_2=\kappa_2 \alpha\,. \label{eq:defkappas}
\end{equation}
Therefore the quantity $\kappa_1 (\kappa_2 )$ parametrises the deviation of the value of $\alpha_1(\alpha_2)$ from the correct value $\alpha$.  It can be shown that the non-vanishing terms that are emerging in the texture of $\rho$ are the same  that are either in blue or green in the matrix of Eq.~$\eqref{NLOrho}$. In particular, such  matrix, which we denote as $\rho_{\rm LO}^{\rm wrong}(\alpha_1,\alpha_2)$ can be obtained simply plugging into the equations of Appendix \ref{sec:rhocylandupdown} the values obtained for the coefficients and one finds:
\begin{multline}
\label{eq:NLOwrongalpha}
\rho_{\rm LO}^{\rm wrong} = \\[3mm]
{\small
\begin{pmatrix}
\teal{\frac{-\rho_{\rm LO}^{33}}{2}\Delta} &  \cdot&  \cdot&  \cdot&  \cdot&  \cdot&  \cdot&  \cdot&  \cdot\\ 
 \cdot& \cdot &  \cdot& \teal{\frac{\rho_{\rm LO}^{35}}{2}\Delta} &  \cdot&  \cdot&  \cdot&  \cdot&  \cdot\\ 
 \cdot&  \cdot& \blue{\frac{\rho_{\rm LO}^{33}}{2}(2+\Delta)} &  \cdot& \blue{\frac{\rho_{\rm LO}^{35}}{2}(2+\Delta)} &  \cdot& \blue{\rho_{\rm LO}^{33}} &  \cdot&  \cdot\\ 
 \cdot& \teal{\frac{\rho_{\rm LO}^{35}}{2}\Delta}  &  \cdot& \cdot &  \cdot&  \cdot&  \cdot&  \cdot&  \cdot\\ 
 \cdot&  \cdot&  \blue{\frac{\rho_{\rm LO}^{35}}{2}(2+\Delta)}  &  \cdot& \blue{1- 2 \rho_{\rm LO}^{33}} &  \cdot& \blue{\frac{\rho_{\rm LO}^{35}}{2}(2+\Delta)}&  \cdot&  \cdot\\ 
 \cdot&  \cdot&  \cdot&  \cdot&  \cdot& \cdot &  \cdot&\teal{\frac{\rho_{\rm LO}^{35}}{2}\Delta}  &  \cdot\\
 \cdot&  \cdot&  \blue{\rho_{\rm LO}^{33}} &  \cdot& \blue{\frac{\rho_{\rm LO}^{35}}{2}(2+\Delta)} &  \cdot& \blue{\frac{\rho_{\rm LO}^{33}}{2}(2+\Delta)} &  \cdot&  \cdot\\
 \cdot&  \cdot&  \cdot&  \cdot&  \cdot& \teal{\frac{\rho_{\rm LO}^{35}}{2}\Delta}  &  \cdot& \cdot &  \cdot\\
 \cdot&  \cdot&  \cdot&  \cdot&  \cdot&  \cdot&  \cdot&  \cdot& \teal{-\frac{\rho_{\rm LO}^{33}}{2}\Delta~} \\
\end{pmatrix} }\,,
\end{multline}
where we $\rho_{\rm LO}$ is the correct $\rho$ matrix\footnote{We used the notation $\rho_{\rm LO}^{33}$ and $\rho_{\rm LO}^{35}$ instead of $(\rho_{\rm LO})_{33}$  and $(\rho_{\rm LO})_{35}$, respectively, for brevity.} and the quantity $\Delta$ is defined as
\begin{equation}
\Delta\equiv \frac{1}{\kappa_1 \kappa_2}-1\,.
\end{equation}
As it can be seen, in the SM limit ($\kappa_1,\kappa_2\to 1$) one has $\Delta\to0$, so  green entries vanish and $\rho_{\rm LO}^{\rm wrong}(\alpha_1,\alpha_2) =\rho_{\rm LO}$.

A simple way to understand the origin of this pattern is that performing the QT step (Eqs.~\eqref{eqn:coeff_defA} and \eqref{eqn:coeff_defC}) with the wrong values of $\alpha$, {\it i.e.} $\alpha_1$ and $\alpha_2$, leads to
 \begin{equation}
C^{\rm wrong}_{L_1,M_1,L_2,M_2}=C_{L_1,M_1,L_2,M_2}\left(1+\left(\frac{1}{\kappa_1}-1\right)\delta_{1,L_1}\right)\left(1+\left(\frac{1}{\kappa_2}-1\right)\delta_{1,L_2}\right)\,,
\end{equation}
which in practice for the $H \rightarrow ZZ^* \rightarrow e^{+} e^{-} \mu^{+} \mu^{-}$ at LO in the SM leads to $C^{\rm wrong}_{1,M,1,-M}=C_{1,M,1,-M}(1+\Delta)$ and no effects on all the other non-vanishing coefficients. As can be seen in Appendix \ref{sec:rhocylandupdown}, $\rho_{37}=\rho_{73} $ and $\rho_{55}$ do not depend on $C_{1,M,1,-M}$, while all the others do depend on them. 

One can easy verify that by setting $\Delta=-0.85$ the bulk of the NLO EW corrections can obtained via Eq.~\eqref{eq:NLOwrongalpha}, {\it i.e.}  $\rho_{\rm NLO}\simeq \rho_{\rm LO}^{\rm wrong}\big|_{\Delta=-0.85}$. This value has been obtained minimising $\sum_{1\le i,j \le 9} ((\rho_{\rm NLO})_{ij}-(\rho_{\rm LO}^{\rm wrong})_{ij})^2 $, which is equal to $\sim0.001$ for  $\Delta=-0.85$. This result is consistent with the NLO/LO value shown in Tab.~\ref{tab:nonzero_coeff} for the $C_{1,M,1,-M}$ coefficients.

\medskip

All the previous argument rely on the fact that the calculation is performed at LO and the QT performed with the wrong $\alpha$. In fact, in the numbers that are discussed in the main text the calculation is not performed at LO but at NLO EW, and the $\alpha$ that is wrong when performing the QT is the one at LO, which does not incorporate the NLO EW corrections to the $\Gamma$ matrix. Thus, if we want to interpret the  $\rho_{\rm NLO}$ in Eq.~\eqref{NLOrho} as  solely induced by the wrong value of $\alpha$ and no effects form NLO EW  on the real $\rho$, we can repeat the previous argument but considering $\kappa_1$($\kappa_2$) as the ratio between the value of $\alpha$ at LO and the  effective $\alpha_1$($\alpha_2$) at NLO EW accuracy,
\begin{equation}
\alpha=\kappa_1 \alpha_1\,, \qquad \alpha=\kappa_2 \alpha_2\, \label{eq:kappadef2},
\end{equation}
so the opposite of the definition in Eq.~\eqref{eq:defkappas}.

If we further simplify the argument, as done in the main text, and assume $\alpha_1=\alpha_2$, the value $\Delta=-0.85$ implies $\kappa_1=\kappa_2\simeq 2.58$, which means that the effective $\alpha_1=\alpha_2=0.39 \,\alpha$ as written in the main text.

\clearpage

\section{Spin density matrices in different set-ups}
\label{app:matr}

In this Appendix we report LO and NLO predictions for the $\rho$ density matrix. As explained at length in the main text, the NLO results cannot be straightforwardly employed for extracting QI. They are not semi-positive defined. We simply document the numerical results obtained. 

\subsection*{Inclusive result with $\mathbf{\Delta R = 1}$ }

$\rho_{\text{LO}}$ is the same as in Eq.~\eqref{eq:rhoLOinclusive}.
\begin{multline}
\rho_{\text{NLO}} = \\[3mm]
{\small
\begin{pmatrix}
\teal{~0.094(4)} & \cdot & \cdot & \cdot & \cdot & \cdot & \cdot & \cdot & \cdot \\
\cdot & \cdot & \cdot & \teal{0.127(4)} & \cdot & \cdot & \cdot & \cdot & \cdot \\
\cdot & \cdot & \blue{0.104(4)} & \cdot & \blue{-0.184(4)} & \cdot & \blue{0.191(2)} & \cdot & \cdot \\
\cdot & \teal{0.127(4)} & \cdot & -0.001(2) & \cdot & \cdot & \cdot & \cdot & \cdot \\
\cdot & \cdot & \blue{-0.184(4)} & \cdot & \blue{0.608(2)} & \cdot & \blue{-0.184(4)} & \cdot & \cdot \\
\cdot & \cdot & \cdot & \cdot & \cdot & -0.001(2) & \cdot & \teal{0.127(4)} & \cdot \\
\cdot & \cdot & \blue{0.191(2)} & \cdot & \blue{-0.184(4)} & \cdot & \blue{0.104(4)} & \cdot & \cdot \\
\cdot & \cdot & \cdot & \cdot & \cdot & \teal{0.127(4)} & \cdot & \cdot & \cdot \\
\cdot & \cdot & \cdot & \cdot & \cdot & \cdot & \cdot & \cdot & \teal{0.094(4)~} 
\end{pmatrix} }.
\end{multline}

\subsection*{Inclusive result with $\mathbf{\Delta R = 0.001}$ }

$\rho_{\text{LO}}$ is the same as in Eq.~\eqref{eq:rhoLOinclusive}.

\begin{multline}
\rho_{\text{NLO}} = \\[3mm]
{\small
\begin{pmatrix}
\teal{~0.102(5)} & \cdot & \cdot & \cdot & \cdot & \cdot & \cdot & \cdot & \cdot \\
\cdot & 0.008(3) & \cdot & \teal{0.135(5)} & \cdot & \cdot & \cdot & -0.001(2) & \cdot \\
\cdot & \cdot & \blue{0.121(5)} & \cdot & \blue{-0.183(5)} & \cdot & \blue{0.185(2)} & \cdot & \cdot \\
\cdot & \teal{0.135(5)} & \cdot & -0.019(3) & \cdot & \cdot & \cdot & \cdot & \cdot \\
\cdot & \cdot & \blue{-0.183(5)} & \cdot & \blue{0.575(2)} & \cdot & \blue{-0.184(5)} & \cdot & \cdot \\
\cdot & \cdot & \cdot & \cdot & \cdot & -0.019(3) & \cdot & \teal{0.135(5)} & \cdot \\
\cdot & \cdot & \blue{0.185(2)} & \cdot & \blue{-0.184(5)} & \cdot & \blue{0.120(5)} & \cdot & \cdot \\
\cdot & -0.001(2) & \cdot & \cdot & \cdot & \teal{0.135(5)} & \cdot & 0.009(3) & \cdot \\
\cdot & \cdot & \cdot & \cdot & \cdot & \cdot & \cdot & \cdot & \teal{0.102(5)~} \\
\end{pmatrix} }.
\end{multline}
\subsection*{$\mathbf{m_{Z_2} > 30}$ GeV}
\begin{equation}
\rho_{\text{LO}} = 
\begin{pmatrix}
\cdot & \cdot & \cdot & \cdot & \cdot & \cdot & \cdot & \cdot & \cdot \\
\cdot & \cdot & \cdot & \cdot & \cdot & \cdot & \cdot & \cdot & \cdot \\
\cdot & \cdot & \blue{0.258(3)} & \cdot & \blue{-0.337(3)} & \cdot & \blue{0.258(1)} & \cdot & \cdot \\
\cdot & \cdot & \cdot & \cdot & \cdot & \cdot & \cdot & \cdot & \cdot \\
\cdot & \cdot & \blue{-0.337(3)} & \cdot & \blue{0.483(1)} & \cdot & \blue{-0.337(3)} & \cdot & \cdot \\
\cdot & \cdot & \cdot & \cdot & \cdot & \cdot & \cdot & \cdot & \cdot \\
\cdot & \cdot & \blue{0.258(1)} & \cdot & \blue{-0.337(3)} & \cdot & \blue{0.258(3)} & \cdot & \cdot \\
\cdot & \cdot & \cdot & \cdot & \cdot & \cdot & \cdot & \cdot & \cdot \\
\cdot & \cdot & \cdot & \cdot & \cdot & \cdot & \cdot & \cdot & \cdot \\
\end{pmatrix}.
\vspace{-5mm}
\end{equation}
\begin{multline}
\rho_{\text{NLO}} = \\[3mm]
{\small
\begin{pmatrix}
\teal{~0.056(5)} & \cdot & \cdot & \cdot & \cdot & \cdot & \cdot & \cdot & \cdot \\
\cdot & 0.003(3) & \cdot & \teal{0.058(4)} & \cdot & \cdot & \cdot & -0.002(1) & \cdot \\
\cdot & \cdot & \blue{0.218(5)} & \cdot & \blue{-0.283(4)} & \cdot & \blue{0.255(2)} & \cdot & \cdot \\
\cdot & \teal{0.058(4)} & \cdot & -0.013(3) & \cdot & 0.001(1) & \cdot & \cdot & \cdot \\
\cdot & \cdot & \blue{-0.283(4)} & \cdot & \blue{0.473(1)} & \cdot & \blue{-0.283(4)} & \cdot & \cdot \\
\cdot & \cdot & \cdot & -0.001(1) & \cdot & -0.013(3) & \cdot & \teal{0.057(4)} & \cdot \\
\cdot & \cdot & \blue{0.255(2)} & \cdot & \blue{-0.283(4)} & \cdot & \blue{0.219(5)} & \cdot & \cdot \\
\cdot & -0.002(1) & \cdot & \cdot & \cdot & \teal{0.057(4)} & \cdot & 0.002(3) & \cdot \\
\cdot & \cdot & \cdot & \cdot & \cdot & \cdot & \cdot & \cdot & \teal{0.056(5)~} \\
\end{pmatrix} }.
\end{multline}
\smallskip
\subsection*{85 GeV $\mathbf{< m_{Z_1} < 95}$ GeV}
\begin{equation}
\rho_{\text{LO}} = 
\begin{pmatrix}
\cdot & \cdot & \cdot & \cdot & \cdot & \cdot & \cdot & \cdot & \cdot \\
\cdot & \cdot & \cdot & \cdot & \cdot & \cdot & \cdot & \cdot & \cdot \\
\cdot & \cdot & \blue{0.202(2)} & \cdot & \blue{-0.318(2)} & \cdot & \blue{0.202(1)} & \cdot & \cdot \\
\cdot & \cdot & \cdot & \cdot & \cdot & \cdot & \cdot & \cdot & \cdot \\
\cdot & \cdot & \blue{-0.318(2)} & \cdot & \blue{0.5950(7)} & \cdot & \blue{-0.319(2)} & \cdot & \cdot \\
\cdot & \cdot & \cdot & \cdot & \cdot & \cdot & \cdot & \cdot & \cdot \\
\cdot & \cdot & \blue{0.202(1)} & \cdot & \blue{-0.319(2)} & \cdot & \blue{0.202(2)} & \cdot & \cdot \\
\cdot & \cdot & \cdot & \cdot & \cdot & \cdot & \cdot & \cdot & \cdot \\
\cdot & \cdot & \cdot & \cdot & \cdot & \cdot & \cdot & \cdot & \cdot \\
\end{pmatrix}.
\end{equation}
\begin{multline}
\rho_{\text{NLO}} = \\[3mm]
{\small
\begin{pmatrix}
\teal{~0.131(4)} & \cdot & \cdot & \cdot & \cdot & \cdot & \cdot & \cdot & \cdot \\
\cdot & \cdot & \cdot & \teal{0.208(4)} & -0.001(2) & \cdot & \cdot & -0.003(1) & \cdot \\
\cdot & \cdot & \blue{0.072(4)} & \cdot & \blue{-0.109(4)} & \cdot & \blue{0.201(1)} & \cdot & \cdot \\
\cdot & \teal{0.208(4)} & \cdot & 0.001(2) & \cdot & \cdot & \cdot & \cdot & \cdot \\
\cdot & -0.001(2) & \blue{-0.109(4)} & \cdot & \blue{0.591(1)} & \cdot & \blue{-0.109(4)} & 0.001(2) & \cdot \\
\cdot & \cdot & \cdot & \cdot & \cdot & 0.001(2) & \cdot & \teal{0.208(4)} & \cdot \\
\cdot & \cdot & \blue{0.201(1)} & \cdot & \blue{-0.109(4)} & \cdot & \blue{0.072(3)} & \cdot & \cdot \\
\cdot & -0.003(1) & \cdot & \cdot & 0.001(2) & \teal{0.208(4)} & \cdot & \cdot & \cdot \\
\cdot & \cdot & \cdot & \cdot & \cdot & \cdot & \cdot & \cdot & \teal{0.131(4)~} \\
\end{pmatrix} }.
\end{multline}
\smallskip
\clearpage

\section{Analytical form of non-zero contributions from momentum structures in Tab.~\ref{tab:momentumstructrues} to the $A$ and $C$ coefficients }
\label{appendixAClist}
\renewcommand{\arraystretch}{1.8}

We provide in the tables of this section the contributions stemming from the momentum structures listed in Tab.~\ref{tab:momentumstructrues} to any $A$ and $C$ coefficient. We show only the non-vanishing contributions using the notation $(A)_X$ or $(C)_X$ for the contribution of the momentum structure $X$ to $A$ or $C$, respectively. Specifically,
\begin{gather}
(C_{L_a,M_a,L_b,M_b})_X \equiv \int \frac{d X}{d\Omega_a d\Omega_b} Y^{*}_{L_a,M_a}(\Omega_a) Y^{*}_{L_b, M_b}(\Omega_b) \, d\Omega_a d\Omega_b \,. 
\label{eq:momentumstruct}
\end{gather}

We show only those entering Eqs.~\eqref{Eftampisq} -- \eqref{Eftijampsq} and if a combination of momenta structure $XY$ is present we denote it directly as $(A)_{XY}$ or $(C)_{XY}$. All the results are provided for fixed values of $m_a$ and $m_b$.

\begin{table}[h!]
    \centering
    \begin{tabular}{c|c}
        $\Pi$ terms & Non-zero contributions to $A$ or $C$ coefficients  \\
        \hline\hline
        $\Pi_0$& All are zero\\
        \hline
        $\Pi_1$&$(A^j_{2,0})_{\Pi_1}=\frac{4\pi}{B^j_2}\left(\frac{-2 m_a^2 m_b^2 (\beta^2-1)\pi^{3/2}}{9\sqrt{5}}\right)$\\
        
               & $(C_{2,2,2,-2})_{\Pi_1}=(C_{2,-2,2,2})_{\Pi_1}=\frac{(4\pi)^2}{B^a_2 B^b_2}\left(\frac{\pi\, m_a^2 m_b^2}{30}\right)$\\
               
            & $(C_{2,1,2,-1})_{\Pi_1}=(C_{2,-1,2,1})_{\Pi_1}=\frac{(4\pi)^2}{B^a_2 B^b_2}\left(\frac{-\pi\, m_a^2 m_b^2 \beta}{30}\right)$\\                  
            & $(C_{2,0,2,0})_{\Pi_1}=\frac{(4\pi)^2}{B^a_2 B^b_2}\left(\frac{\pi m_a^2 m_b^2(2\beta^2+1)}{90}\right)$\\ 
        
            & $(C_{1,0,1,0})_{\Pi_1}=\frac{(4\pi)^2}{B^a_1 B^b_1}\left(\frac{-\pi\, m_a^2 m_b^2}{6}\right)$\\  

            & $(C_{1,1,1,-1})_{\Pi_1}=(C_{1,-1,1,1})_{\Pi_1}=\frac{(4\pi)^2}{B^a_1 B^b_1}\left(\frac{\pi\, m_a^2 m_b^2\beta}{6}\right)$\\
         
    \hline
    
    $\Pi_2$ & $(A^j_{2,0})_{\Pi_1}=(A^j_{2,0})_{\Pi_2}$\\
        
               & $(C_{2,2,2,-2})_{\Pi_2}=(C_{2,-2,2,2})_{\Pi_2}=(C_{2,-2,2,2})_{\Pi_1}$\\
               
            & $(C_{2,1,2,-1})_{\Pi_2}=(C_{2,-1,2,1})_{\Pi_2}=(C_{2,-1,2,1})_{\Pi_1}$\\                  
            & $(C_{2,0,2,0})_{\Pi_2}=(C_{2,0,2,0})_{\Pi_1}$\\ 
        
            & $(C_{1,0,1,0})_{\Pi_2}=-(C_{1,0,1,0})_{\Pi_1}$\\  

            &$(C_{1,1,1,-1})_{\Pi_2}=(C_{1,-1,1,1})_{\Pi_2}=-(C_{1,1,1,-1})_{\Pi_1}$\\
            
            \hline
            
        $\Pi_\epsilon$ & $(C_{1,1,1,-1})_{\Pi_\epsilon}=(C^*_{1,-1,1,1})_{\Pi_\epsilon}=\frac{(4\pi)^2}{B^a_1 B^b_1}\frac{i\pi\,m_a^2 m_b^2\sqrt{\beta^2-1}}{3}$ \\
        
         \hline
    \end{tabular}
    \caption{$\Pi$ momentum structures.}
    \label{tab:tab1coffi}
\end{table}

\begin{table}[h]
    \centering
    \begin{tabular}{c|c}
        $H$ terms & Non-zero contributions to $A$ or $C$ coefficients  \\
        \hline\hline
                
         $H_1$&$(A^j_{2,0})_{H_1}=\frac{4\pi}{B^j_2}\frac{2\pi^{3/2}m_a^4 m_b^4(\beta^2-1)}{9\sqrt{5}}$\\
        
               & $(C_{2,2,2,-2})_{H_1}=(C_{2,-2,2,2})_{H_1}=\frac{(4\pi)^2}{B^a_2 B^b_2}\frac{(\pi m_a^4 m_b^4 \beta^2) }{30}$\\
               
            & $(C_{2,1,2,-1})_{H_1}=(C_{2,-1,2,1})_{H_1}=\frac{(4\pi)^2}{B^a_2 B^b_2}\frac{(-\pi m_a^4 m_b^4 \beta) }{30}$\\                  
            & $(C_{2,0,2,0})_{H_1}=\frac{(4\pi)^2}{B^a_2 B^b_2}\frac{\pi m_a^4 m_b^4(\beta^2+2)}{90}$\\ 
        
            & $(C_{1,0,1,0})_{H_1}=\frac{(4\pi)^2}{B^a_1 B^b_1}\frac{(-\pi m_a^4 m_b^4\beta^2) }{6}$\\  

            & $(C_{1,1,1,-1})_{H_1}=(C_{1,-1,1,1})_{H_1}=\frac{(4\pi)^2}{B^a_1 B^b_1}\frac{(\pi m_a^4 m_b^4 \beta) }{6}$\\
        
    \hline
        
    $H_2$ & $(A^j_{2,0})_{H_2}=(A^j_{2,0})_{H_1}$\\
               & $(C_{2,2,2,-2})_{H_2}=(C_{2,-2,2,2})_{H_2}=(C_{2,-2,2,2})_{H_1}$\\
            & $(C_{2,1,2,-1})_{H_2}=(C_{2,-1,2,1})_{H_2}=(C_{2,-1,2,1})_{H_1}$\\    
            & $(C_{2,0,2,0})_{H_2}=(C_{2,0,2,0})_{H_1}$\\ 
      
            &$(C_{1,0,1,0})_{H_2}=-(C_{1,0,1,0})_{H_1}$\\  

            &$(C_{1,1,1,-1})_{H_2}=(C_{1,-1,1,1})_{H_2}=-(C_{1,1,1,-1})_{H_1}$\\
        
            \hline
            
         $H_3$&$(A^j_{2,0})_{H_3}=(A^j_{2,0})_{H_1}$\\
        
               & $(C_{2,2,2,-2})_{H_3}=(C_{2,-2,2,2})_{H_3}=\frac{(4\pi)^2}{B^a_2 B^b_2}\frac{(-\pi m_a^4 m_b^4(\beta^2-1)) }{30}$\\ 

            & $(C_{2,0,2,0})_{H_3}=\frac{(4\pi)^2}{B^a_2 B^b_2}\frac{\pi m_a^4 m_b^4(\beta^2-1)}{90}$\\ 
        
            & $(C_{1,0,1,0})_{H_3}=\frac{(4\pi)^2}{B^a_1 B^b_1}\frac{(-\pi m_a^4 m_b^4(\beta^2-1)) }{6}$\\ 
                
         \hline
    
         $H_4$&$(A^j_{2,0})_{H_4}=(A^j_{2,0})_{H_1}$\\
               & $(C_{2,2,2,-2})_{H_4}=(C_{2,-2,2,2})_{H_4}=(C_{2,-2,2,2})_{H_3}$\\ 
            & $(C_{2,0,2,0})_{H_4}=(C_{2,0,2,0})_{H_3}$,\quad\quad$(C_{1,0,1,0})_{H_4}=-(C_{1,0,1,0})_{H_3}$ \\ 
              
         \hline
    \end{tabular}
      \caption{$H$ momentum structures.}
    \label{tab:tab2coeff}
\end{table}

\begin{table}[h]
    \centering
    \begin{tabular}{c|c}
        $\Sigma K$ terms & Non-zero contributions to $A$ or $C$ coefficients  \\
        \hline\hline
         $\Sigma_1 K_1$& $(C_{2,2,2,-2})_{\Sigma_1 K_1}=(C_{2,-2,2,2})_{\Sigma_1 K_1}=\frac{(4\pi)^2}{B^a_2 B^b_2}\frac{(\pi m_a^3 m_b^3\beta) }{15}$\\
               
            & $(C_{2,1,2,-1})_{\Sigma_1 K_1}=(C_{2,-1,2,1})_{\Sigma_1 K_1}=\frac{(4\pi)^2}{B^a_2 B^b_2}\frac{(-\pi m_a^3 m_b^3(\beta^2+1)) }{30}$\\                  
            & $(C_{2,0,2,0})_{\Sigma_1 K_1}=\frac{(4\pi)^2}{B^a_2 B^b_2}\frac{(\pi m_a^3 m_b^3\beta)}{15}$\\ 
        
            & $(C_{1,0,1,0})_{\Sigma_1 K_1}=\frac{(4\pi)^2}{B^a_1 B^b_1}\frac{(-\pi m_a^3 m_b^3\beta) }{3}$\\  

            & $(C_{1,1,1,-1})_{\Sigma_1 K_1}=(C_{1,-1,1,1})_{\Sigma_1 K_1}=\frac{(4\pi)^2}{B^a_1 B^b_1}\frac{(\pi m_a^3 m_b^3(\beta^2+1)) }{6}$\\
    \hline
    
    $\Sigma_2 K_2$ &  $(C_{2,2,2,-2})_{\Sigma_2 K_2}=(C_{2,-2,2,2})_{\Sigma_2 K_2}=(C_{2,-2,2,2})_{\Sigma_1 K_1}$\\
            & $(C_{2,1,2,-1})_{H_2}=(C_{2,-1,2,1})_{\Sigma_2 K_2}=(C_{2,-1,2,1})_{\Sigma_1 K_1}$\\                
            & $(C_{2,0,2,0})_{\Sigma_2 K_2}=(C_{2,0,2,0})_{\Sigma_1 K_1}$\\ 
            & $(C_{1,0,1,0})_{\Sigma_2 K_2}=-(C_{1,0,1,0})_{\Sigma_1 K_1}$\\  
            &$(C_{1,1,1,-1})_{\Sigma_2 K_2}=(C_{1,-1,1,1})_{\Sigma_2 K_2}=-(C_{1,1,1,-1})_{\Sigma_1 K_1}$\\

            \hline
         $\Sigma_3 K_1$& $(A^a_{1,0})_{\Sigma_3 K_1}=\frac{4\pi}{B^a_1}\frac{(4\pi^{3/2}m_a^3 m_b^3\sqrt{\beta^2-1})}{3\sqrt{3}}$ \\
         
        & $(A^b_{1,0})_{\Sigma_3 K_1}=\frac{4\pi}{B^b_1}\frac{(-4\pi^{3/2}m_a^3 m_b^3 \sqrt{\beta^2-1})}{3\sqrt{3}}$\\
         
         & $(C_{1,-1,2,1})_{\Sigma_3 K_1}=(C_{1,1,2,-1})_{\Sigma_3 K_1}=\frac{(4\pi)^2}{B^a_1 B^b_2}\frac{(-\pi m_a^3 m_b^3\beta\sqrt{\beta^2-1})}{6\sqrt{5}}$\\
         
         &$(C_{2,-1,1,1})_{\Sigma_3 K_1}=
        (C_{2,1,1,-1})_{\Sigma_3 K_1}=\frac{(4\pi)^2}{B^a_2 B^b_1}\frac{(\pi m_a^3 m_b^3\beta\sqrt{\beta^2-1})}{6\sqrt{5}}$\\
        
        & $(C_{1,0,2,0})_{\Sigma_3 K_1}=\frac{(4\pi)^2}{B^a_1 B^b_2}\frac{(\pi m_a^3 m_b^3\sqrt{\beta^2-1})}{3\sqrt{15}}$\\
                
        & $(C_{2,0,1,0})_{\Sigma_3 K_1}=\frac{(4\pi)^2}{B^a_2 B^b_1}\frac{(-\pi m_a^3 m_b^3 \sqrt{\beta^2-1})}{3\sqrt{15}}$\\
         \hline
         
 $\Sigma_4 K_2$ & $(A^a_{1,0})_{\Sigma_4 K_2}= (A^a_{1,0})_{\Sigma_3 K_1},\quad\quad (A^b_{1,0})_{\Sigma_4 K_2}=-(A^b_{1,0})_{\Sigma_3 K_1}$\\
 & $(C_{1,-1,2,1})_{\Sigma_4 K_2}=(C_{1,1,2,-1})_{\Sigma_4 K_2}=(C_{1,-1,2,1})_{\Sigma_3 K_1}$\\
 & $(C_{2,-1,1,1})_{\Sigma_4 K_2}=
        (C_{2,1,1,-1})_{\Sigma_4 K_2}=-(C_{2,-1,1,1})_{\Sigma_3 K_1}$\\
& $(C_{1,0,2,0})_{\Sigma_4 K_2}=(C_{1,0,2,0})_{\Sigma_3 K_1}$\\
& $(C_{2,0,1,0})_{\Sigma_4 K_2}=-(C_{2,0,1,0})_{\Sigma_3 K_1}$\\
         \hline
    \end{tabular}
       \caption{$\Sigma K$ momentum structures.}
    \label{tab:tab3coeff}
\end{table}

\begin{table}[h]
    \centering
    \begin{tabular}{c|c}
        $\Pi_\epsilon \Sigma$ terms & Non-zero contributions to $A$ or $C$ coefficients  \\
        \hline\hline
           
         $\Pi_\epsilon \Sigma_1$& $(C_{2,2,2,-2})_{\Pi_\epsilon \Sigma_1}=(C_{2,-2,2,2})^*_{\Pi_\epsilon \Sigma_1}=\frac{(4\pi)^2}{B^a_2 B^b_2}\frac{(-i\pi m_a^3 m_b^3\sqrt{\beta^2-1}) }{15}$\\
               
            & $(C_{2,1,2,-1})_{\Pi_\epsilon \Sigma_1}=(C_{2,-1,2,1})^*_{\Pi_\epsilon \Sigma_1}=\frac{(4\pi)^2}{B^a_2 B^b_2}\frac{(i\pi m_a^3 m_b^3 \beta\sqrt{\beta^2-1}) }{30}$\\                  
            & $(C_{1,1,1,-1})_{\Pi_\epsilon \Sigma_1}=(C_{1,-1,1,1})^*_{\Pi_\epsilon \Sigma_1}=\frac{(4\pi)^2}{B^a_1 B^b_1}\frac{(-i\pi m_a^3 m_b^3\sqrt{\beta^2-1}) }{6}$\\
         
    \hline
    
    $\Pi_\epsilon \Sigma_2$ &  $(C_{2,2,2,-2})_{\Pi_\epsilon \Sigma_2}=(C_{2,-2,2,2})^*_{\Pi_\epsilon \Sigma_2}=(C_{2,2,2,-2})^*_{\Pi_\epsilon \Sigma_1}$\\
               
            & $(C_{2,1,2,-1})_{\Pi_\epsilon \Sigma_2}=(C_{2,-1,2,1})_{\Pi_\epsilon \Sigma_2}^*=(C_{2,1,2,-1})^*_{\Pi_\epsilon \Sigma_1}$\\    

            & $(C_{1,1,1,-1})_{\Pi_\epsilon \Sigma_2}=(C_{1,-1,1,1})^*_{\Pi_\epsilon \Sigma_2}=(C_{1,1,1,-1})_{\Pi_\epsilon \Sigma_1}$\\
            
            \hline
           
         $\Pi_\epsilon \Sigma_3$& $(C_{1,-1,2,1})_{\Pi_\epsilon \Sigma_3}=-(C_{1,1,2,-1})_{\Pi_\epsilon \Sigma_3}=\frac{(4\pi)^2}{B_1^a B_2^b}\frac{(-i\pi m_a^3 m_b^3 (\beta^2-1))}{6\sqrt{5}}$\\
         
         &  $(C_{2,-1,1,1})_{\Pi_\epsilon \Sigma_3}=-(C_{2,1,1,-1})_{\Pi_\epsilon \Sigma_3}=\frac{(4\pi)^2}{B_2^a B_1^b}\frac{(i\pi m_a^3 m_b^2 (\beta^2-1) )}{6\sqrt{5}}$\\
         
         \hline
         
          $\Pi_\epsilon \Sigma_4$ &$(C_{1,-1,2,1})_{\Pi_\epsilon \Sigma_4}=-(C_{1,1,2,-1})_{\Pi_\epsilon \Sigma_4}=-(C_{1,-1,2,1})_{\Pi_\epsilon \Sigma_3}$\\
          & $(C_{2,-1,1,1})_{\Pi_\epsilon \Sigma_4}=-(C_{2,1,1,-1})_{\Pi_\epsilon \Sigma_4}=(C_{2,-1,1,1})_{\Pi_\epsilon \Sigma_3}$\\
          
         \hline
    \end{tabular}
    \caption{$\Pi_\epsilon \Sigma$ momentum structures.}
    \label{tab:tab4ceff}
\end{table}

\begin{table}[h]
    \centering
    \begin{tabular}{c|c}
       $\Pi_\epsilon K$ or $\Sigma\Sigma$ terms & Non-zero contributions to $A$ or $C$ coefficients  \\
        \hline\hline
           
         $\Pi_\epsilon K_1$& $(C_{2,2,2,-2})_{\Pi_\epsilon K_1}=(C_{2,-2,2,2})^*_{\Pi_\epsilon K_1}=\frac{(4\pi)^2}{B^a_2 B^b_2}\frac{(-i\pi m_a^4 m_b^4 \beta \sqrt{\beta^2-1}) }{30}$\\
               
            & $(C_{2,1,2,-1})_{\Pi_\epsilon K_1}=(C_{2,-1,2,1})^*_{\Pi_\epsilon K_1}=\frac{(4\pi)^2}{B^a_2 B^b_2}\frac{(i\pi m_a^4 m_b^4\sqrt{\beta^2-1}) }{60}$\\                  
            & $(C_{1,1,1,-1})_{\Pi_\epsilon K_1}=(C_{1,-1,1,1})^*_{\Pi_\epsilon K_1}=\frac{(4\pi)^2}{B^a_1 B^b_1}\frac{(-i\pi m_a^4 m_b^4\sqrt{\beta^2-1}) }{12}$\\
         
    \hline
    
    $\Pi_\epsilon K_2$ & $(C_{2,2,2,-2})_{\Pi_\epsilon K_2}=(C_{2,-2,2,2})^*_{\Pi_\epsilon K_2}=(C_{2,2,2,-2})^*_{\Pi_\epsilon K_1}$\\
            &  $(C_{2,1,2,-1})_{\Pi_\epsilon K_2}=(C_{2,-1,2,1})_{\Pi_\epsilon K_2}^*=(C_{2,1,2,-1})^*_{\Pi_\epsilon K_1}$\\    
            & $(C_{1,1,1,-1})_{\Pi_\epsilon K_2}=(C_{1,-1,1,1})^*_{\Pi_\epsilon K_2}=(C_{1,1,1,-1})_{\Pi_\epsilon K_1}$\\
            
            \hline
           
         $\Sigma_1 \Sigma_3$ & $(A^a_{1,0})_{\Sigma_1\Sigma_3}=\frac{4\pi}{B^a_1}\frac{(8\pi^{3/2}m_a^2 m_b^2\beta \sqrt{\beta^2-1})}{3\sqrt{3}}$ \\
         
        & $(A^b_{1,0})_{\Sigma_1\Sigma_3}=\frac{4\pi}{B^b_1}\frac{(-8\pi^{3/2}m_a^2 m_b^2 \beta \sqrt{\beta^2-1})}{3\sqrt{3}}$\\
         
         & $(C_{1,-1,2,1})_{\Sigma_1\Sigma_3}=(C_{1,1,2,-1})_{\Sigma_1\Sigma_3}=\frac{(4\pi)^2}{B^a_1 B^b_2}\frac{(-\pi m_a^2 m_b^2\sqrt{\beta^2-1})}{3\sqrt{5}}$\\
         
         &$(C_{2,-1,1,1})_{\Sigma_1\Sigma_3}=
        (C_{2,1,1,-1})_{\Sigma_1\Sigma_3}=\frac{(4\pi)^2}{B^a_2 B^b_1}\frac{(\pi m_a^2 m_b^2 \sqrt{\beta^2-1})}{3\sqrt{5}}$\\
        
        & $(C_{1,0,2,0})_{\Sigma_1\Sigma_3}=\frac{(4\pi)^2}{B^a_1 B^b_2}\frac{(2\pi m_a^2 m_b^2\beta \sqrt{\beta^2-1})}{3\sqrt{15}}$\\
                
        & $(C_{2,0,1,0})_{\Sigma_1\Sigma_3}=\frac{(4\pi)^2}{B^a_2 B^b_1}\frac{(-2\pi m_a^2 m_b^2 \beta \sqrt{\beta^2-1})}{3\sqrt{15}}$\\
        
         \hline
         
  $\Sigma_2 \Sigma_4$ & $(A^a_{1,0})_{\Sigma_2\Sigma_4}=(A^a_{1,0})_{\Sigma_1\Sigma_3}$\\
  &$(A^b_{1,0})_{\Sigma_2\Sigma_4}=-(A^b_{1,0})_{\Sigma_1\Sigma_3}$\\
  & $(C_{1,-1,2,1})_{\Sigma_2\Sigma_4}=(C_{1,1,2,-1})_{\Sigma_2\Sigma_4}=C_{1,-1,2,1})_{\Sigma_1\Sigma_3}$\\
  &$(C_{2,-1,1,1})_{\Sigma_2\Sigma_4}=
        (C_{2,1,1,-1})_{\Sigma_4\Sigma_4}=-(C_{2,-1,1,1})_{\Sigma_1\Sigma_3}$\\
        & $(C_{1,0,2,0})_{\Sigma_2\Sigma_4}=(C_{1,0,2,0})_{\Sigma_1\Sigma_3}$\\
          & $(C_{2,0,1,0})_{\Sigma_2\Sigma_4}=-(C_{2,0,1,0})_{\Sigma_1\Sigma_3}$\\
         
         \hline
    \end{tabular}
       \caption{$\Pi_\epsilon K$ or $\Sigma\Sigma$ momentum structures.}
    \label{tab:tab5coeff}
\end{table}

\section{SMEFT operators leading to the anomalous $HVV$ couplings}
\label{appendixSMEFTOp}

The effective couplings of the Higgs Characterisation model that have been analysed in Sec.~\ref{sec:EFTnum} can be related to  SMEFT operators in the Warsaw basis~\cite{Grzadkowski:2010es}. In Tab.~\ref{Tab:EFT} we list a few examples, denoting directly the  effective couplings with the corresponding $\kappa$'s.

\begin{table}[!h]
    \centering
    \begin{tabular}{c|c|c}
    $\kappa$& CP & SMEFT Operators \\
    \hline
    \hline
    $\kappa_{\rm SM}$ & even & $\mathcal{O}_{\varphi D} ,\quad \mathcal{O}_{\varphi \Box}$  \\
    $\kappa_{H\gamma\gamma}$ & even & $\mathcal{O}_{\varphi W} ,\quad
    \mathcal{O}_{\varphi B} , \quad \mathcal{O}_{\varphi WB}$ \\
    $\kappa_{HZ\gamma}$ & even & $\mathcal{O}_{\varphi W},\quad \mathcal{O}_{\varphi B},\quad \mathcal{O}_{\varphi WB}$ \\
    $\kappa_{HZZ}$ & even & $\mathcal{O}_{\varphi W},\quad \mathcal{O}_{\varphi B},\quad \mathcal{O}_{\varphi WB}$ \\
    $\kappa_{AZZ}$ & odd & ${\mathcal{O}}_{\varphi \widetilde{W}},\quad {\mathcal{O}}_{\varphi \widetilde{B}},\quad {\mathcal{O}}_{\varphi \widetilde{W}B}$ \\
    \end{tabular}
    \caption{SMEFT operators~\cite{Grzadkowski:2010es} related to each Higgs Characterisation effective coupling considered in Sec.~\ref{sec:EFTnum}.
    \label{Tab:EFT}}
\end{table}

\bibliographystyle{JHEP}
\bibliography{refs}

\end{document}